\documentclass[11pt,a4,twoside]{article}
\usepackage[dvips,a4paper,left=3cm,right=3cm,top=2.1cm,bottom=2.1cm,headsep=1em]{geometry}
\usepackage{fancyhdr}
\usepackage{titlesec}
\usepackage[latin1]{inputenc}
\usepackage{amsmath,amsfonts}
\usepackage{rotating}
\usepackage[font={small}]{caption}
\usepackage{natbib}
\usepackage{lmodern,slantsc}
\usepackage[breaklinks,
colorlinks=true, linkcolor=blue, citecolor=black, urlcolor=blue,
pdfborder={0 0 0}, pdfpagelabels]{hyperref}
\usepackage{tikz}
\def\vx{\mathbf x}

\def\vc{\mathbf c}

\def\vf{\mathbf f}
\def\vk{\mathbf k}
\def\vn{\mathbf n}

\def\vu{\mathbf u}
\def\vv{\mathbf v}

\def\vA{\mathbf A}
\def\vB{\mathbf B}

\def\vF{\mathbf F}

\def\vH{\mathbf H}

\def\vR{\mathbf R}

\def\vT{\mathbf T}
\def\vU{\mathbf U}

\def\v0{\boldsymbol{0}}
\def\vnabla{\boldsymbol{\nabla}}
\def\valpha{\boldsymbol{\alpha}}

\def\vrho{\boldsymbol{\rho}}

\def\vtau{\boldsymbol{\tau}}

\def\L{\langle}
\def\R{\rangle}
\newlength{\FigureHeight}
\newlength{\FigureHeightHalf}
\newcommand{\FigureXYLabel}[5]
{\settoheight{\FigureHeight}{#1}
\setlength{\FigureHeightHalf}{0.5\FigureHeight}
\begin{center}
\raisebox{\FigureHeightHalf}{\makebox{#4\makebox[#5]{}}}
#1\\
\vspace{#3}
#2\\
\end{center}}
\numberwithin{equation}{section}
\titleformat{\section}
{\large\bfseries}{\thetitle.}{0.5em}{}
\titleformat{\subsection}
{\normalfont\itshape\filcenter}{\normalfont\thetitle.}{0.5em}{}
\begin{document}

\title{On the physical inconsistency of a new statistical scaling
symmetry in incompressible Navier-Stokes turbulence}
\author{M. Frewer$\,^1$\thanks{Email address for correspondence:
frewer.science@gmail.com}$\:\,$, G. Khujadze$\,^2$ \& H. Foysi$\,^2$\\ \\
\small $^1$ Tr\"ubnerstr. 42, 69121 Heidelberg, Germany\\
\small $^2$ Chair of Fluid Mechanics, Universit\"at Siegen, 57068
Siegen, Germany}
\date{{\small\today}}
\clearpage \maketitle \thispagestyle{empty}

\begin{abstract} \noindent A detailed theoretical investigation
is given which demonstrates that a recently proposed statistical
scaling symmetry is physically void. Although this scaling is
mathematically admitted as a unique symmetry transformation by the
underlying statistical equations for incompressible Navier-Stokes
turbulence on the level of the functional Hopf equation, by closer
inspection, however, it leads to physical inconsistencies and
erroneous conclusions in the theory of
turbulence.\footnote[2]{This present investigation has been
peer-reviewed by four independent referees. Their reports have
been\linebreak \indent\indent\hspace{-1.5mm} published in
\cite{Frewer14.0}.}

The new statistical symmetry is thus misleading in so far as it
forms within an unmodelled theory an analytical result which at
the same time lacks physical consistency. Our investigation will
expose this inconsistency on different levels of statistical
description, where on each level we will gain new insights for its
non-physical transformation behavior. With a view to generate
invariant turbulent scaling laws, the consequences will be
fin\-ally discussed when trying to analytically exploit such a
symmetry. In fact, a mismatch between theory and numerical
experiment is conclusively~quantified.

We ultimately propose a general strategy on how to not only track
unphysical statistical symmetries, but also on how to avoid
generating such misleading invariance results from the outset. All
the more so as this specific study on a physically inconsistent
scaling symmetry only serves as a representative example within
the broader context of statistical invariance analysis. In this
sense our investigation is applicable to all areas of statistical
physics in which symmetries get determined in order to either
characterize complex dynamical systems, or in order to extract
physically useful and meaningful information from the underlying
dynamical process itself.

\vspace{0.5em}\noindent{\footnotesize{\bf Keywords:} {\it
Symmetries and Equivalences, Lie Groups, Scaling Laws,
Deterministic and Statistical Systems, Principle of Causality,
Turbulence, Closure Problem, Boundary Layer~Flow}}$\,$;
\end{abstract}

\newpage
\thispagestyle{empty}
\tableofcontents
\newpage
\pagenumbering{arabic}\setcounter{page}{1}

\section{Introduction\label{Sec1}}

With the aid of today's modern computer algebra systems, the
method of symmetry analysis is one of the most prominent and
efficient tools to investigate differential equations arising in
various sciences
\citep{Ovsiannikov82,Stephani89,Olver93,Ibragimov94,Andreev98,
Bluman96,Meleshko05}. A considerable number of special techniques
for simplifying, reducing, mapping and solving differential
equations have been developed and enhanced so far.

The natural language for symmetry transformations is that of a
mathematical group, which either can be discrete or continuous. If
an invariant transformation group involves one or more parameters
which can vary continuously it is called a Lie symmetry group,
named after Sophus Lie who first developed the theory of
continuous transformation groups at the end of the nineteenth
century \citep{Lie93}.

In fact, most differential equations of the sciences possess
nontrivial Lie symmetry groups. Under favorable conditions these
symmetries can be exploited for various purposes, e.g. performing
integrability tests and complete integration of ODEs, finding
invariant and asymptotic solutions for ODEs and PDEs, constructing
conservation laws and dynamical invariants, etc. Not to forget
that Lie-groups are also successfully utilized in the `opposite'
direction in modelling dynamical behavior itself, i.e.~used for
constructing dynamical equations which should admit a certain
given set of symmetries. The most impressive results to date were
gained by gauge theory for quantum fields
\citep{Weinberg00,Penrose05}. Hence the existence of symmetries
thus has a profound and far-reaching impact on solution properties
and modelling of differential equations in general. Their presence
very often simplifies our understanding of physical phenomena.

Of particular interest are {\it scaling} symmetries as they lead
to concepts as scale invariance of dynamical laws or
self-similarity of solution manifolds. A scaling symmetry of a
physical system can either be associated with a finite dimensional
Lie-group (global scaling symmetry) in which all group parameters
are strict constants, or with an infinite dimensional Lie-group
(local scaling symmetry) in which at least one group parameter is
not constant, e.g. by showing a space-time coordinate dependence
of the considered system.

Most physical processes, however, only admit global scaling
symmetries since the requirement for a local scaling symmetry is
too restrictive. In fact, a physical process which admits a local
scaling symmetry also admits this symmetry globally. For example,
for a local scaling symmetry exhibiting space-time dependent group
parameters (which essentially forms the cornerstone of every
quantum gauge theory) the corresponding global symmetry is then
just given by the same symmetry where only the group parameters
are identically fixed at every point in space-time. The opposite,
in which a global symmetry automatically implies a local symmetry,
is, of course, not the rule.

The purpose of this article is to show that in general caution has
to be exercised when interpreting and exploiting symmetries if
they act in a purely {\it statistical} manner. Although being
mathematically admitted as statistical symmetries by the
underlying statistical system of dynamical equations, they
nevertheless can lead to physical inconsistencies. Without loss of
generality, we will demonstrate this issue at the example of a new
and recently proposed global statistical scaling symmetry for the
incompressible Navier-Stokes equations. Our study and its
conclusion can then be easily transferred to any other statistical
symmetry within the Navier-Stokes theory, or, more generally, to
any other theory within physics which necessitates a statistical
description in the thermodynamical sense.

The current study is organized as follows: Section \ref{Sec2}
opens the investigation by introducing the single and only
continuous (Lie-point) scaling {\it symmetry} which the {\it
deterministic} incompressible Navier-Stokes equations can admit.
Although being the only true scaling {\it symmetry}, it is yet not
the only scaling transformation which leaves these equations
invariant when viewed in a broader context. Regarding the class of
all possible invariant Lie-point scaling transformations, a brief
outline is given to distinguish between the concept of a {\it
symmetry} transformation and that of an {\it equivalence}
transformation. A careful distinction between these two concepts
is surely necessary in to order fully grasp the spirit of this
article.

Section \ref{Sec3} then changes from the deterministic to the
statistical description. By choosing the functional Hopf
formulation we are dealing with a formally closed and thus
complete statistical approach to turbulence. Instead of the weaker
invariant class of equivalence transformations, this enables us to
generate true statistical {\it symmetry} transformations, in
particular a new scaling symmetry is considered which first got
mentioned in the study of \cite{Oberlack13X}.

Section \ref{Sec4} is at the heart of the article's line of
reasoning. It not only demonstrates that the new Hopf scaling
symmetry induces a disguised symmetry, which, on a lower level of
statistical description, only acts as an equivalence
transformation, but also gives a mathematical proof that both the
Hopf symmetry and its induced equivalence transformation are
essentially unphysical.

Section \ref{Sec5} presents the consequences when generating
statistical scaling laws from such a misleading symmetry
transformation. These laws will be matched to DNS data at the
example of a zero-pressure-gradient (ZPG) turbulent boundary layer
flow for the high Reynolds number case of $Re_\theta=6000$ (based
on the momentum thickness $\theta$ of the flow). Best curve fits
are generated with the aid of using basic tools from statistical
data analysis, as the chi-square method to quantitatively measure
the quality of the fits relative to the underlying DNS error. As a
result, a mismatch between theory and numerical experiment is
clearly quantified.

Section \ref{Sec6} concludes and completes the investigation.
Theoretically as well as graphically we will conclude that all
recently proposed statistical scaling laws which are based on this
new unphysical symmetry have no predictive value and, in our
opinion, should be discarded to avoid any further misconceptions
in future work when generating turbulent scaling laws according to
the invariance method of Lie-groups. In a brief historical outline
we finally point out that even if this method of Lie-groups in its
full extent is applied and interpreted correctly, it nevertheless
faces strong natural limits which prevents the effect of achieving
a significant breakthrough in the theory of turbulence.

A large but indispensable part of this investigation has been
devoted to the appendix. All appendices stand for their own and
can be read independently from the main text. In particular
Appendix \ref{SecA} \& \ref{SecC} are written in the form of a
compendium to serve as an aid and to accompany the reader through
the main text. Their purpose is to mathematically support the
criticism we put forward in our first part, the theoretical part
of our study from Section \ref{Sec2} to Section \ref{Sec4}.

\section{The deterministic incompressible Navier-Stokes equations\label{Sec2}}

For reasons of simplicity we will in the following only consider
the general solution ma\-ni\-fold of the incompressible
Navier-Stokes equations in the infinite domain without specifying
any initial or boundary conditions
\citep{Batchelor67,Pope00,Davidson04}.

The corresponding deterministic equations can either be written in
local differential form as
\begin{equation}
\left. \begin{aligned}
\nabla\cdot \vu =0,\text{\hspace{1.51cm}}\\
\partial_t \vu+(\vu\cdot\nabla)\vu=-\nabla p+\nu\Delta\vu,
\end{aligned}
~~~ \right \} \label{140529:1736}
\end{equation}
or equivalently, when using the continuity equation to eliminate
the pressure from the momentum equations, in nonlocal
integro-differential form as
\begin{equation}
\partial_t
\vu+(\vu\cdot\nabla)\vu=-\nabla\int\frac{\nabla^\prime\cdot
\big[(\vu^\prime\cdot\nabla^\prime)\vu^\prime\big]}{4\pi\cdot\vert
\vx-\vx^\prime\vert}d^3\vx^\prime +\nu\Delta\vu.
\label{140529:1657}
\end{equation}
By construction, equation (\ref{140529:1657}) has the property
that if the initial velocity field $\vu$ is solenoidal, i.e. if
$\nabla\cdot \vu$ is initially zero, then it will be solenoidal
for all times.

The single and only continuous (Lie-point) scaling symmetry which
the deterministic incompressible Navier-Stokes equations
(\ref{140529:1736}), or in the form (\ref{140529:1657}), can admit
is given by \citep{Olver93,Fushchich93,Frisch95,Andreev98}
\begin{align}
\mathsf{S}: & \;\;\; \tilde{t}=e^{2\varepsilon}t,\;\;\;
\tilde{\vx}=e^{\varepsilon}\vx,\;\;\;
\tilde{\vu}=e^{-\varepsilon}\vu,\;\;\;\tilde{p}=e^{-2\varepsilon}p,
\label{130813:1900}
\end{align}
being just a global scaling symmetry with constant group parameter
$\varepsilon$. That (\ref{130813:1900}) really acts as a symmetry
transformation can be easily verified due its globally uniform
structure: By inserting transformation (\ref{130813:1900}) into
system (\ref{140529:1736}), or into (\ref{140529:1657}), will
leave the equations in each case fully indifferent.

Before we turn in the next section to a complete (fully
determined) statistical description of the Navier-Stokes
equations, it is essential at this stage to make a careful
distinction between two different kinds of invariant
transformations. Those being true {\it symmetry} transformations
and those being only {\it equivalence} transformations
\citep{Ovsiannikov82,Ibragimov94,Ibragimov04}.

Although both types of invariant transformations form a Lie-group,
they each have a completely different impact when trying to
extract valuable information from a given dynamical system. The
knowledge of symmetry transformations is mainly used to construct
special or general solutions of differential equations, while
equivalence transformations are used to solve the equivalence
problem for a certain class of differential equations by group
theory, that is, to find general criteria whether two or more
different differential equations are connected by a change of
variables drawn from a transformation group. Hence, the quest for
a symmetry transformation is thus fundamentally different to that
for an equivalence transformation. The difference between these
two kinds of transformations is defined as:

\begin{itemize}
\item A {\it symmetry} of a differential equation is a
transformation which maps every solution of the differential
equation to another solution of the {\it same equation}. As a
consequence a symmetry transformation leads to complete form-{\it
indifference} of the equation. It results as an invariant
transformation if the considered equation is {\it
closed}.\footnote[2]{A set of equations is defined as {\it closed}
if the number of equations involved is either equal to or more
than the number of dependent variables to be solved for.}
\item An {\it equivalence transformation} for a differential
equation in a given class is a change of variables which only maps
the equation to another equation in the {\it same class}. As a
consequence an equivalence transformation {\it only} leads to a
weaker form-{\it invariance} of the equation. It results as an
invariant transformation either if existing parameters of the
considered equation get identified as own independent variables,
or if the considered equation itself is {\it
unclosed}.\footnote[3]{A set of equations is defined as {\it
unclosed} if the number of equations involved is less than the
number of unknown dependent variables.}
\end{itemize}

\noindent Hence, although both transformations are {\it invariant}
transformations and both form a {\it Lie-group}, they yet lead to
different implications. Let us illustrate this decisive difference
at two simple examples:

{\bf Example 1:} By considering the viscosity $\nu$ in
(\ref{140529:1736}) not as a parameter, but rather, next to the
space-time coordinates, as an own independent variable, a detailed
invariance analysis will give the following additional scaling
group, which in infinitesimal form reads as \citep{Unal94,Unal95}
\begin{align}
\mathsf{X}_{\mathsf{E_1}(\mathsf{f})}: & \;\;\;
f(\nu)\cdot\Big(t\partial_t+x^i\partial_{x^i}+\nu\partial_\nu\Big),
\label{140119:1743}
\end{align}
being an infinite dimensional Lie-group with a group parameter $f$
depending on the viscosity variable $\nu$. Specifying for example
$f(\nu)=1$ will reduce to a finite dimensional subgroup, for which
the non-infinitesimal form can then be explicitly determined to
\begin{align}
\mathsf{E}_\mathsf{1}: & \;\;\;
\tilde{t}=e^{\varepsilon_1}t,\;\;\;
\tilde{\vx}=e^{\varepsilon_1}\vx,\;\;\; \tilde{\vu}=\vu,\;\;\;
\tilde{p}=p,\;\;\; \tilde{\nu}=e^{\varepsilon_1}\nu.
\label{140529:2033}
\end{align}
Hence, just by considering the viscosity, or alternatively the
Reynolds number $Re\sim 1/\nu$, as an own independent variable, we
see that next to the global scaling symmetry $\mathsf{S}$
(\ref{130813:1900}) we gained an additional global scaling
invariance $\mathsf{E}_\mathsf{1}$ (\ref{140529:2033}): The
viscosity as well as the space-time coordinates scale in exactly
the same manner respective to the constant group parameter
$\varepsilon_1$. However, this additional invariant transformation
(\ref{140529:2033}) does not act as a true symmetry, but only in
the weaker sense as an equivalence transformation, in that it only
maps the Navier-Stokes equation in the class of different
viscosities to another equation in that same class. Indeed,
inserting transformation (\ref{140529:2033}) into form
(\ref{140529:1736}), will not leave it form-{\it indifferent}, but
only form-{\it invariant}
\begin{equation}
\left. \begin{aligned}
\tilde{\nabla}\cdot \tilde{\vu} =0,\text{\hspace{1.51cm}}\\
\partial_{\tilde{t}} \tilde{\vu}+(\tilde{\vu}\cdot\tilde{\nabla})\tilde{\vu}
=-\tilde{\nabla}\tilde{p}+\tilde{\nu}\tilde{\Delta}\tilde{\vu},
\end{aligned}
~~~ \right \} \label{140529:2208}
\end{equation}
since the parametric value changed to $\tilde{\nu}\neq \nu$.
Particularly in this simple case, however, we can alternatively
also say that transformation (\ref{140529:2033}) actually maps a
solution of equation (\ref{140529:1736}), with a certain value in
viscosity $\nu$, to another solution of the same equation
(\ref{140529:2208}), but with a different value in viscosity
$\tilde{\nu}$. Yet, note that irrespective of the functional
choice for the continuous group parameter $f$, the invariant
transformation (\ref{140119:1743}) will never reduce to a true
symmetry transformation. Every specific functional choice of $f$
will give a different global equivalence scaling transformation.

{\bf Example 2:} Taking the statistical ensemble average of the
deterministic Navier-Stokes equations in the form
(\ref{140529:1736}), we get, due to the existence of the nonlinear
convective term, the following unclosed (underdetermined) set of
equations \citep{Pope00,Davidson04}
\begin{equation}
\left. \begin{aligned}
\nabla\cdot \L\vu\R =0,\text{\hspace{1.51cm}}\\
\partial_t \L\vu\R+\nabla\cdot\vT=-\nabla \L p\R+\nu\Delta\L\vu\R,
\end{aligned}
~~~ \right \} \label{140529:2236}
\end{equation}
where the second rank tensor $\vT=\L\vu\otimes\vu\R$ is the
unclosed second velocity moment based on the full instantaneous
velocity field $\vu$. In the most general case $\vT$ is to be
identified as an unknown and thus arbitrary functional of the
space-time coordinates $(\vx,t)$ and of the mean fields of
velocity $\L\vu\R$ and pressure $\L p\R$ along with its
spatiotemporal variations, either in local, nonlocal or mixed
form.

For reasons of simplicity let us consider $\vT$ for the moment as
an arbitrary function which only shows an explicit dependence on
the space-time coordinates, i.e. $\vT=\vT(\vx,t)$. If we now
perform an invariance analysis of the underdetermined system
(\ref{140529:2236}), by extending, next to the mean velocity
$\L\vu\R$ and the mean pressure $\L p\R$, the list of dependent
variables with the unclosed and thus arbitrary function $\vT$ as
an own dependent variable, we immediately gain the following
invariant {\it statistical} scaling\footnote[2]{Note that in the
general case a careful distinction must be made between the
transformed expression $\widetilde{\L\vu\R}$, which directly
refers to the transformed mean velocity field, and the transformed
expression $\L\tilde{\vu}\R$, which, on the other hand, refers to
the transformed instantaneous (fluctuating) velocity field being
averaged only after its transformation. However, in the specific
and simple case as (\ref{140529:2336}) both transformed fields are
identical $\widetilde{\L\vu\R}=\L\tilde{\vu}\R$. The obvious
reason is that since transformation $\mathsf{E}_\mathsf{2}$
(\ref{140529:2336}) only represents a globally uniform scaling
with the constant factor $e^{\varepsilon_2}$, it will commute with
every averaging operator $\L,\R$ (for a more detailed discussion
on this subject, see Appendix \ref{SecD.1}).}
\begin{align}
\mathsf{E}_\mathsf{2}: & \;\;\; \tilde{t}=t,\;\;\;
\tilde{\vx}=\vx,\;\;\;
\L\tilde{\vu}\R=e^{\varepsilon_2}\L\vu\R,\;\;\;
\tilde{\vT}=e^{\varepsilon_2}\vT,\;\;\;
\L\tilde{p}\R=e^{\varepsilon_2}\L p\R, \label{140529:2336}
\end{align}
which globally only scales the system's dependent variables while
the coordinates stay invariant. It is clear that this invariant
transformation cannot act as a symmetry transformation. It can
only act in the weaker sense as an equivalence transformation,
since in the considered functional class of arbitrary second
moment functions $\vT=\vT(\vx,t)$ it only maps the unclosed first
moment equation (\ref{140529:2236}) into another equation of the
same class:
\begin{equation}
\left. \begin{aligned}
\tilde{\nabla}\cdot \L\tilde{\vu}\R =0,\text{\hspace{1.51cm}}\\
\partial_{\tilde t} \L\tilde{\vu}\R+\tilde{\nabla}\cdot\tilde{\vT}
=-\tilde{\nabla} \L \tilde{p}\R+\nu\tilde{\Delta}\L\tilde{\vu}\R,
\end{aligned}
~~~ \right \} \label{140530:0028}
\end{equation}
where the unclosed and thus arbitrary function $\vT$ itself gets
mapped to a new and different, but still unclosed and thus
arbitrary function $\tilde{\vT}\neq \vT$. However, since
$\tilde{\vT}$ is from the same considered functional class as
$\vT$, it thus also exhibits an explicit dependence only on the
coordinates: $\tilde{\vT}=\tilde{\vT}(\tilde{\vx},\tilde{t})$.

Again, the invariant transformation (\ref{140529:2336}) only
represents an equivalence and {\it not} a symmetry transformation
of the unclosed system (\ref{140529:2236}), since it turns this
system only into an equivalent but {\it not} identical form. To
see this explicitly, imagine we would specify the unclosed moment
function $\vT=\vT(\vx,t)$, say by
\begin{equation}
\vT(\vx,t)=\nabla \phi(\vx)\otimes \nabla
\phi(\vx),\;\;\text{with}\;\; \phi(\vx)=e^{-\vx^2}.
\label{140922:1124}
\end{equation}
Then according to (\ref{140529:2336}) the transformed moment is
{\it defined} or given by
\begin{equation}
\left. \begin{aligned} \tilde{\vT}(\tilde{\vx},\tilde{t})
& = e^{\varepsilon_2}\vT(\vx,t)\\
&= e^{\varepsilon_2} \Big(\nabla \phi(\vx)\otimes \nabla \phi(\vx)\Big) \\
&= e^{\varepsilon_2} \Big(\tilde{\nabla} \phi(\tilde{\vx})\otimes
\tilde{\nabla} \phi(\tilde{\vx})\Big),\;\;\text{since
$\tilde{\vx}=\vx$.}
\end{aligned}
~~~ \right \}\label{r10}
\end{equation}
Hence, while system (\ref{140529:2236}) turns into the closed form
\begin{equation}
\left. \begin{aligned}
\nabla\cdot \L\vu\R =0,\text{\hspace{2.85cm}}\\
\partial_t \L\vu\R+\nabla\cdot\big[\nabla \phi(\vx)\otimes \nabla
\phi(\vx)\big]=-\nabla \L p\R+\nu\Delta\L\vu\R,
\end{aligned}
~~~ \right \}\label{r11}
\end{equation}
the transformed system (\ref{140530:0028}), according to
\eqref{r10}, will turn into
\begin{equation}
\left. \begin{aligned}
\tilde{\nabla}\cdot \L\tilde{\vu}\R =0,\text{\hspace{2.85cm}}\\
\partial_{\tilde t} \L\tilde{\vu}\R
+e^{\varepsilon_2}\tilde{\nabla}\cdot\big[\tilde{\nabla}
\phi(\tilde{\vx})\otimes \tilde{\nabla} \phi(\tilde{\vx})\big]
=-\tilde{\nabla} \L \tilde{p}\R+\nu\tilde{\Delta}\L\tilde{\vu}\R,
\end{aligned}
~~~ \right \}\label{r12}
\end{equation}
which obviously, due to the explicit factor $e^{\varepsilon_2}$,
is {\it not identical} to the corresponding untransformed
differential system \eqref{r11}. Instead, we can only say that
system \eqref{r12} is {\it equivalent} to system \eqref{r11} in
that they originate from the {\it same} class of functions
$\tilde{\vT}$ and $\vT$ which both only show an explicit
dependence on the coordinates.

This of course stands in strong contrast to any given symmetry
transformation of a closed system. For example, the scaling
symmetry $\mathsf{S}$ (\ref{130813:1900}) of the deterministic
Navier-Stokes equations (\ref{140529:1736}), which, if we would
specify a certain solution $\vu=\vu_0$ and $p= p_0$, it will be
mapped according to $\mathsf{S}$ (\ref{130813:1900}) to another
solution $\vu_0\rightarrow \tilde{\vu}=\tilde{\vu}_0$ and
$p_0\rightarrow \tilde{p}=\tilde{p}_0$ of the {\it same} and thus
to (\ref{140529:1736}) identical equation:
\begin{equation}
\left. \begin{aligned}
\tilde{\nabla}\cdot \tilde{\vu} =0,\text{\hspace{1.51cm}}\\
\partial_{\tilde{t}} \tilde{\vu}+(\tilde{\vu}\cdot\tilde{\nabla})
\tilde{\vu}=-\tilde{\nabla}
\tilde{p}+\nu\tilde{\Delta}\tilde{\vu}.
\end{aligned}
~~~ \right \} \label{140922:1053}
\end{equation}
Furthermore, the statistical symmetry
\begin{align}
\mathsf{S}: & \;\;\; \tilde{t}=e^{2\varepsilon}t,\;\;\;
\tilde{\vx}=e^\varepsilon\vx,\;\;\;
\L\tilde{\vu}\R=e^{-\varepsilon}\L\vu\R,\;\;\;
\tilde{\vT}=e^{-2\varepsilon}\vT,\;\;\;
\L\tilde{p}\R=e^{-2\varepsilon}\L p\R, \label{140922:1106}
\end{align}
which corresponds to $\mathsf{S}$ (\ref{130813:1900}) when
reformulated for the mean fields up to the second velocity moment,
leaves only the unclosed system (\ref{140529:2236}) invariant, but
{\it not} the specified closed system (\ref{r11}). That means that
the specification (\ref{140922:1124}) is not compatible with the
statistical symmetry $\mathsf{S}$ (\ref{140922:1106}), thus
showing that the specific functional choice (\ref{140922:1124}) on
the averaged level is inconsistent to the underlying deterministic
(fluctuating) level (\ref{140529:1736}). In strong contrast to the
statistical equivalence transformation $\mathsf{E}_\mathsf{2}$
(\ref{140529:2336}) which is compatible to both the {\it
unspecified} system (\ref{140529:2236}) and the {\it specified}
system (\ref{r11}).

This explicit demonstration clearly shows that a Lie {\it
symmetry} transformation induces a far more stronger invariance
than a Lie {\it equivalence} transformation. Hence, the
consequences which can be drawn from a symmetry transformation are
by far more richer than for any equivalence transformation.

Three things should be noted here. Firstly, since the
transformation (\ref{140529:2336}) only scales the system's
dependent variables by keeping the coordinates invariant, it is a
typical scaling invariance which only linear systems can admit.
Indeed, due to the identification of the unclosed function $\vT$
as an own dependent variable, we turned the underdetermined
statistical system (\ref{140529:2236}) formally into a linear set
of equations. As we will discuss in more detail in the next
sections, such an identification is misleading, since it is hiding
essential information about the underlying deterministic theory.
In other words, although transformation (\ref{140529:2336})
correctly acts as a mathematical equivalence transformation for
the statistical system (\ref{140529:2236}), we will demonstrate
that it nevertheless leads to a physical inconsistency.

Secondly, the type and particular structure of an equivalence
transformation strongly depends on the explicit variable
dependence of $\vT$ itself. Allowing for various different
functional dependencies, as e.g. for $\vT=\vT(\vx,t;\vu)$, or more
generally for $\vT=\vT(\vx,t;\vu_{\{n\}})$ where $\vu_{\{n\}}$
denotes the collection of functions $\vu$ together with all their
derivatives up to order $n$, can cause different equivalence
groups in each case
\citep{Meleshko96,Ibragimov04,Bila11,Chirkunov12}.

Thirdly, the equivalence transformation (\ref{140529:2336}) given
in this example has a much weaker impact when trying to extract
information from the solution manifold of its underlying dynamical
set of equations than the equivalence transformation
(\ref{140529:2033}) given in the previous example. In contrast to
$\mathsf{E}_\mathsf{1}$ (\ref{140529:2033}), which at least could
map between specific solutions of different viscosity, the
equivalence transformation $\mathsf{E}_\mathsf{2}$
(\ref{140529:2336}) is completely unable to map between specific
solutions. The reason is that the considered system of equations
is unclosed\linebreak and thus underdetermined, however not
arbitrarily, but in the specified sense that the unclosed term
$\vT=\L\vu\otimes\vu\R$ can be physically and uniquely determined
from the underlying but analytically non-accessible deterministic
velocity field $\vu$. In other words, this circumstance, in having
an underlying theory from which the unclosed term $\vT$ physically
emerges, opens the high possibility that physical solutions get
mapped into unphysical ones when employing an equivalence
transformation as $\mathsf{E}_\mathsf{2}$ (\ref{140529:2336}).
This problem will be~discussed~next.

\vspace{-0.05em}
\subsection{The concept of an invariant solution\label{Sec2.1}}

In order to understand and recognize the subtle difference between
a symmetry and an equivalence transformation in its full spectrum,
we will discuss this difference again, however, from a different
perspective, from the perspective of generating invariant
solutions.

First of all, one should recognize that the Lie algorithm to
generate invariant transformations for differential equations can
be equally applied in the same manner without any restrictions to
{\it under-}, {\it fully-} as well as {\it overdetermined} systems
of equations
\citep{Ovsiannikov82,Stephani89,Olver93,Ibragimov94,Andreev98,
Bluman96,Meleshko05}, even if the considered system is infinite
dimensional \citep{Frewer15.1,Frewer15.2}.~However, only for {\it
fully} or {\it overdetermined} systems these invariant\linebreak
Lie transformations are called and have the effect of
\emph{symmetry} transformations, while for {\it underdetermined}
systems these invariant Lie transformations are called and have
the effect of \emph{equivalence} transformations.

In other words, although both a symmetry as well as an equivalence
transformation form a Lie-group which by construction leave the
considered equations invariant, the action and the consequence of
each transformation is absolutely different. While a {\it
symmetry} transformation always maps a solution to another
solution of the {\it same equation}, an\linebreak
equivalence\hfill transformation,\hfill in contrast,\hfill
generally\hfill only\hfill maps\hfill a\hfill {\it possible}\hfill
solution\hfill of\hfill one
\newpage
\newgeometry{left=3cm,right=3cm,top=2.1cm,bottom=1.85cm,headsep=1em}
\noindent underdetermined equation to a {\it possible} solution of
{\it another underdetermined equation}, where in the latter case
we assume of course that a solution of an underdetermined equation
can be somehow constructed or is somehow given beforehand.

Now, it is clear that {\it if} for an unclosed and thus
underdetermined equation, or a set of equations, the unclosed
terms are {\it not} correlated to an existing underlying theory,
then the construction of an invariant solution will only be a
particular and non-privileged solution within an infinite set of
other possible and equally privileged solutions. But if, on the
other hand, the unclosed terms are in fact correlated to an
underlying theory, either in that they underly a specific but
analytically non-accessible process or in that they show some
existing but unknown substructure, then the construction of an
invariant {\it solution} is misleading and essentially
ill-defined, in particular if no prior modelling assumptions for
the unclosed terms are made. To follow this conclusion in more
detail we refer to Appendix \ref{SecA} for an extensive discussion
on this subject.

Hence, for an unclosed and thus underdetermined system of
equations either infinitely many and equally privileged solutions
(including all possible invariant solutions) can be constructed,
or, depending on whether the unclosed terms are correlated to an
underlying but analytically non-accessible theory as turbulence,
no true solutions and thus also no true invariant solutions can be
determined as long as no prior modelling procedure is invoked to
close the system of equations. Therefore, since {\it closed}
systems do not face this problem, the construction of invariant
solutions from symmetry transformations is well-defined, while for
equivalence transformations, which are admitted by {\it unclosed}
systems, the construction of invariant solutions is misleading and
can be even ill-defined as in the statistical theory of
turbulence. Thus using for example the equivalence
transformation~(\ref{140529:2336}) to generate a {\it privileged}
statistical invariant solution for the unclosed system
(\ref{140529:2236}) is basically ill-defined, if no prior
modelling assumptions for the underlying substructure of~$\vT$ is
made to close the equations (see first part of
Appendix~\ref{SecA.2}).

However, if nevertheless within the theory of turbulence such
invariant results are generated, they must be carefully
interpreted as only being functional relations or functional
complexes which stay invariant under the derived equivalence
group, and not as being privileged {\it solutions} of the
associated underdetermined system, as done, for example, in
\cite{Oberlack03,Khujadze04,Guenther05,Oberlack06,Oberlack10,
She11,Oberlack13.1,Oberlack14} and \cite{Oberlack14.1}. In all
these studies the underlying statistical system of dynamical
equations is unclosed and thus underdetermined, however, not
arbitrarily underdetermined, but underdetermined in the sense that
all unclosed terms can be physically and uniquely determined from
the underlying but analytically non-accessible instantaneous
(fluctuating) velocity field. In particular the system considered
in \cite{Oberlack10}, although formally infinite in dimension,
reveals itself by closer inspection as such an underdetermined
system, for which, as was already said before, the determination
of invariant solutions is ill-defined (see last part of Appendix
\ref{SecA.2}). This study of \cite{Oberlack10}, which serves as a
key study for the recent results made in
\cite{Oberlack13.1,Oberlack14} and \cite{Oberlack14.1}, will be
analyzed in more detail in the next sections.

Important to note is that up to now only in the specific case of
homogeneous isotropic turbulence \citep{Davidson04,Sagaut08} all
those invariant functional complexes which are gained from {\it
equivalence} scaling groups can be further used to yield more
valuable results, in particular the explicit values for the decay
rates \citep{Oberlack02}, since one has exclusive access to
additional nonlocal invariants such as the Birkhoff-Saffman or the
Loitsyansky integral. However, for {\it wall-bounded} flows it is
not clear yet how to use or exploit such invariant functional
complexes in a meaningful way, since up to now no additional
nonlocal invariants are known.\footnote[2]{This aspect also needs
to be addressed in Oberlack's earlier work
\citep{Oberlack99.1,Oberlack01}, where also only equivalence
transformations were obtained, but which, in addition, were
specifically obtained as a~result of an incorrect conclusion
\citep{Frewer14.2}.}
\restoregeometry

Finally it is worthwhile to mention that for example the work of
\cite{Khabirov02.1,Khabirov02.2} clearly shows in how {\it
equivalence} transformations within the theory of turbulence can
be exploited in a correct manner, which stands in strong contrast
to the misleading approach of Oberlack et al. The major difference
to the Oberlack et al. approach is that in
\cite{Khabirov02.1,Khabirov02.2} the invariant functions for the
unclosed term (which are generated within different optimal Lie
subalgebras for all possible Lie-point equivalence transformations
of the unclosed K{\'a}rm{\'a}n-Howarth equation) are {\it not}
identified as {\it true solutions} of the underlying unclosed
equation itself, but, instead, are identified as {\it possible
model terms} which then in each case consequently leads to a
closed model equation. This is done in \cite{Khabirov02.1}, while
in \cite{Khabirov02.2} these {\it closed} K{\'a}rm{\'a}n-Howarth
model equations are then solved in each case by the {\it now}
well-defined technique of {\it invariant solutions}, which
\cite{Khabirov02.1,Khabirov02.2} then call {\it physical}
invariant solutions. Of course, in how far these solutions then
describe reality must be checked in each case by experiment or
DNS. But that's a different problem!

We want to close this section by giving a citation from
\cite{Khabirov02.2} which exactly describes the behavior and
effect of an equivalence transformation when trying to exploit it
in order to gain insight into the solution manifold of an unclosed
and thus underdetermined equation: ``Equivalence transformations
may affect the behavior of solutions in physical sense. In other
words, they may transform {\it physical} solutions into {\it
unphysical} ones. But {\it inverse} equivalence transformations
may act better in physical sense. These properties of the
equivalence transformations will be made use of in the sequel."

\section{A complete statistical description: The Hopf equation\label{Sec3}}

In order to determine new statistical {\it symmetry
transformations}, and not {\it equivalence} transformations, we
have to operate within a framework which offers a complete and
fully determined statistical description of Navier-Stokes
turbulence. Any statistical description which is not formally
closed, that is, every statistical description which from the
outset would involve unclosed and thus arbitrary functionals, is
not suited for this purpose. As was shown in the previous section
(Example 2), every invariance analysis would then only generate
very weak equivalence transformations.

Currently there are only two statistical approaches to
incompressible and spatially unbounded Navier-Stokes turbulence
which independently offer a complete and fully determined
statistical description. Both approaches {\it formally} circumvent
the explicit closure problem of turbulence in that they not only
overcome the local differential framework in favor of a consistent
nonlocal integral framework, but also in that they operate on a
higher statistical level which goes beyond the level of the
statistical moments. In each case the consequence is a linearly
infinite but {\it formally} closed statistical approach.

These two approaches are the Lundgren-Monin-Novikov chain of
equations \citep{Lundgren67,Monin67,Friedrich12} and the Hopf
equation \citep{Hopf52,McComb90,Shen91}. While the former operates
on the high statistical level of the {\it probability density
functions} for the $n$-point velocity moments
\begin{equation}
\vH_n= \big\L \vu(\vx_1,t)\otimes\cdots \otimes\vu(\vx_n,t)\big\R
,\quad n\geq 1, \label{140530:1156}
\end{equation}
the Hopf equation operates on the even higher level of the {\it
probability density functionals} for these moments
(\ref{140530:1156}). As shown in \cite{Monin67}, the
Lundgren-Monin-Novikov system is just the discrete version of the
functional Hopf equation. The former is iteratively given as an
infinite but fully determined hierarchy of linearly coupled
equations, while the latter is given as a single fully determined
linear functional equation of infinite dimension. Since in both
cases no arbitrary functions are involved, they both can be {\it
formally} identified as closed systems.

To note is that a third statistical approach exists, which also
leads to a linearly infinite hierarchy of equations, the so-called
Friedmann-Keller chain of equations \citep{Monin71}, which, in
contrast to the other two approaches, operates directly on the
lower level of the $n$-point velocity moments (\ref{140530:1156}).
This chain can either be formulated in the local differential
framework, as presented in \cite{Oberlack10} and also recently in
\cite{Oberlack14.1}, or in the nonlocal integral framework as
presented in \cite{Fursikov99} and re-derived in Appendix
\ref{SecB}.

However, in contrast to the Lundgren-Monin-Novikov chain or the
Hopf equation, the Friedmann-Keller chain is not closed, not even
in a formal sense. This matter is extensively discussed in
Appendix \ref{SecC}. Irrespective of the analytical framework and
in the sense as explained in detail in Appendix \ref{SecC}, the
Friedmann-Keller chain always involves more unknown functions than
determining equations. For both the integral framework as
presented in Appendix \ref{SecB}, as well as for the differential
framework as presented in \cite{Oberlack10} and
\cite{Oberlack14.1}, this can be easily confirmed by just
explicitly counting the number of equations versus the number of
functions to be determined. In this sense the Friedmann-Keller
chain, although infinite in dimension, does not serve as a fully
determined statistical description of Navier-Stokes turbulence.
Any invariance analysis performed upon this chain will only
generate the weaker class of equivalence transformations, simply
because the chain is always permanently underdetermined and thus
involving arbitrary functions.

Now, in order to prove our statement that a new statistical
scaling symmetry is physically inconsistent with the underlying
deterministic Navier-Stokes equations (\ref{140529:1657}), either
the Lundgren-Monin-Novikov chain or Hopf equation can be used.
They are equivalent in so far as they both lead to the same
conclusion. However, to prove this statement in the next section
as efficiently as possible, we will only choose the functional
Hopf-approach.

The functional Hopf equation (HEq)
\begin{equation}
\frac{\partial\Phi}{\partial t}=\int\! d^3\vx\, \alpha_k \left(
i\frac{\partial}{\partial x_l}
\frac{\delta^2}{\delta\alpha_k\delta\alpha}_l
+\nu\Delta\frac{\delta}{\delta\alpha_k}\right)\Phi,
\label{140122:1250}
\end{equation}
describes the dynamical evolution of the characteristic or
moment-generating functional
\begin{equation}
\Phi[\valpha(\vx);t]=\int\! P[\vu(\vx);t]\, e^{i\int\! d^3\vx\,
\valpha(\vx)\cdot\vu(\vx)}\mathcal{D}\vu(\vx), \label{140125:1515}
\end{equation}
which is the functional Fourier transform \citep{Klauder10,
Kleinert13} of the probability density functional $P[\vu(\vx);t]$
for the velocity field $\vu$ sampled for each time step in
infinitely non-denumerable (con\-ti\-nuum) number of points $\vx$,
which itself plays the role of a continuous index inside the
functionals $\Phi$ and $P$, but nonetheless still to be
interpreted next to the coordinate $t$ and the field
$\valpha(\vx)$ as an own independent and active variable in the
underlying dynamical equation (\ref{140122:1250}). In other words,
both functionals $\Phi$ and $P$ do not explicitly depend on $\vx$,
i.e. in equation (\ref{140122:1250}) the variable $\vx$ only
appears implicitly in the dependent variable $\Phi$ upon which the
coordinate operators can then act on. The functional variable
$\valpha(\vx)$, however, is an arbitrary but real, integrable and
time-independent solenoidal external source function with
vanishing normal component at the (infinite far) boundary. In
order to guarantee for physical consistency, a mathematical
solution of the Hopf equation (\ref{140122:1250}) is only admitted
if {\it for all times} the following conditions are fulfilled
\begin{equation}
\Phi^*[\valpha(\vx);t]=\Phi[-\valpha(\vx);t],\quad
\Phi[0;t]=1,\quad\big\vert\Phi[\valpha(\vx);t]\big\vert\leq 1,
\label{140124:2059}
\end{equation}
which stem from the fact that the probability density functional
is real, non-negative, and normalized to one in sample space, i.e.
$\int P[\vu(\vx);t]\mathcal{D}\vu(\vx)=1$, with $P[\vu(\vx);t]\geq
0$. This then defines the (infinite) physical dimension of the
probability density functional as $[P]=[1/\mathcal{D}\vu]$ with
$\mathcal{D}\vu=\prod_\vx\, (\sqrt{d^3\vx/2\pi})^3\cdot
d^3\vu(\vx)$, while the characteristic functional $\Phi$ is
dimensionless.

Finally note that the above-presented functional integration
element $\int \mathcal{D}\vu(\vx)$ for the Fourier transform
(\ref{140125:1515}) is symmetrically defined as the following
infinite product of one-dim\-en\-sio\-nal integrals over
$\vu(\vx_\vn)=(u^i(\vx_\vn))\in\mathbb{R}^3$ at every point
$\vx_\vn=(x^i)_\vn\in\mathbb{R}^3$ for every component and
coordinate $i=1,2,3$ (see e.g. \cite{Kleinert13}, Chapter 13):
\begin{equation}
\int \mathcal{D}\vu(\vx)=\prod_{i=1}^3\;\int \mathcal{D}u^i(\vx)
=\prod_{i=1}^3\;\prod_{(n_1,n_2,n_3)=-\infty}^\infty\, \int
\frac{d u^i(\epsilon\cdot \vn)}{\sqrt{2\pi/\epsilon^3}}.
\label{140920:1915}
\end{equation}
Hereby, space-time is grated into a fine {\it equidistant}
lattice, where for every coordinate $x^i$ the following discrete
lattice points were introduced
\begin{equation}
\vx \rightarrow \vx_\vn=\epsilon\cdot\vn =\epsilon\cdot
(n_1,n_2,n_3)\in\mathbb{R}^3,\;\; n_i\in\mathbb{Z},
\label{140920:1917}
\end{equation}
with a very small lattice spacing $\epsilon$.

\section{The new statistical scaling symmetry and its
inconsistency\label{Sec4}}

It is straightforward to recognize that the linear Hopf equation
(\ref{140122:1250}) admits the following (functional) Lie-point
scaling symmetry
\begin{align}
\mathsf{Q}: & \;\;\; \tilde{t}=t,\;\;
\tilde{\vx}=\vx,\;\;\tilde{\valpha}=\valpha,\;\;
\tilde{\Phi}=e^q\left(\Phi-1\right)+1, \label{140124:1833}
\end{align}
where $q$ is a globally constant and real group parameter. This
invariance first got mentioned in \cite{Oberlack13X}. Note that
symmetry $\mathsf{Q}$ is only compatible with the first two
physical constraints given in (\ref{140124:2059}). The third,
non-holonomic constraint in (\ref{140124:2059}) gets violated if
the characteristic functional is generally transformed as
(\ref{140124:1833}).

But, if the values of the group constant $q$ are restricted to
$q\leq 0$ then symmetry $\mathsf{Q}$ (\ref{140124:1833}) is fully
compatible with all three physical constraints
(\ref{140124:2059}). However, by restricting the values of $q$ to
$q\leq 0$, the symmetry (\ref{140124:1833}) turns into a
semi-group since no inverse element can be defined or constructed
anymore. In other words, the third constraint in
(\ref{140124:2059}) breaks the Lie-group structure of the symmetry
(\ref{140124:1833}) down to a semi-group.

The connection of the moment generating functional $\Phi$
(\ref{140125:1515}) to the multi-point velocity correlation
functions $\vH_n$ (\ref{140530:1156}) is given as
\citep{Hopf52,McComb90,Shen91}
\begin{equation}
\vH_n=\big\L \vu(\vx_1,t)\otimes\cdots \otimes\vu(\vx_n,t)\big\R
=(-i)^n\frac{\delta^n\Phi[\valpha(\vx);t]}{\delta
\valpha(\vx_1)\cdots\delta\valpha(\vx_n)}\bigg\vert_{\valpha=\v0}.
\label{140127:1615}
\end{equation}
By inserting $\mathsf{Q}$ (\ref{140124:1833}) into the above
functional relation (\ref{140127:1615}) of the transformed domain
\begin{align}
(-i)^n\frac{\delta^n\tilde{\Phi}[\tilde{\valpha}(\tilde{\vx});\tilde{t}]}{\delta
\tilde{\valpha}(\tilde{\vx}_1)\cdots
\delta\tilde{\valpha}(\tilde{\vx}_n)}\bigg\vert_{\tilde{\valpha}=\v0}
&= (-i)^n\frac{\delta^n\big[e^{q}\Phi[\valpha(\vx);t]
+\left(1-e^{q}\right)\big]}{\delta \valpha(\vx_1)\cdots
\delta\valpha(\vx_n)}\bigg\vert_{\valpha=\v0}\nonumber\\
&= e^{q}\cdot(-i)^n\frac{\delta^n\Phi[\valpha(\vx);t]}{\delta
\valpha(\vx_1)\cdots\delta\valpha(\vx_n)}\bigg\vert_{\valpha=\v0},
\label{140704:0841}
\end{align}
we can see how the symmetry transformation $\mathsf{Q}$
(\ref{140124:1833}) induces the following invariant transformation
for the $n$-point velocity correlation functions
(\ref{140530:1156})
\begin{align}
\mathsf{Q}_\mathsf{E}: & \;\;\; \tilde{t}=t,\;\;\;
\tilde{\vx}_n=\vx_n,\;\;\;\tilde{\vH}_n=e^q\vH_n,\quad n\geq 1,
\label{140530:2112}
\end{align}
which for the first time was derived in \cite{Khujadze04} as a
``new statistical symmetry" of Navier-Stokes turbulence. For their
derivation they however only considered the unclosed multi-point
equations for the velocity moments up to order $n=2$ in the limit
of an inviscid parallel shear flow in ZPG turbulent boundary layer
flow\footnote[2]{In \cite{Khujadze04} the iterative sequence of
the ``new statistical scaling symmetry" (\ref{140530:2112}) begins
only from $n=2$ onwards, i.e only for all $n\geq 2$. The
transformation for $n=1$ is excluded, i.e. the mean velocity
$\vH_1=\L\vu\R$ stays invariant.}, while recently in
\cite{Oberlack10} this result (\ref{140530:2112}) was re-derived
most generally without any flow restrictions by using the full
infinite chain of Friedmann-Keller equations. In both derivations
the statistical invariance analysis was performed in the local
differential framework based on the corresponding deterministic
form (\ref{140529:1736}), in which the pressure field is
explicitly present. Thus in both derivations their results also
include, next to the $n$-point velocity moments $\vH_n$, all
invariant transformations for the
velocity-pressure~correlations.~Of course, these correlations are
not part of our result given here, since we derived
(\ref{140530:2112}) from the Hopf equation (\ref{140122:1250})
which is based on the underlying nonlocal deterministic integral
form (\ref{140529:1657}), in which the pressure field has been
consistently eliminated from the dynamical equations.

Indeed, it can be easily verified that transformation
$\mathsf{Q}_\mathsf{E}$ (\ref{140530:2112}) is admitted as an
invariant transformation also by the nonlocal integro-differential
Friedmann-Keller equations
\begin{equation}
\partial_t \vH_n +\boldsymbol{\mathcal{A}}_n \cdot\vH_n
+\boldsymbol{\mathcal{B}}_n\cdot\widehat{\vH}_{n+1}=\v0,\quad
n\geq 1, \label{140601:1140}
\end{equation}
which are defined and derived in Appendix \ref{SecB}. However, as
noted in Appendix \ref{SecB} and discussed in more detail in
Appendix \ref{SecC}, the invariant transformation
$\mathsf{Q}_\mathsf{E}$ (\ref{140530:2112}) does not act as a
symmetry transformation, but only in the weaker form as an
equivalence transformation. The reason is that the hierarchy
(\ref{140601:1140}) forms an unclosed system. The still missing
transformation rule for the unclosed $n$-point function of
$(n+1)$-th moment $\widehat{\vH}_{n+1}$, which formally stands for
$\widehat{\vH}_{n+1}=\vH_{n+1}\vert_{\vx_{n+1}=\vx_n}$, is then
dictated by the given transformation rule (\ref{140530:2112}) for
the corresponding $(n+1)$-point function $\vH_{n+1}$ as
\vspace{-0.25em}
\begin{equation}
\widetilde{\widehat{\vH}}_{n+1}=\widetilde{\vH}_{n+1}\big\vert_{\tilde{\vx}_{n+1}
=\tilde{\vx}_n}= \big(e^q\vH_{n+1}\big)\big\vert_{\vx_{n+1}
=\vx_n}=e^q\big(\vH_{n+1}\big\vert_{\vx_{n+1}=\vx_n}\big)
=e^q\widehat{\vH}_{n+1}. \label{140607:2309}
\end{equation}

\vspace{-0.25em}\noindent It is clear that this simple
transformation rule (\ref{140607:2309}) is only due to the global
and uniform nature of $\mathsf{Q}_\mathsf{E}$ (\ref{140530:2112}),
in which all system variables transform uniformly by the same
constant scaling exponent $q$ and independent from the
coordinates, which themselves stay invariant.

Important to note here is that in \cite{Oberlack10} the invariant
transformation $\mathsf{Q}_\mathsf{E}$ (\ref{140530:2112}) is
considered as a true {\it symmetry} transformation. However, as
already discussed in Section \ref{Sec3} and explained in Appendix
\ref{SecC}, this claim is not correct. Transformation
$\mathsf{Q}_\mathsf{E}$ (\ref{140530:2112}) can only act as an
{\it equivalence} transformation and not as a {\it symmetry}
transformation. Hence, the invariance analysis performed in
\cite{Oberlack10} is based on equivalence and not on symmetry
groups, simply because unclosed and thus arbitrary functions are
permanently involved within the considered analysis.

But this insight now has consequences in the interpretation of
their newly derived statistical scaling laws, because these laws
as presented in \cite{Oberlack10} may not be interpreted as being
privileged {\it solutions} of the underlying statistical set of
equations as was done therein. They may only be interpreted as
being functional relations or functional complexes which stay
invariant under the derived equivalence group, nothing more!
Therefore these new relations derived in \cite{Oberlack10} {\it
only possibly but not necessarily} can serve as useful turbulent
scaling functions. Moreover, a comparison to DNS results reveals
that these new statistical scaling laws presented in
\cite{Oberlack10} are unphysical as they clearly fail to fulfil
the most basic predictive requirements of a scaling law. For ZPG
turbulent boundary layer flow this investigation is presented and
further discussed in Section \ref{Sec5}.

The reason for this failure is twofold: Next to the reason just
discussed, that the invariance analysis in \cite{Oberlack10} was
performed upon an underdetermined statistical system which cannot
admit true invariant {\it solutions} without establishing a
correct link to the underlying deterministic equations, the second
and more stronger reason is that the symmetry $\mathsf{Q}$
(\ref{140124:1833}) itself is unphysical. This physical
inconsistency of course transfers down to $\mathsf{Q}_\mathsf{E}$
(\ref{140530:2112}), as it is induced by $\mathsf{Q}$. This will
also explain why on the higher level of the probability density
functionals a true symmetry transformation, such as $\mathsf{Q}$
(\ref{140124:1833}), only induces an equivalence transformation,
such as $\mathsf{Q}_\mathsf{E}$ (\ref{140530:2112}), and not a
corresponding symmetry transformation on the lower level of the
$n$-moment functions $\vH_n$.

Before we proceed with the proof, it is worthwhile to see that
when considering the chain only up to the second moment ($n\leq
2$), the {\it general} equivalence transformation\pagebreak[4]
$\mathsf{Q}_\mathsf{E}$~(\ref{140530:2112}) will reduce in the
smooth limit of zero correlation length to the equivalence
transformation $\mathsf{E}_\mathsf{2}$ (\ref{140529:2336})
discussed in Section \ref{Sec2}:
\begin{equation}
\lim_{\vert\vx_2-\vx_1\vert\to 0} \mathsf{Q}_\mathsf{E}^{(n\leq
2)} = \mathsf{E}_\mathsf{2}\backslash\L\tilde{p}\R,
\end{equation}
where, due to the eliminated pressure field in the underlying Hopf
equation, the missing transformation rule $\L\tilde{p}\R=e^q\L
p\R$ for the mean pressure field in $\mathsf{Q}_\mathsf{E}$
(\ref{140530:2112}) is consistently dictated by the mean
solenoidal velocity field $\tilde{\vH}_1=\L\tilde{\vu}\R$ through
the one-point momentum equation (\ref{140529:2236}) as given in
$\mathsf{E}_\mathsf{2}$ (\ref{140529:2336}). Hence, the above
mentioned and still to be proven physical inconsistency of
$\mathsf{Q}$ (\ref{140124:1833}) will thus even fully transfer
down to $\mathsf{E}_\mathsf{2}$ (\ref{140529:2336}).

\subsection{Proof of the physical inconsistency of symmetry
$\mathsf{Q}$\label{Sec4.1}}

The physical inconsistency of symmetry $\mathsf{Q}$
(\ref{140124:1833}) can be readily observed when connecting the
averaged (statistical) level back to the fluctuating
(deterministic) level. For the Hopf equation (\ref{140122:1250})
the transition rule in going from the fine-grained (fluctuating)
to the coarse-grained (averaged) level is defined by the path
integral (\ref{140125:1515}), which in each time step $t$ sums up
all coarse-grained probabilities $P$ for all possible realizations
in the fine-grained velocity field $\vu$. Now, in order to see the
inconsistency, it is necessary to consider the inverse functional
Fourier transform of (\ref{140125:1515}) in the transformed domain
of $\mathsf{Q}$ (\ref{140124:1833}):
\begin{equation}
\tilde{P}[\tilde{\vu}(\tilde{\vx});\tilde{t}]=\int\!
\tilde{\Phi}[\tilde{\valpha}(\tilde{\vx});\tilde{t}]\, e^{-i\int\!
d^3\tilde{\vx}\,
\tilde{\valpha}(\tilde{\vx})\cdot\tilde{\vu}(\tilde{\vx})}
\mathcal{D}\tilde{\valpha}(\tilde{\vx}). \label{140126:1607}
\end{equation}
By inserting the transformation $\mathsf{Q}$ (\ref{140124:1833})
into the right-hand side of (\ref{140126:1607}) we get
\begin{align}
\tilde{P}[\tilde{\vu};\tilde{t}]&= \int\! \Big[\,
e^{q}\Phi[\valpha;t] +\left(1-e^{q}\right)\Big]\, e^{-i\int\! d^3
\vx\,
\valpha\cdot\vu}\mathcal{D}\valpha\nonumber\\
&= e^{q}P[\vu;t]+\left(1-e^{q}\right)\cdot\delta [\vu],
\label{140127:1625}
\end{align}
which is the corresponding transformation rule for $P$, if $\Phi$
transforms as given in (\ref{140124:1833}), where for better
readability we suppressed the implicit dependence
$\tilde{\vx}=\vx$ on both sides. The function $\delta[\cdot]$ is
the functional analog of the Dirac $\delta$-function, and is
called the $\delta$-functional \citep{Klauder10, Kleinert13}. In
the lattice approximation, corresponding to (\ref{140920:1915})
with (\ref{140920:1917}), they are defined as an infinite product
of ordinary one-dimensional $\delta$-functions
\begin{equation}
\delta[\vu]=\prod_{i=1}^3\,\delta[u^i] \,
=\,\prod_{i=1}^3\,\prod_{\vn=-\infty}^\infty\sqrt{2\pi/\epsilon^3}
\cdot \delta(u^i(\epsilon\cdot\vn)),
\end{equation}
and thus having the obvious property
\begin{equation}
\int \mathcal{D}\vu\,\delta[\vu]=1,\quad \int
\mathcal{D}\valpha\,\delta[\valpha]=1.
\end{equation}
Note that since the variables $\vx$ and $\valpha$ transform
invariantly under (\ref{140124:1833}), the velocity field $\vu$
must stay invariant too, otherwise we would loose the definition
(\ref{140125:1515}) of a functional Fourier transform in the
transformed domain.

However, on the other hand, if we insert $\mathsf{Q}$
(\ref{140124:1833}) into the functional relation
(\ref{140127:1615}) in order to explicitly generate the
transformation rule for the $n$-point velocity moments, as was
already exercised in (\ref{140704:0841}), we will get again
\begin{equation}
(-i)^n\frac{\delta^n\tilde{\Phi}[\tilde{\valpha}(\tilde{\vx});\tilde{t}]}{\delta
\tilde{\valpha}(\tilde{\vx}_1)\cdots
\delta\tilde{\valpha}(\tilde{\vx}_n)}\bigg\vert_{\tilde{\valpha}=\v0}=
e^{q}\cdot(-i)^n\frac{\delta^n\Phi[\valpha(\vx);t]}{\delta
\valpha(\vx_1)\cdots\delta\valpha(\vx_n)}\bigg\vert_{\valpha=\v0},
\end{equation}
which, if using the representation of $\Phi$ (\ref{140125:1515}),
turns into the following relations
\begin{align}
n=1:\!\!\! &\;\;\; \int\!
\tilde{P}[\tilde{\vu}(\tilde{\vx});\tilde{t}]\cdot\tilde{\vu}(\tilde{\vx}_1)\cdot
\mathcal{D}\tilde{\vu}(\tilde{\vx}) =e^{q}\int\!
P[\vu(\vx);t]\cdot\vu(\vx_1)\cdot\mathcal{D}\vu(\vx),\nonumber\\
n=2:\!\!\! &\;\;\; \int\!
\tilde{P}[\tilde{\vu}(\tilde{\vx});\tilde{t}]\cdot\tilde{\vu}(\tilde{\vx}_1)\otimes
\tilde{\vu}(\tilde{\vx}_2)\cdot
\mathcal{D}\tilde{\vu}(\tilde{\vx})=e^{q}\int\!
P[\vu(\vx);t]\cdot\vu(\vx_1)\otimes\vu(\vx_2)\cdot\mathcal{D}\vu(\vx)
,\nonumber\\
\phantom{,}\vdots\;\;\;\,
&\;\;\;\qquad\quad\qquad\qquad\qquad\vdots
\qquad\qquad\qquad\qquad\qquad\qquad\qquad\qquad\quad\;\;\vdots
\label{140127:1623}
\end{align}
\newpage
\newgeometry{left=3cm,right=3cm,top=2.1cm,bottom=1.45cm,headsep=1em}
\noindent By recognizing again the already mentioned fact that the
velocity field (along with its continuous index) is an invariant
under the considered transformation $\mathsf{Q}$
(\ref{140124:1833}), we can replace the variable
$\tilde{\vu}(\tilde{\vx})$ in (\ref{140127:1623}) with $\vu(\vx)$
and vice versa for all points. Then, by equating in its present
and already irreducible form the left-hand side with the
right-hand side for each order $n$, we obtain from
(\ref{140127:1623}) the transformation relation\footnote[2]{Note
that a local relation can only be identified correctly from an
integral relation if it's formulated irreducibly, i.e., in a form
such that it cannot be reduced or simplified any further.}
\begin{equation}
\tilde{P}[\tilde{\vu};\tilde{t}]=e^{q}P[\vu;t],
\end{equation}
which is in conflict with the previously found transformation rule
(\ref{140127:1625}) for $P$, i.e.~there is no unique
transformation rule for the probability density functional $P$.
Consequently, via the fine- to coarse-grained transition rule
(\ref{140126:1607}) the symmetry transformation $\mathsf{Q}$
(\ref{140124:1833}) induces an inconsistency. This conflict,
however, can only be solved if $q=0$, but which then turns the
symmetry transformation $\mathsf{Q}$ (\ref{140124:1833}) into a
trivial identity transformation.

Worthwhile to note is that by physical intuition alone one already
can recognize this conflict just by solely observing relation
(\ref{140127:1625}) more closely. Because, since the variables
$\vx$, $\vu$ and $t$ transform invariantly under $\mathsf{Q}$
(\ref{140124:1833}) we can identically write the transformation
rule (\ref{140127:1625}) also as
\begin{equation}
\tilde{P}[\vu;t]=e^{q}P[\vu;t] +\left(1-e^{q}\right)\cdot\delta
[\vu], \label{140127:1649}
\end{equation}
which states that although the system on the fine-grained level
$\vu$ stays unchanged, it nevertheless undergoes a {\it global}
change $P\rightarrow\tilde{P}$ on the coarse-grained level, which
is completely unphysical and not realized in nature.

We thus have a classical violation of cause and effect, as the
system would experience an effect (change in averaged dynamics)
without a corresponding cause (change in fluctuating dynamics).
Note that the opposite conclusion is not the rule, i.e. a change
on the fluctua\-ting level can occur without inducing an effect on
the averaged level. A macroscopic or coarse-grained (averaged)
observation might be insensitive to many microscopic or
fine-grained (fluctuating) details, a property of nature widely
known as universality (see e.g. \cite{Marro14}). For example, a
high-level complex coherent turbulent structure, though a
consequence of the low-level fluctuating description, does not
depend on all its details on its lowest level. The opposite again,
however, is not realized in nature, i.e., stated differently, if
the coherent structure experiences a {\it global} change, e.g. in
scale or in a translational shift, it definitely must have a cause
and thus {\it must} go along with a corresponding change~on the
lower fluctuating level --- see also the discussions, e.g., in
\cite{Frewer15.3,Frewer16.1}.

Exactly this non-physical behavior can also be independently
observed in the induced transformation rule
$\mathsf{Q}_\mathsf{E}$ (\ref{140530:2112}) for the $n$-point
velocity moments $\vH_n$ (\ref{140530:1156}). It can either be
exposed directly on the fluctuating level as an unphysical
equivalence transformation, or indirectly on the averaged level as
a superfluous or artificial equivalence transformation. In any
case, on each level we will gain different insights for this
non-physical transformation behavior.

Hence, fully detached from the finding that the equivalence
transformation $\mathsf{Q}_\mathsf{E}$ (\ref{140530:2112}) for the
velocity moments is induced by an unphysical symmetry
transformation  $\mathsf{Q}$ (\ref{140124:1833}) on the higher
statistical level of the corresponding probability densities, we
will now repeat our investigation on the lower statistical level
of the velocity moments themselves, by only focussing on the link
between $\mathsf{Q}_\mathsf{E}$ (\ref{140530:2112}) and the
unclosed Friedmann-Keller equations (\ref{140601:1140}).

\vspace{-0.2em}
\subsection{The unphysical behavior of equivalence $\mathsf{Q}_\mathsf{E}$
on the fluctuating level\label{Sec4.2}}

In the case of the Friedmann-Keller chain, especially when used in
the oversimplified form (\ref{140601:1140}), particular care has
to be taken when actually performing a systematic invariance
analysis on these equations. The problem is that in contrast to
the other two statistical approaches, i.e. the
Lundgren-Monin-Novikov chain or the Hopf equation, the
Friedmann-Keller chain does not naturally come along with
additional physical constraints which are necessary in order to
reveal the nonlinear and nonlocal connection between all
constituents (see Appendix \ref{SecC}).
\restoregeometry
\newpage
\newgeometry{left=3cm,right=3cm,top=2.1cm,bottom=1.2cm,headsep=1em}
This circumstance can easily lead to misleading results, as it is
the case for $\mathsf{Q}_\mathsf{E}$ (\ref{140530:2112}). Here it
is necessary to recognize that $\mathsf{Q}_\mathsf{E}$
(\ref{140530:2112}) is an invariant scaling transformation which
only linear systems can admit, since only the system's dependent
variables get uniformly scaled, while the coordinates $(t, \vx)$
themselves stay invariant. Indeed, the corresponding dynamical
system which admits $\mathsf{Q}_\mathsf{E}$ (\ref{140530:2112}) is
the Friedmann-Keller chain of equations (\ref{140601:1140}), which
is a linear system, since $\boldsymbol{\mathcal{A}}_n$ and
$\boldsymbol{\mathcal{B}}_n$ are both linear operators (see
Appendix \ref{SecB}).

However, this result, that the hierarchical system
(\ref{140601:1140}) admits $\mathsf{Q}_\mathsf{E}$
(\ref{140530:2112}) as an equivalence transformation, is
misleading, since it suggests that all correlation functions
$\vH_n$ scale uniformly with the same scaling factor, which really
is not the case as the underlying theory dictates a nonlinear
correlation between all these quantities. The problem clearly lies
in the notation: Using a formal symbol as $\vH_n$, where only an
external index~$n$ allows to distinguish between different
multi-point functions, hides the actual underlying correlation
information among them. In this sense the notation used in
equation (\ref{140601:1140}) is counterproductive from the
perspective of an analysis on invariance, in that it
oversimplifies the physical situation by representing the dynamics
for the $n$-point velocity correlations as a linear PDE system,
while it is actually based on a nonlinear theory.\footnote[2]{Note
that also the Hopf equation, and its discrete version, the
Lundgren-Monin-Novikov equations are linear systems, but at the
expense of operating on a higher statistical level than the
moments of the Friedmann-Keller chain of equations, which, by
definition, are all uniquely correlated in a nonlinear~manner.}

It is this misleading aspect which was not recognized and taken
into account in \cite{Oberlack10}. That is to say, by {\it
explicitly} revealing the underlying nonlinear structure behind
the formal symbol $\vH_n$ in (\ref{140601:1140}), namely that
$\vH_{n+1}$ contains one deterministic velocity field $\vu$ more
than $\vH_n$, will ultimately break the equivalence scaling
$\mathsf{Q}_\mathsf{E}$~(\ref{140530:2112}),\linebreak as will be
shown next.

Since the velocity correlation function $\vH_n$ in
$\mathsf{Q}_\mathsf{E}$ (\ref{140530:2112}) is nonlinearly built
up by $n$ multiplicative evaluations of the instantaneous
(fluctuating) velocity field $\vu$ according to
(\ref{140530:1156}), the following chain of reasoning instantly
emerges: Since for $n=1$ the averaged function $\vH_1=\L \vu_1\R$
scales as $e^q$ for all points $\vx_1=\vx$ in the domain, the
corresponding fluctuating quantity $\vu_1$ has to scale in the
same manner, since every averaging operator $\L,\R$ is linearly
commuting with any {\it constant} scale factor. But this implies
that any product of $n$ fluctuating fields $\vu_1\otimes\cdots
\otimes\vu_n$ has to scale as $e^{n\cdot q}$, which again implies
that also the corresponding averaged quantity $\vH_n=\L
\vu_1\otimes\cdots \otimes\vu_n\R$ then has to scale as $e^{n\cdot
q}$. Symbolically the chain reads as
\begin{align}
\tilde{\vH}_1=e^q \vH_1 \; & \Rightarrow\;
\big\L\tilde{\vu}(\tilde{\vx}_1,\tilde{t})\big\R=e^q\big\L
\vu(\vx_1,t)\big\R,\;\text{for all points
$\vx_1=\vx$}\nonumber\\
& \Rightarrow\;
\big\L\tilde{\vu}(\tilde{\vx}_k,\tilde{t})\big\R=e^q\big\L
\vu(\vx_k,t)\big\R,\;\text{for all}\; k\geq 1\nonumber\\
& \Rightarrow\;
\big\L\tilde{\vu}(\tilde{\vx}_k,\tilde{t})\big\R=\big\L e^q
\vu(\vx_k,t)\big\R,\;\text{for all possible configurations $\vu$} \nonumber\\
& \Rightarrow\;\tilde{\vu}(\tilde{\vx}_k,\tilde{t})=e^q
\vu(\vx_k,t)\nonumber\\
& \Rightarrow\;\tilde{\vu}(\tilde{\vx}_1,\tilde{t})\otimes\cdots
\otimes\tilde{\vu}(\tilde{\vx}_n,\tilde{t})=e^{n\cdot q}
\vu(\vx_1,t)\otimes\cdots \otimes\vu(\vx_n,t)\nonumber\\
&
\Rightarrow\;\big\L\tilde{\vu}(\tilde{\vx}_1,\tilde{t})\otimes\cdots
\otimes\tilde{\vu}(\tilde{\vx}_n,\tilde{t})\big\R=\big\L e^{n\cdot
q}
\vu(\vx_1,t)\otimes\cdots \otimes\vu(\vx_n,t)\big\R\nonumber\\
&
\Rightarrow\;\big\L\tilde{\vu}(\tilde{\vx}_1,\tilde{t})\otimes\cdots
\otimes\tilde{\vu}(\tilde{\vx}_n,\tilde{t})\big\R=e^{n\cdot
q}\big\L
\vu(\vx_1,t)\otimes\cdots \otimes\vu(\vx_n,t)\big\R\qquad\quad\nonumber\\
& \Rightarrow\; \tilde{\vH}_n=e^{n\cdot q} \vH_n
.\label{140127:1405}
\end{align}
For a detailed explanation of this proof in all its steps, please
refer to~Appendix \ref{SecD}. Conclusion (\ref{140127:1405})
clearly demonstrates that if the one-point function $\vH_1$
globally scales as $e^q$ then the $n$-point function $\vH_n$ has
to scale accordingly, namely as $e^{n\cdot q}$ and not as $e^q$ as
dictated by $\mathsf{Q}_\mathsf{E}$ (\ref{140530:2112}). Only the
former scaling $e^{n\cdot q}$ will guarantee for an all-over
consistent relation between the fluctuating and averaged level of
the dynamical Navier-Stokes system. In other words, if a dynamical
system experiences a {\it global} transformational change on the
averaged level then there must exist {\it at least} one
corresponding change on the fluctuating level \citep{Frewer16.1}.
But exactly this is not the case for $\mathsf{Q}_\mathsf{E}$
(\ref{140530:2112}) as both $\vH_1$ and $\vH_n$ scale therein with
the {\it same} global factor, for which, thus, a corresponding
fluctuating scaling cannot be derived or constructed, neither as a
symmetry nor as any regular transformation, meaning that the
system experiences a {\it global} change on the averaged level
with {\it no} corresponding change on the fluctuating level \hfill
--- \hfill again\hfill the\hfill classical\hfill violation\hfill
of\hfill cause
\restoregeometry
\newpage\noindent and effect as was already discussed
before. Hence, on the lower statistical level of the velocity
moments, too, the physical consistency can only be restored if
$q=0$, i.e. if the equivalence transformation
$\mathsf{Q}_\mathsf{E}$ (\ref{140530:2112}) gets broken.

To conclude, it should be pointed out that this proof
(\ref{140127:1405}) clearly shows that the transformation
(\ref{140530:2112}) itself, i.e.~detached from any transport
equations, leads to contradictions as soon as one considers more
than one point ($n\geq 2$). However, for $n=1$, i.e. for the mean
velocity $\vH_1=\L \vu\R$ itself, no contradiction exits. Only as
from $n\geq 2$ onwards, the contradiction starts, which also can
be clearly observed when comparing to DNS data as will be
demonstrated in Section \ref{Sec5}: The mismatch of the
corresponding scaling laws which involve this contradictive
scaling group (\ref{140530:2112}) gets more strong as the order of
the moments $n$ increases.

Also finally note again that in order to perform the proof
(\ref{140127:1405}) we basically used the {\it consistency} of
$n=1$ (the first four lines of (\ref{140127:1405})) to show the
{\it inconsistency} for all $n\geq 2$ (the remaining lines of
(\ref{140127:1405})). Hence, irrelevant of whether
$\mathsf{Q}_\mathsf{E}$ (\ref{140530:2112}) represents an
invariance or not, the transformation itself leads for $n\geq 2$
to contradictions.

\subsection{The superfluous behavior of equivalence $\mathsf{Q}_\mathsf{E}$ on
the averaged level\label{Sec4.3}}

The immediate consequence on the averaged level in using an
oversimplified statistical representation is that
$\mathsf{Q}_\mathsf{E}$ (\ref{140530:2112}) will show a
superfluous or artificial transformation behavior as soon as one
changes to a more detailed representation which reveals more
information about the underlying theory. From the perspective of
an invariance analysis, it is intuitively clear that changing the
statistical description for example to the Reynolds decomposed
representation will be superior to the oversimplified notation
used in (\ref{140601:1140}), as it explicitly will reveal the
nonlinearity within the system on the averaged lower level of the
moments. Performing a Reynolds decomposition, for example, of the
instantaneous 2-point velocity field $\vH_2$ into its mean and
fluctuating part, will thus lead to
\begin{align}
\vH_2 &= \big\L \vu_1\otimes\vu_2\big\R
=\big\L\big(\vU_1+\vu^\prime_1\big)
\otimes\big(\vU_2+\vu^\prime_2\big)\big\R\nonumber\\
&= \vR_{12}+\vU_1\otimes\vU_2. \label{140115:1947}
\end{align}
This relation explicitly unfolds its nonlinear connection to the
one-point velocity fields, where $\vR_{12}=\L
\vu_1^\prime\otimes\vu_2^\prime\R$ is the corresponding 2-point
correlation function for the (zero-mean) fluctuating field
$\vu^\prime$ evaluated in the points $\vx=\vx_1$ and $\vx_2$
respectively, while $\vU_1$ and $\vU_2$ is the mean velocity
evaluated in the same points. Next, the decomposition for the
instantaneous 3-point velocity field $\vH_3$
\begin{align}
\vH_3 &= \big\L \vu_1\otimes\vu_2\otimes\vu_3\big\R
=\big\L\big(\vU_1+\vu^\prime_1\big)\otimes
\big(\vU_2+\vu^\prime_2\big)\otimes
\big(\vU_3+\vu^\prime_3\big)\big\R\nonumber\\
&= \vR_{123}+\vU_1\otimes\vU_2\otimes\vU_3
\nonumber\\
&\qquad\quad\;\,
+\vR_{12}\otimes\vU_3+\vR_{13}\otimes\vU_2+\vR_{23}\otimes\vU_1,
\label{140603:2126}
\end{align}
will not only nonlinearly connect to 1-point, but also to 2-point
functions. This nonlinear connection will then iteratively
continue for all higher multi-point functions. Hence, in a
bijective, one-to-one manner the equivalence transformation
$\mathsf{Q}_\mathsf{E}$ in the oversimplified (linear)
representation (\ref{140530:2112}) then changes to the following
more detailed (nonlinear) representation
\begin{align}
\mathsf{Q}_\mathsf{E}: & \;\;\; \tilde{t}=t,\;\; \tilde{\vx}_n =
\vx_n,\;\; \tilde{\vU}_n=e^q\,
\vU_n,\;\;\cdots\nonumber\\
&\;\;\; \tilde{\vR}_{nm}=e^{q}\,
\vR_{nm}+\big(e^q-e^{2q}\big)\cdot \vU_n\otimes\vU_m,
\;\;\cdots\nonumber\\
&\;\;\; \tilde{\vR}_{nml}=e^{q}\,
\vR_{nml}\nonumber\\
&\hspace{1.5cm}
+\big(e^q-e^{2q}\big)\cdot\big(\vR_{nm}\otimes\vU_l+
\vR_{nl}\otimes\vU_m+\vR_{ml}\otimes\vU_n\big)
\qquad\quad\nonumber\\
&\hspace{1.5cm} + \big(e^q-3e^{2q}+2e^{3q}\big)\cdot
\vU_n\otimes\vU_m\otimes\vU_l,\;\;\cdots \label{140222:1123}
\end{align}
where we explicitly expressed the transformation only up to third
order in the velocity field for all point-indices $n,m,l\geq 1$ in
all possible combinations.\pagebreak[4]

In contrast to representation (\ref{140530:2112}), the above
representation (\ref{140222:1123}) of $\mathsf{Q}_\mathsf{E}$
immediately reveals its superfluous or artificial behavior as an
equivalence transformation. In (\ref{140222:1123}) one can see
that the aim of all terms containing the prefactor $e^q$ is to
enforce a linear system scaling invariance which attempts to scale
all field variables uniformly and independently from its
coordinates. But since the underlying Navier-Stokes theory is
inherently nonlinear, typical error terms proportional to $e^{2q}$
and $e^{3q}$ then emerge in (\ref{140222:1123}) which need to be
subtracted accordingly in order to allow for a misleading linear
invariance property within a true nonlinear system of moments. In
other words, although transformation (\ref{140222:1123}) acts as a
true equivalence transformation in the correspondingly Reynolds
decomposed representation of the instantaneously averaged system
(\ref{140601:1140}), it acts artificially in that it interprets
the nonlinear terms $\vU_n\otimes\vU_m$,
$\vU_n\otimes\vU_m\otimes\vU_l$, etc. as error terms which all are
corrected for in order to achieve the desired linear system
scaling invariance, but which, as was demonstrated before in
(\ref{140127:1405}), ultimately cannot exist physically as it
induces inconsistencies already on the fluctuating level.

In order to avoid a misconception on this subtle issue, we will
repeat the above argumentation again by using different words and
by viewing it from a different perspective.

Our claim here is that for the moments the $\vH$ notation as used
by Oberlack et al.~should not be used when performing an analysis
on invariance, because, due to its high notational simplification,
it can easily lead to misleading results, in particular when
ignoring its connection to the underlying deterministic
theory.~Careful, our statement is only to be interpreted as a
precautionary measure to avoid possible misguidance from the
outset. We do not say that the $\vH$ notation is wrong, we just
say that it is counterproductive to use, because when working with
this oversimplified notation without making a direct connection to
the underling deterministic theory, one clearly has a higher risk
to produce non-physical results than when working in the classical
$\vR$ notation.~The self-evident~reason is that the oversimplified
$\vH$ notation hides essential information of the underlying
deterministic Navier-Stokes equations, while the $\vR$ notation,
in contrast, reveals it.~In~other~words, when not connecting the
notation to the underlying deterministic theory,
the~$\vR$~notation~is {\it physically} more transparent and
helpful than the mathematically equivalent $\vH$ notation.

Of course, as both notations are just linked by a bijective
(one-to-one) mapping, the classical $\vR$ notation is not free of
the risk to induce a non-physical result, too. But such a
non-physical result will be more easy to track in the detailed
$\vR$ notation than in the oversimplified $\vH$ notation, where
it's even not noticeable without properly connecting the notation
to the underlying theory. In the $\vR$ notation, however,
unphysical results immediately reveal themselves by showing an
artificial functional behavior, as in the case of the new
unphysical scaling invariance (\ref{140222:1123}).

It is clear that since this scaling invariance is unphysical in
the $\vH$ notation (\ref{140530:2112}) it is also unphysical in
the $\vR$ notation (\ref{140222:1123}). But in contrast to
relation (\ref{140530:2112}), the corresponding relation
(\ref{140222:1123}) {\it immediately} indicates that it's
unphysical. To be explicit, let's consider the new scaling
invariance in the $\vR$ notation (\ref{140222:1123}) for the
one-point correlations up to second~order
\begin{equation}
\tilde{t}=t,\qquad \tilde{\vx}=\vx,\qquad \tilde{\vU}=e^q
\vU,\qquad\tilde{\vtau}=e^q\vtau+(e^q-e^{2q})\cdot\vU\otimes\vU,
\quad \cdots\label{r75}
\end{equation}
where $\vtau=\L \vu^\prime\otimes\vu^\prime\R$ is the
Reynolds-stress tensor, and compare it to the single and only
scaling symmetry of the deterministic Navier-Stokes equations
$\mathsf{S}$ (\ref{130813:1900}), which, when transcribed into the
statistical form of the $\vR$ notation, will read
\begin{equation}
\tilde{t}=e^{2\varepsilon}t,\qquad
\tilde{\vx}=e^{\varepsilon}\vx,\qquad \tilde{\vU}=e^{-\varepsilon}
\vU,\qquad\tilde{\vtau}=e^{-2\varepsilon}\vtau ,\quad\cdots
\label{r76}
\end{equation}
Although both \eqref{r75} and \eqref{r76} are mathematically
admitted as invariant transformations of the underlying
Reynolds-stress transport equations up to second order
\citep{Pope00,Davidson04}, it is only transformation \eqref{r75}
which on this level of description immediately shows an artificial
and thus a physically non-useful transformation behavior. Thus,
without making a connection to the underlying fluctuating dynamics
we already can observe that \eqref{r75} is actually a physically
non-useful transformation just by solely inspecting expression
\eqref{r75}. This is definitely not possible in the oversimplified
$\vH$ notation (\ref{140530:2112}), and hence, therefore, one has
the higher risk of being misguided when using this notation.

The reason why on this level of description \eqref{r75} acts
artificially and \eqref{r76} not, is that in order to explicitly
scale the values of the Reynolds-stress tensor
$\vtau\rightarrow\tilde{\vtau}$ one has to involve the mean
velocity field itself (in the quadratic form $\vU\otimes\vU$). But
such a transformation (\ref{r75}) is not in accord with the idea
of a Reynolds decomposition which has the intention to study
turbulence statistics {\it relative} to the mean velocity field
$\vU$. The problem is that since the {\it untransformed}
Reynolds-stress $\vtau$ is quadratically built up by a (zero-mean)
fluctuating field $\vu^\prime$ which measures the mean stress {\it
relative} to the mean velocity $\vU$, the {\it transformed}
quantity $\tilde{\vtau}$ in \eqref{r75} doesn't have this
`relative measure'-property anymore because the values are now
mixed with mean-velocity values. In other words, {\it within the
transformed system} the quantity $\tilde{\vtau}$ {\it cannot} be
identified as a Reynolds-stress anymore, which actually should
measure the stress {\it relative} to the transformed mean velocity
$\tilde{\vU}$.

In this sense, transformation \eqref{r75} is not physically
useful, which we directly can also observe when fitting the
resulting scaling laws to DNS data (see Section \ref{Sec5}).~The
observed result will be a clear mismatch between theory and
experiment, but, as soon as the unphysical structure of
transformation \eqref{r75} is excluded or removed from the scaling
laws, the matching will improve again by several orders of
magnitude which, ultimately, is a clear indication that the
scaling \eqref{r75} is unphysical.

Moreover, when returning back to the previously mentioned
perspective where the additional scaling terms in \eqref{r75} are
only required to restore a misleading linear scaling within a
nonlinear theory of moments, the artificial transformation
behavior of \eqref{r75} can also be immediately seen when
generating invariant functions. Consider the following invariant
one-point function of transformation \eqref{r75}
\begin{equation}
\vF(\vx)=\frac{\vtau+\vU\otimes\vU}{\sqrt{\vU\cdot\vU}}.
\label{r77}
\end{equation}
Now, when explicitly performing this invariant transformation
\begin{align}
\tilde{\vF}(\tilde{\vx})&=\frac{\tilde{\vtau}
+\tilde{\vU}\otimes\tilde{\vU}}{\sqrt{\tilde{\vU}\cdot\tilde{\vU}}}\nonumber\\
&=\frac{e^q\vtau+(e^q-e^{2q})\cdot\vU\otimes\vU+(e^q\vU)\otimes
(e^q\vU)}{\sqrt{(e^q\vU)\cdot
(e^q\vU)}}=\frac{\vtau+\vU\otimes\vU}{\sqrt{\vU\cdot\vU}}=\vF(\vx),
\end{align}
we see how the transformation rule for the Reynolds-stress $\vtau$
acts artificially, in that one of its direct aims is to only
cancel the disturbing nonlinear terms. Hence, it's highly
questionable whether, and in which sense, the invariant function
\eqref{r77} is actually physically relevant, since its
corresponding invariant transformation \eqref{r75} is not
incorporating the nonlinear terms into the transformation process
itself but instead only treats them as `error terms' which must be
cancelled accordingly.

Finally, the reader should note that such a superfluous
construction is not specific to the Navier-Stokes theory, i.e. the
construction principle itself to yield the misleading type of
invariance (\ref{140222:1123}) is not unique or particular to the
Navier-Stokes equations but can be established basically in any
statistical system of any nonlinear theory. In other words, the
superfluous type of linear scaling invariance (\ref{140222:1123})
inherently also exists for example in any unclosed statistical
model of the nonlinear Maxwell or the nonlinear Schrödinger
equations (see Appendix \ref{SecE}), by just reformulating the
corresponding expressions accordingly. Hence, if one is not
careful enough wrong conclusions will be the general consequence.

In a more general sense we can thus conclude that systematically
ignoring any information about the functional structure of an
either closed or an unclosed model equation, which is directly
linked to an underlying theory in using an oversimplified
representa-\linebreak tion\hfill (instead\hfill of\hfill an\hfill
appropriate\hfill representation\hfill which\hfill
explicitly\hfill reveals\hfill this\hfill information),
\pagebreak[4]
\newgeometry{left=3cm,right=3cm,top=2.1cm,bottom=1.3cm,headsep=1em}
\noindent unawarely allows for generating unphysical and thus
useless results when performing an analysis on invariance. This
conclusion can be stated as the following general principle:

$\mathcal{P}$:$\;$ \emph{For every invariance analysis to be
performed on an equation-based model which is linked to an
underlying physical theory, it is crucial how the model equations
are represented. It is necessary to reveal all information
available for the system. If the notation tends to be
oversimplified by not revealing all essential information, the
analysis runs the risk to generate non-physical results without
knowing.}

In other words, caution has to be exercised in knowing that
mathematical notation, even if formally correct, always has the
unfortunate ability to simplify or even oversimplify the actual
physical situation and thus causing misguidance, or suggesting an
intuition which by closer inspection is not supported.

\subsection{An example of a physically consistent statistical
scaling symmetry\label{Sec4.4}}

We want to close Section \ref{Sec4} with a contrasting (positive)
example of a statistical scaling symmetry admitted by the Hopf
equation (\ref{140122:1250}), which not only is compatible with
all three constraints (\ref{140124:2059}), but which also acts
fully consistent on the coarse-grained (averaged) as well as on
the fine-grained (fluctuating) level. The symmetry is
\begin{equation}
\mathsf{S}^{\mathsf{HEq}}:  \;\;\;
\tilde{t}=e^{2\varepsilon}t,\;\;
\tilde{\vx}=e^\varepsilon\vx,\;\;\tilde{\valpha}=e^{-2\varepsilon}\valpha,\;\;
\tilde{\Phi}=\Phi, \label{140127:1001}
\end{equation}
which is the only admitted physical (global) scaling symmetry
$\mathsf{S}$ (\ref{130813:1900}) of the Navier-Stokes equations
just reformulated here for the Navier-Stokes-Hopf equation
(\ref{140122:1250}).\footnote[2]{The `official' theoretical
development of extending classical point symmetry analysis from
partial to functional differential equations is provided in
\cite{Oberlack06-X}, and recently also in \cite{Waclawczyk13}
adjusting it to Fourier space. However, it still lacks
completeness, since the extension is based on an incomplete set of
variables, in that the continuous index point $\vx$ (in coordinate
space) or $\vk$ (in wavenumber space) are considered as being
unchangeable quantities, which is {\it not} the case, simply
because both variables carry a physical dimension which always, at
least, must allow for a (re-)scaling in the units. A clear
counter-example is given by (\ref{140127:1001}). But also from a
pure mathematical perspective, the independent variables $\vx$ or
$\vk$ have to be included into the transformation process, even if
they at most only act as integration (dummy) variables,
nevertheless, their transformational change is always ruled by the
Jacobian. Hence, by making sole use of the extended Lie algorithm
developed in \cite{Oberlack06-X} and \cite{Waclawczyk13}, the
fundamental scaling symmetry (\ref{140127:1001}) can {\it not} be
generated and essentially an important symmetry as
(\ref{140127:1001}) will thus be missed. For more details, we
refer to our comments \cite{Frewer16.22,Frewer16.21} and to our
reactions \cite{Frewer.X4,Frewer.X3}, respectively.} Note that
(\ref{140127:1001}) then induces the transformation rule for the
\begin{itemize}
\item {\it velocity field}: since the exponent inside the kernel
of (\ref{140125:1515}) should stay invariant, i.e.
$d^3\tilde{\vx}\cdot\tilde{\valpha}\cdot\tilde{\vu}
=d^3\vx\cdot\valpha\cdot\vu$, in order to consistently define a
functional Fourier transform also in the transformed (scaled)
domain, the velocity field must scale as
\begin{equation}
\tilde{\vu}=e^{-\varepsilon}\vu, \label{140127:1651}
\end{equation}
\item {\it functional derivative}: since the functional derivative
carries the physical dimension of the considered field variable
per volume, it must scale as
\begin{equation}
\frac{\delta}{\delta\tilde{\valpha}}=\frac{\partial}
{\partial\tilde{\valpha}d^3\tilde{\vx}}=e^{-\varepsilon}\frac{\partial}
{\partial\valpha
d^3\vx}=e^{-\varepsilon}\frac{\delta}{\delta\valpha},
\end{equation}
\item {\it functional volume element}: since for path integrals
the measure is of infinite size, it will scale accordingly for
each field as
\begin{align}
\mathcal{D}\tilde{\valpha}
&=\prod_{\tilde{\vx}}\big(\sqrt{d^3\tilde{\vx}/2\pi}\big)^3 \cdot
d^3\tilde{\valpha}(\tilde{\vx})\nonumber\\ &=\prod_\vx
e^{-3\varepsilon/2}\big(\sqrt{d^3\vx/2\pi}\big)^3 \cdot
d^3\valpha(\vx) =\Big(\prod_\vx
e^{-3\varepsilon/2}\Big)\mathcal{D}\valpha,\\
\mathcal{D}\tilde{\vu}&=\prod_{\tilde{\vx}}\big(\sqrt{d^3\tilde{\vx}/2\pi}\big)^3
\cdot d^3\tilde{\vu}(\tilde{\vx})\nonumber\\ &=\prod_\vx
e^{+3\varepsilon/2}\big(\sqrt{d^3\vx/2\pi}\big)^3 \cdot
d^3\vu(\vx) =\Big(\prod_\vx
e^{+3\varepsilon/2}\Big)\mathcal{D}\vu,\,
\end{align}
\end{itemize}
\newgeometry{left=3cm,right=3cm,top=2.1cm,bottom=2.1cm,headsep=1em}
\begin{itemize}
\item[] where it should be noted that the continuous counting
index (not the variable itself) stays invariant under
transformation (\ref{140127:1001}), i.e.
$\prod_{\tilde{\vx}}=\prod_\vx$, since by
$\tilde{\vx}=e^{\varepsilon}\vx$ any set of real numbers is mapped
again in a one-to-one manner to a set of real numbers of the same
measure,
\item {\it probability density functional}: since the physical
constraint $\int P[\vu;t]\mathcal{D}\vu=1$ must stay invariant, it
has to scale as
\begin{equation}
\tilde{P}[\tilde{\vu};\tilde{t}]=\Big(\prod_\vx
e^{-3\varepsilon/2}\Big)P[\vu;t], \label{140127:1652}
\end{equation}
which, in contrast to (\ref{140127:1649}), makes an intuitive
physical statement, in that if the system experiences a {\it
global} change in scale on the fine-grained level
$\vu\rightarrow\tilde{\vu}$ of type (\ref{140127:1651}), then the
system will experience this change in scale also on the
coarse-grained level accordingly $P\rightarrow \tilde{P}$
(\ref{140127:1652}).
\item {\it $n$-point velocity correlation functions}: since the
construction of all $\vH_n$ in the Hopf framework are given
according to rule (\ref{140127:1615}), they scale as
\begin{equation}
\tilde{\vH}_n=e^{-n\cdot\varepsilon}\,\vH_n,
\end{equation}
which is the only correct possible scaling behavior for the
incompressible Navier-Stokes $n$-point velocity correlation
functions. To date, no other statistical scaling symmetry exists!
\end{itemize}

\section{Comparing to DNS results\label{Sec5}}

This section will investigate if all statistical scaling laws
which are based on the ``new statistical scaling symmetry"
$\mathsf{Q}_\mathsf{E}$ (\ref{140530:2112}) qualify as useful
scaling laws. For geometrically simple wall-bounded flows the
general construction principle to generate these laws as
``first-principle results" in the inertial region is given in
\cite{Oberlack10,Oberlack11.3}, which recently in
\cite{Oberlack14} got extended to include more sophisticated
wall-bounded flows.\footnote[2]{In \cite{Oberlack10,Oberlack11.3}
as well as in \cite{Oberlack14} all scaling laws for wall-bounded
flows are actually based on two ``new statistical symmetries".
Next to the ``new scaling symmetry" (\ref{140530:2112}) also a
``new translation symmetry" is involved, but which, as the scaling
symmetry too, turns out to be completely unphysical. This can be
easily demonstrated by using the same procedure as shown and
developed in this article.} Of particular interest are those laws,
which according to \cite{Oberlack10,Oberlack11.3} should scale all
higher order velocity moments beyond the log-law of the
mean-velocity profile. The corresponding derivation procedure is
revisited in Appendix~\ref{SecF}. Up to third moment, the explicit
functional structure for all one-point velocity moments is derived
in (\ref{130821:1404}) and given as
\begin{equation}
\left. \begin{aligned} U(y) & = \alpha_U\cdot\ln(y+c)+\beta_U,\\
\tau^{ij}(y) & =
\alpha_H^{ij}+\beta_H^{ij}\cdot(y+c)^{\gamma}-\delta^{1i}\delta^{1j}U(y)^2,\\
T^{111} & = \alpha_H^{111}+\beta_H^{111}\cdot(y+c)^{2\gamma}
-3U(y)\cdot\tau^{11}(y)-U(y)^3,\\
T^{112} & = \alpha_H^{112}+\beta_H^{112}\cdot(y+c)^{2\gamma}
-2U(y)\cdot\tau^{12}(y),\\
T^{ij1} & = \alpha_H^{ij1}+\beta_H^{ij1}\cdot(y+c)^{2\gamma}
-U(y)\cdot\tau^{ij}(y),\;\text{for}\;
(i,j)=(2,2),(3,3),\\
T^{ij2} & =
\alpha_H^{ij2}+\beta_H^{ij2}\cdot(y+c)^{2\gamma},\;\text{for}\;
(i,j)=(2,2),(3,3).
\end{aligned}
~~~ \right \} \label{140929:2347}
\end{equation}
At the example of ZPG turbulent boundary layer flow these ``new
statistical scaling laws", which by construction all apply in the
inertial region of the flow, will be matched to DNS data. The
investigation itself will be based on the method of least square
fits (chi-square methods) for DNS data with a Reynolds number as
high as $Re_\theta=6000$. An open source software package which
runs in Mathematica was used for all fitting needs, designed and
programmed by \cite{Zielesny11}. The DNS data was made available
to us on the courtesy of Jiménez et al.
\citep{Simens09,Borrell13}.

\newpage
The curve fitting strategy is organized as follows: The complete
fitting procedure will be based on an experimentally obtained fit
for the mean streamwise velocity profile in the form of the
modified ($y$-shifted) log-law according to (\ref{140929:2347}):
\begin{equation}
U^+_{\text{exp}}=\alpha_{U\! ,\text{exp}}\cdot\ln(y^+
+c_{\text{exp}})+\beta_{U\! ,\text{exp}}, \label{140115:2140}
\end{equation}
where all variables and parameters were normalized to wall
units.\footnote[2]{The normalization into wall units is based on
the kinematic viscosity $\nu$ and the mean streamwise friction
velocity $U_\tau=\sqrt{\nu\cdot\partial_y U\vert_{y=0}}$, which
needs to be extracted from DNS data. In order to avoid an
overloading of notation, the `+'-index in the parameters will be
suppressed.} This $y$-shifted (non-classical) log-law
(\ref{140115:2140}) was first derived and proposed in
\cite{Oberlack01}, and then later re-derived in \cite{Oberlack10}
as shown in Appendix \ref{SecF}. The corresponding three
parameters $\kappa_{\text{exp}}=1/\alpha_{U\! , \text{exp}}=0.38$,
$\beta_{U\! , \text{exp}}=4.1$ and $c_{\text{exp}}=5.0$ are taken
from \cite{Lindgren04}, which were determined as an average-fit
over an ensemble of experimental data sets ranging in $Re_\theta$
from 2500 up to 27000. Hence, this particular specification in the
log-law parameters should also apply to our chosen DNS data set,
as its Reynolds number value of $Re_\theta=6000$ lies within that
ensemble range. The idea behind this strategy is to make use of
the universal scaling behavior of the log-law, under the critical
assumption, of course, that it's valid. In other words, if we
assume the universality of the log-law to be correct, it is more
than reasonable to take a well-established average-fit over an
ensemble of experiments within a wide range of different Reynolds
numbers and to imbed it into a numerical data set for a Reynolds
number which lies inside that range. In this way, as will be done
herein, one obtains a robust reference for all upcoming fits to be
generated in this section.

\begin{figure}[t]
\centering
\begin{minipage}[c]{.48\linewidth}
\FigureXYLabel{\includegraphics[width =
.91\textwidth]{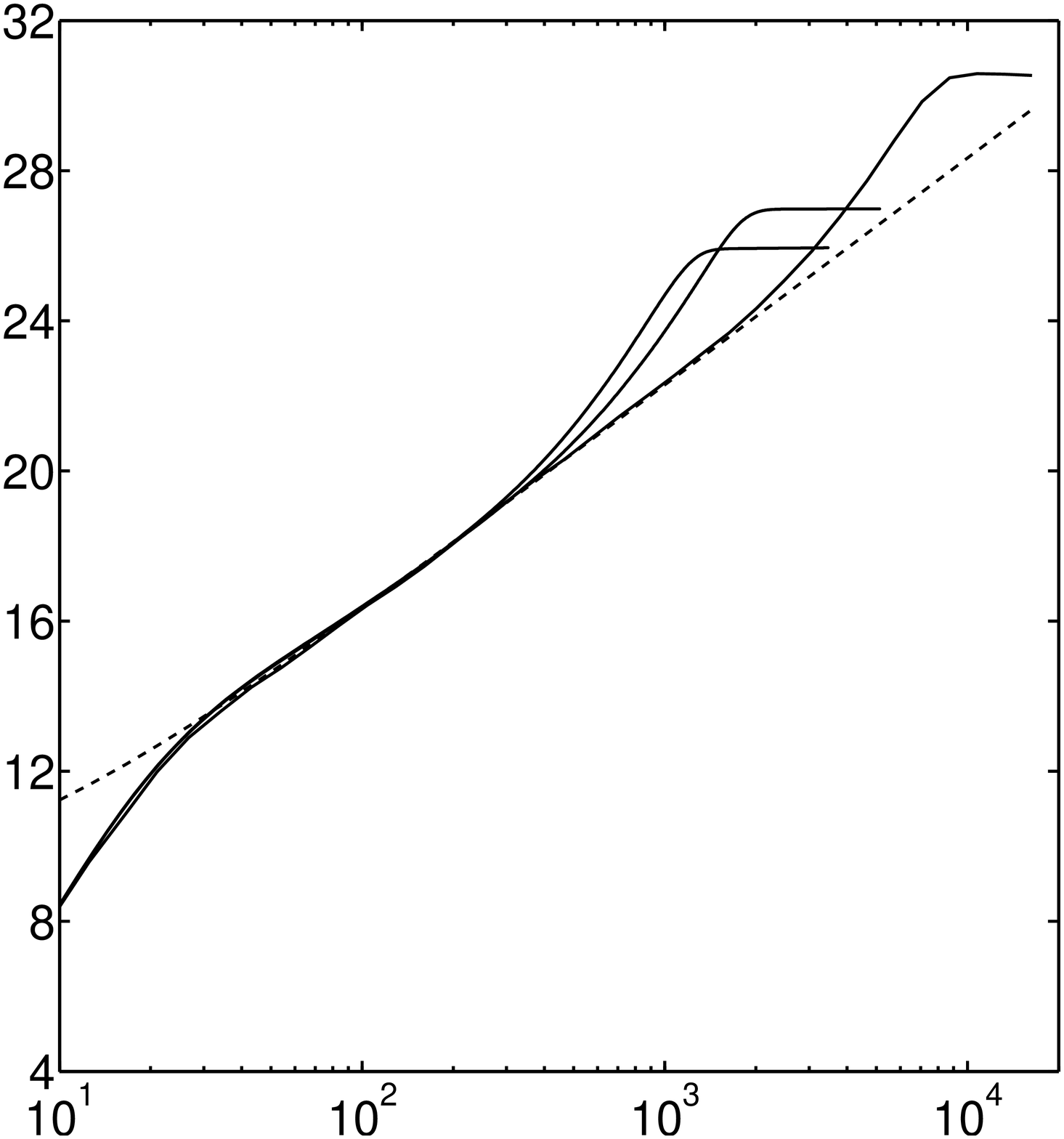}} {$y^+$}{.5mm}{\begin{rotate}{90}
$U^+$
\end{rotate}}{1mm}
\end{minipage}
\hfill
\begin{minipage}[c]{.505\linewidth}
\FigureXYLabel{\includegraphics[width=.91\textwidth]{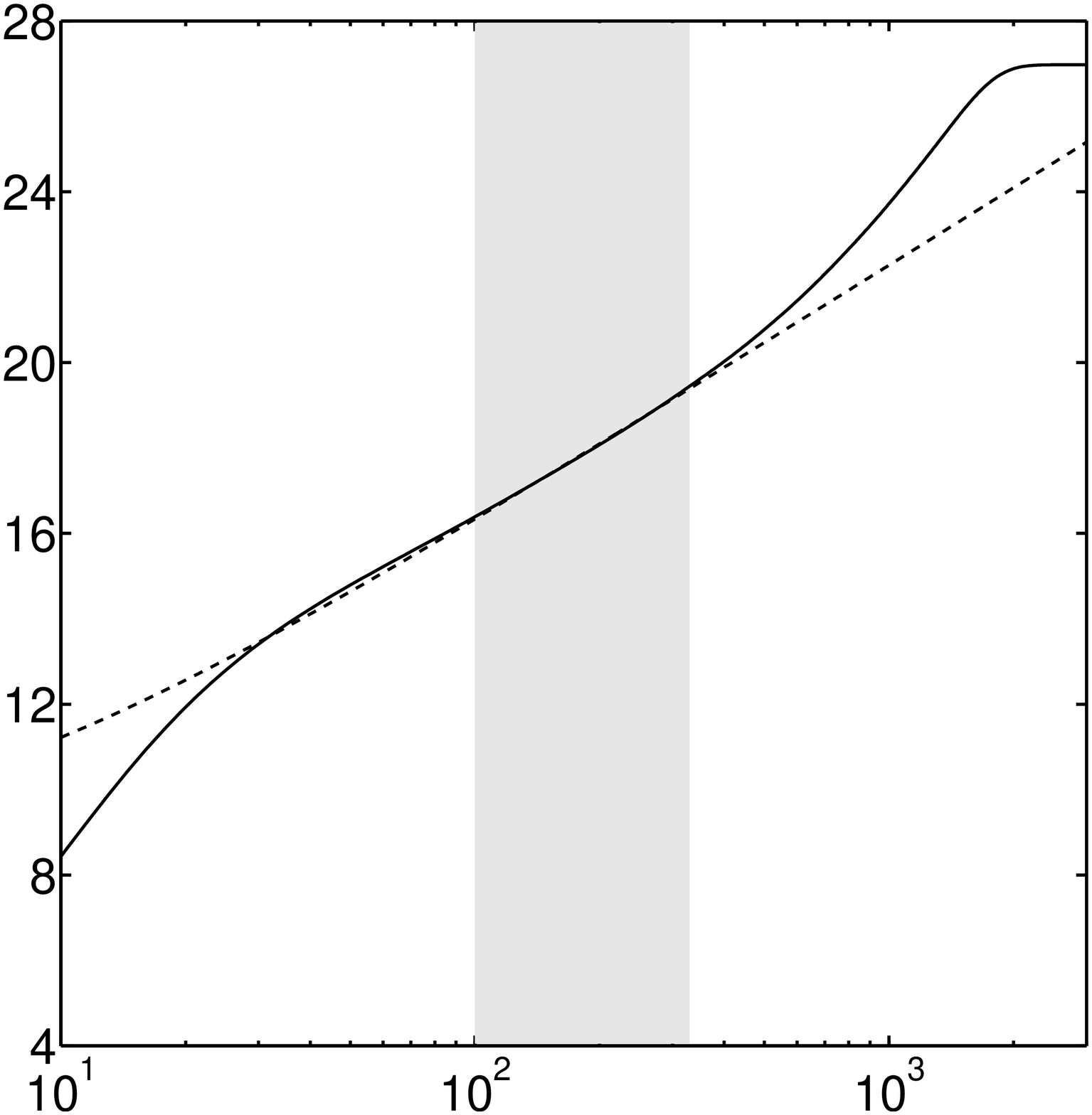}}
{$y^+$}{.5mm}{\begin{rotate}{90} $U^+$ \end{rotate}}{1mm}
\end{minipage}
\caption{{\it Left plot:} Comparison between DNS data from
\cite{Borrell13} of the mean streamwise velocity profile for
$Re_\theta=6000$ (middle solid line from top end), and from
\cite{Schlatter10} for $Re_\theta=4500$ (lower solid line from top
end), and the experimental data set from \cite{Osterlund00} for
$Re_\theta=27000$ (upper solid line from top end). The dashed line
shows the log-law (\ref{131203:0854}), which was experimentally
gained as an average-fit over an ensemble of $Re_\theta$ ranging
from $2500$ up to $27000$ \citep{Lindgren04}.
\\
{\it Right plot:} The solid line displays the DNS data of the mean
streamwise velocity profile for $Re_\theta=6000$
\citep{Borrell13}, while the dashed line shows again the
experimentally fitted log-law (\ref{131203:0854}). Inside the
grey-shaded region, ranging from $y^+= 100$ to $330$, the
statistical fitting error is $\delta_{\text{fit}}= 1.3\cdot
10^{-3}$ and $\chi^2_\text{red}= 7.5\cdot 10^{1}$ when referred to
the underlying DNS error. Hereby the shaded region visualizes the
inertial range, which from hereon will now consistently define the
fitting domain in all plots.} \label{fig:umean}
\end{figure}

Figure \ref{fig:umean}, left plot (L), convinces with a good
comparison for our chosen DNS data set of $Re_\theta=6000$ to
other sets with different Reynolds numbers. Also the above
universal modified log-law (\ref{140115:2140}) is included to get
an impression for its range of validity.

As the general aim of this section is to perform proper fits by
using basic methods from statistical data analysis, the error of
the considered data needs to be known. Fortunately the numerical
error for some simulated quantities in the case of ZPG turbulent
boundary layer flow are easily determined. Due to statistical
symmetries in the flow the mean spanwise velocity component as
well as all moments involving an uneven number of fluctuating
spanwise velocity fields should be exactly zero by theory.
However, every numerical simulation is and always will be unable
to resolve these zero-fields exactly to zero. Hence the difference
can be interpreted as the error of the simulation for that
particular field, i.e. the mean spanwise velocity component $W$
then serves as the error for $U$, and e.g. the Reynolds stress
$\tau^{23}$ can serve as the error for $\tau^{22}$ and
$\tau^{33}$, etc.

\subsection{Curve fitting for $Re_\theta=6000$ in the inertial region\label{Sec5.1}}

The course of action will follow the underlying statistical
hierarchy, by first fitting the lowest order moments to be then
used as input information to fit the next higher order ones. The
process starts with Figure \ref{fig:umean}, right plot (R), for
the mean streamwise velocity profile. It shows the result of
comparison between the above in (\ref{140115:2140}) specified and
discussed log-law
\begin{equation}
U^+_{\text{exp}}=1/0.38\cdot\ln(y^+ +5)+4.1, \label{131203:0854}
\end{equation}
and the corresponding DNS data at $Re_\theta= 6000$. According to
\cite{Lindgren04} the fit should be valid down to about $y^+\sim
100$. The upper limit is then fixed symmetrically by taking the
same residual as found at the lower limit. This then gives a
convincing fit ranging from $y^+\sim 100$ to $y^+\sim 330$, shown
in Figure \ref{fig:umean}R as the shaded region, with a rather
good reduced chi-square ($\chi^2_\text{red}$) value of 75. Here
the mean spanwise velocity $W$ locally served in each point of the
considered range as the underlying error field to determine the
necessary quality of that fit.

As a result the log-law in Figure \ref{fig:umean}R thus slightly
underfits the DNS data in this range, as its
$\chi^2_\text{red}$-value is above the optimal value of one
\citep{Zielesny11}. In particular, the rule is that {\it relative
to the underlying numerical error of the data}, any
$\chi^2_\text{red}$-value smaller than one indicates a so-called
{\it overfit} of the data meaning that the considered fitting
range of the domain is possibly too small, while a value larger
than one indicates an {\it underfit} of the data showing that
either the considered fitting range of the domain is too large or
that the considered function itself is inappropriate. However,
despite this small underfit residing in Figure \ref{fig:umean}R we
preferably gained an universal (Reynolds number independent)
reference for all remaining scaling laws still to be fitted in
this section
--- however only under the critical assumption of course that the
universality principle holds.

Note that we excluded, on purpose, the lower end range from
$y^+\sim 30$ to $y^+\sim 100$ as part of the log-region. Our aim
here is to operate and to continue the investigation with a
convincing fit for the log-law (\ref{131203:0854}). Including the
range below $y^+\sim 100$ would only deteriorate the quality of
the fit (by nearly one order of magnitude in $\chi^2_\text{red}$),
since this range clearly shows a small overshoot which noticeably
deviates from a typical log-behavior in both the experimental and
the DNS data as seen in Figures \ref{fig:umean}L and
\ref{fig:umean}R.

Finally to note is that in {\it all} plots to be discussed from
hereon, the solid line represents the DNS data for $Re_\theta=
6000$ from \cite{Borrell13}, while the dashed line represents the
corresponding {\it best-fit} according to the functional form as
given in (\ref{140929:2347}). To enforce consistency, all fits
will only be performed within the shaded region covering here the
inertial range between $y^+\sim 100$ and $y^+\sim 330$.
Additionally, all fits will be compared relative to the underlying
DNS error in a consistent manner by employing either the
corresponding zero-fields or the budget residuals. The explicit
values of the fitted parameters will be stated up to three digits
precise in the corresponding caption of each figure.\newpage

\begin{figure}
\centering
\begin{minipage}[c]{.48\linewidth}
\FigureXYLabel{\includegraphics[width =
.91\textwidth]{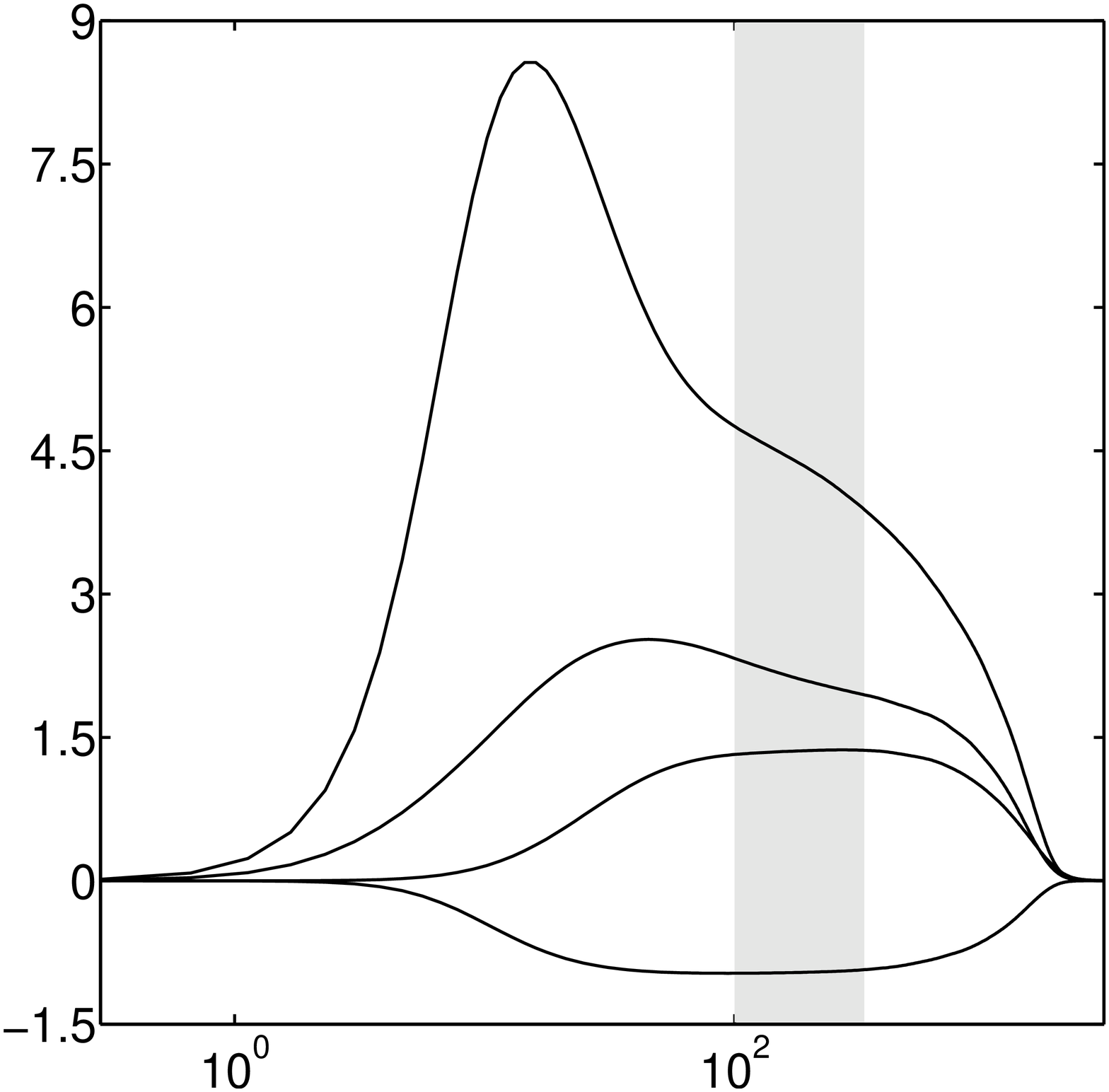}} {$y^+$}{.5mm}{\begin{rotate}{90}
$\tau^{ij+}$ \end{rotate}}{0mm}
\end{minipage}
\begin{tikzpicture}[overlay]
\fill[fill=white] (-4.65, 2.50) node[fill=white] {$\tau^{11}$};
\fill[fill=white] (-4.65,-2.10) node[fill=white] {$\tau^{12}$};
\fill[fill=white] (-3.05,-1.50) node[fill=white] {$\tau^{22}$};
\fill[fill=white] (-3.05, 0.15) node[fill=white] {$\tau^{33}$};
\end{tikzpicture}
\hfill
\begin{minipage}[c]{.48\linewidth}
\FigureXYLabel{\includegraphics[width=.91\textwidth]{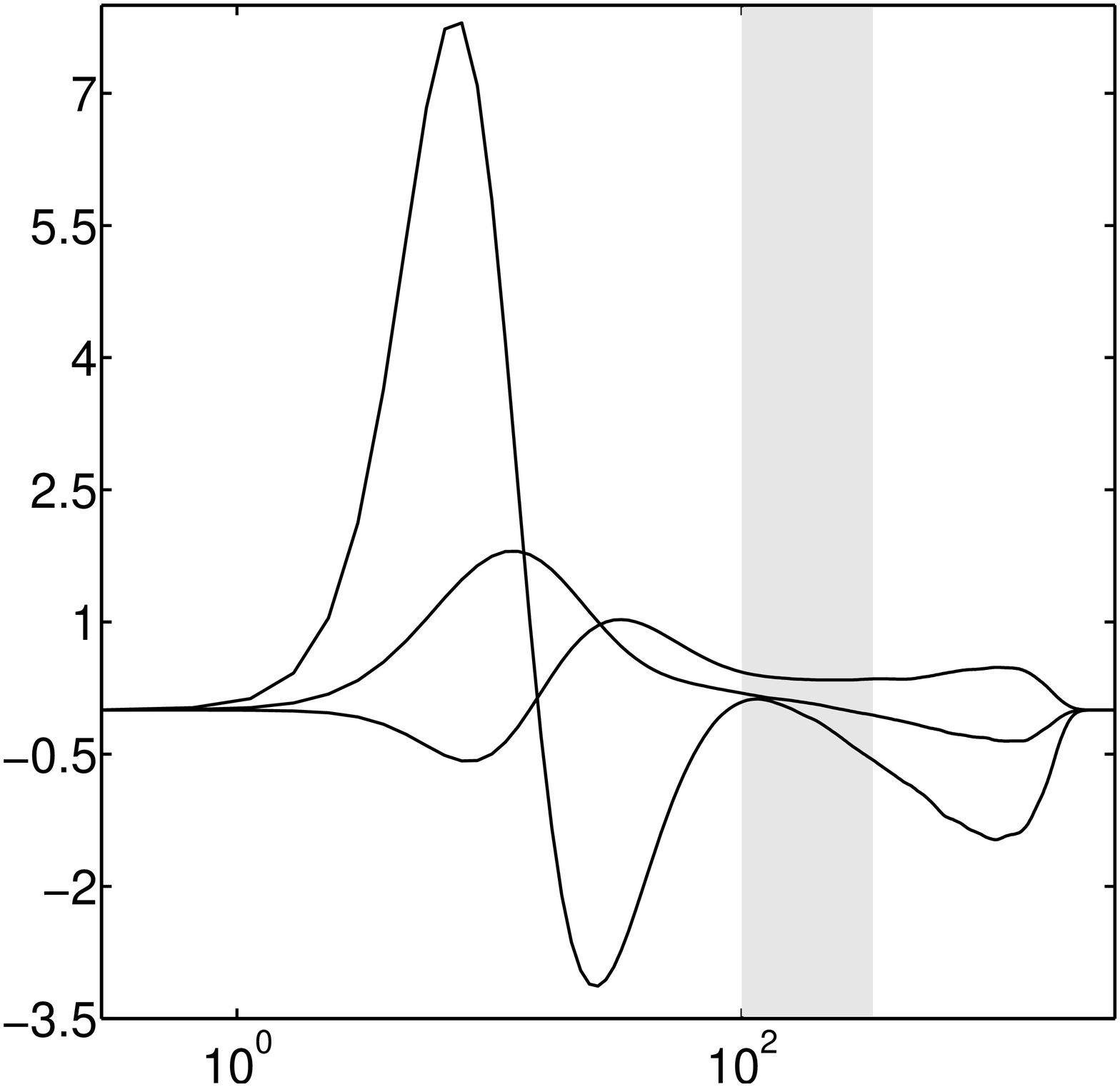}}
{$y^+$}{.5mm}{\begin{rotate}{90} $T^{ijk+}$
\end{rotate}}{2mm}
\end{minipage}
\begin{tikzpicture}[overlay]
\fill[fill=white] (-3.35,2.5) node[fill=white] {$T^{111}$};
\fill[fill=white] (-1.0,0.0) node[fill=white] {$T^{112}$};
\fill[fill=white] (-3.05,0.475) node[fill=white] {$T^{331}$};
\end{tikzpicture}
\caption{{\it Left plot:} All Reynolds stresses for $Re_\theta=
6000$ \citep{Borrell13} plotted over the complete wall-normal
range.\\
{\it Right plot:} A selection of third order velocity moments for
$Re_\theta= 6000$ \citep{Borrell13} plotted in the same
wall-normal range as left. In both plots the grey-shaded area
shows the inertial region.} \label{fig:triple}
\end{figure}

Before we start investigating the higher order moments inside the
shaded inertial range, it is helpful to first visualize the second
and third order velocity moments in the complete wall-normal
domain of the flow. While Figure \ref{fig:triple}L presents all
second moments, Figure \ref{fig:triple}R only shows those third
moments which were selected later on for a detailed investigation.
In both plots the shaded inertial range was made visible in order
to get a strong comparative impression of the complete functional
structure inside and outside that domain.

\begin{figure}
\begin{minipage}[c]{.48\linewidth}
\centering
\FigureXYLabel{\includegraphics[width=.91\textwidth]{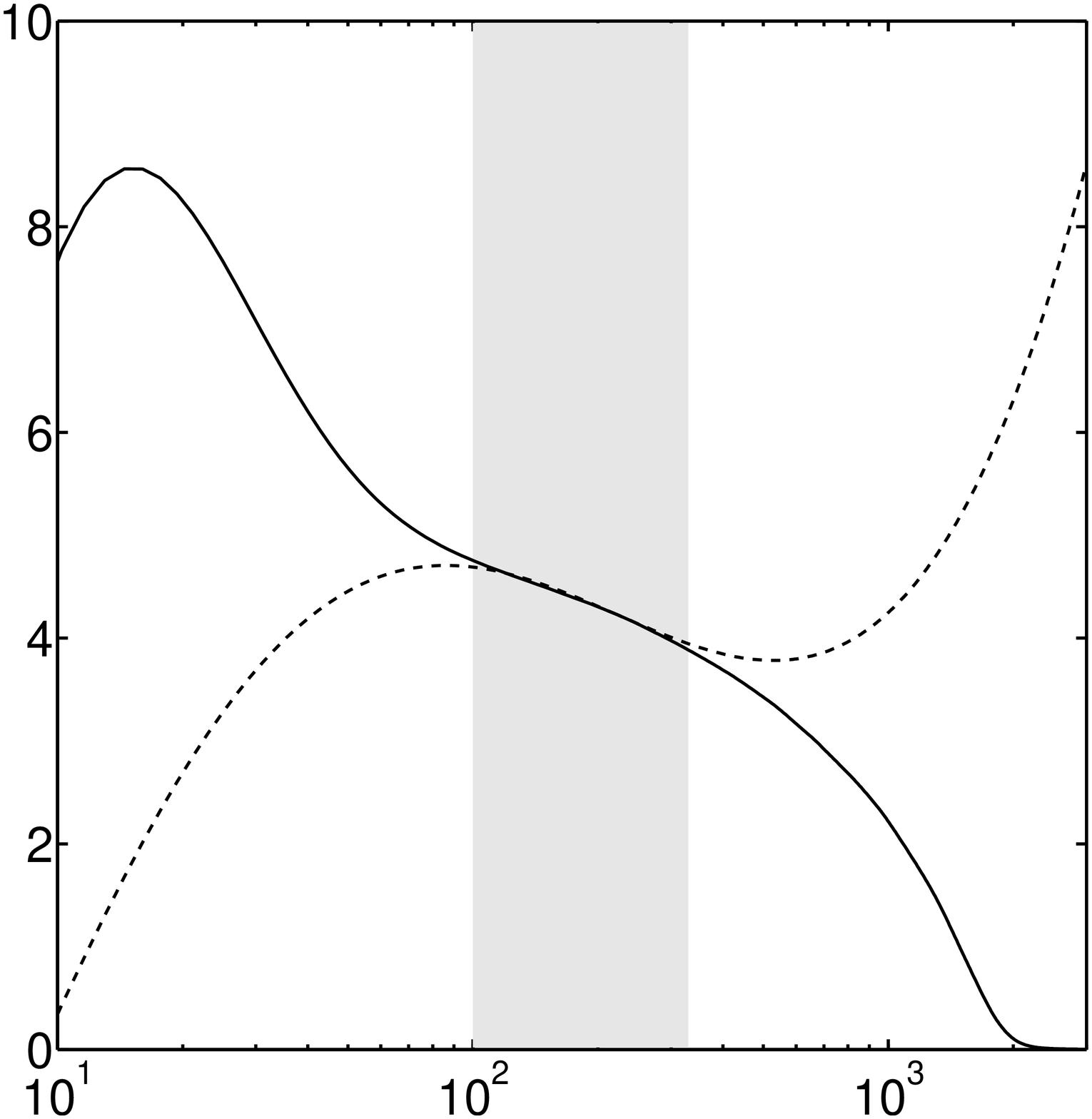}}
{$y^+$}{.5mm}{\begin{rotate}{90} $\tau^{11+}$ \end{rotate}}{1mm}
\end{minipage}
\hfill
\begin{minipage}[c]{.48\linewidth}
\centering
\FigureXYLabel{\includegraphics[width=.91\textwidth]{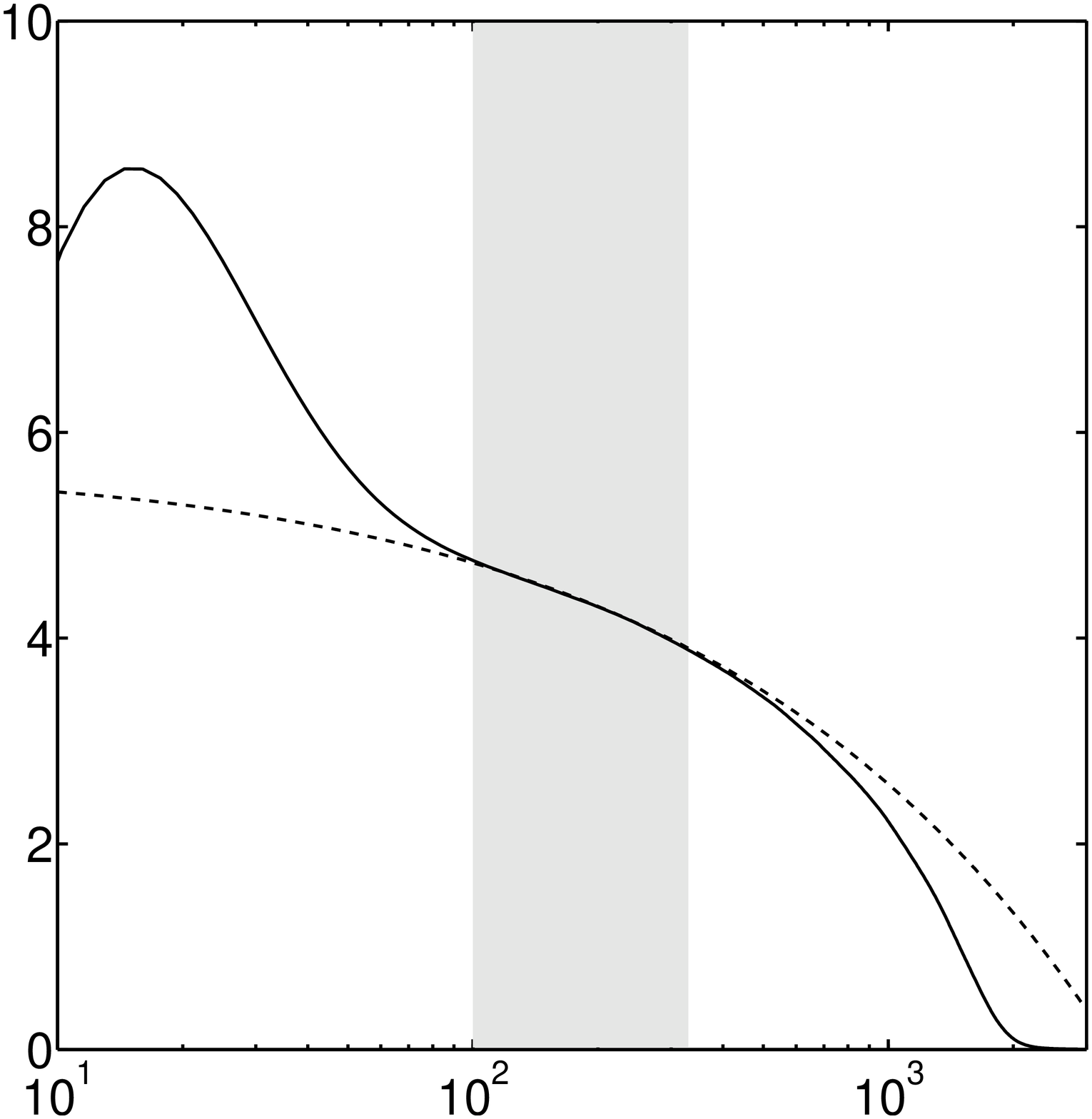}}
{$y^+$}{.5mm}{\begin{rotate}{90} $\hat{\tau}^{11+}$
\end{rotate}}{1mm}
\end{minipage}
\caption{{\it Left plot:} Best-fit of $\tau^{11+}$ according to
the new statistical scaling law (\ref{131203:1022}). The
corresponding parameters are $\alpha_H^{11}=-321.3$,
$\beta_H^{11}=302.6$, $\gamma=0.145$, with the quality measures
$\delta_{\text{fit}}= 5.5\cdot 10^{-3}$, $\chi^2_\text{red}=
3.3\cdot 10^{2}$.
\\
{\it Right plot:} Best-fit of $\hat{\tau}^{11+}$ according to the
pure power-law (\ref{131203:1020}), without the unphysical
log-squared-term $U^{+2}_{\text{exp}}$ appearing in the new
scaling law (\ref{131203:1022}). The corresponding parameters are
$\hat{\alpha}_H^{11}=5.893$, $\hat{\beta}_H^{11}=-0.134$,
$\hat{\gamma}=0.464$, with the quality measures
$\delta_{\text{fit}}= 1.9\cdot 10^{-3}$, $\chi^2_\text{red}=
3.1\cdot 10^{1}$.} \label{fig:tau11}
\end{figure}

Figure \ref{fig:tau11}L shows the {\it best-fit} for the
Reynolds-stress component (\ref{140929:2347})
\begin{equation}
\tau^{11+}=\alpha_H^{11}+\beta_H^{11}\cdot(y^+
+c_{\text{exp}})^{\gamma}-U^{+2}_{\text{exp}}, \label{131203:1022}
\end{equation}
the new mixed scaling law first derived in \cite{Oberlack10} as
the zero-correlation-length limit of the corresponding 2-point
function, which next to a power-law also includes the universal
log-law $U^{+}_{\text{exp}}$ (\ref{131203:0854}). Figure
\ref{fig:tau11}R however shows the {\it best-fit} for
\begin{equation}
\hat{\tau}^{11+}=\hat{\alpha}_H^{11}+\hat{\beta}_H^{11}\cdot(y^+
+c_{\text{exp}})^{\hat{\gamma}}, \label{131203:1020}
\end{equation}
an alternative three-parametric scaling law, which, just for the
sake of interest, excludes the unphysical log-part
$U^{+2}_{\text{exp}}$ from the newly proposed scaling law
(\ref{131203:1022}). Our motivation to use (\ref{131203:1020}) as
comparison to (\ref{131203:1022}) stems from the idea to observe
the difference in functional behavior when a {\it physical}
subgroup of
$\mathsf{T}\circ\mathsf{S}_\mathsf{1}\circ\mathsf{S}_\mathsf{2}
\circ\mathsf{Q}_\mathsf{1}\circ\mathsf{Q}_\mathsf{2}$
(\ref{140303:1246})-(\ref{140110:1749}) is being considered, that
is, to compare function (\ref{131203:1022}) with an alternative
function which is not linked to the unphysical ``new statistical
symmetries" $\mathsf{Q}_\mathsf{1}$ (\ref{140110:1748}) and
$\mathsf{Q}_\mathsf{2}$ (\ref{140110:1749}).

\begin{figure}
\begin{minipage}[c]{.48\linewidth}
\centering
\FigureXYLabel{\includegraphics[width=.91\textwidth]{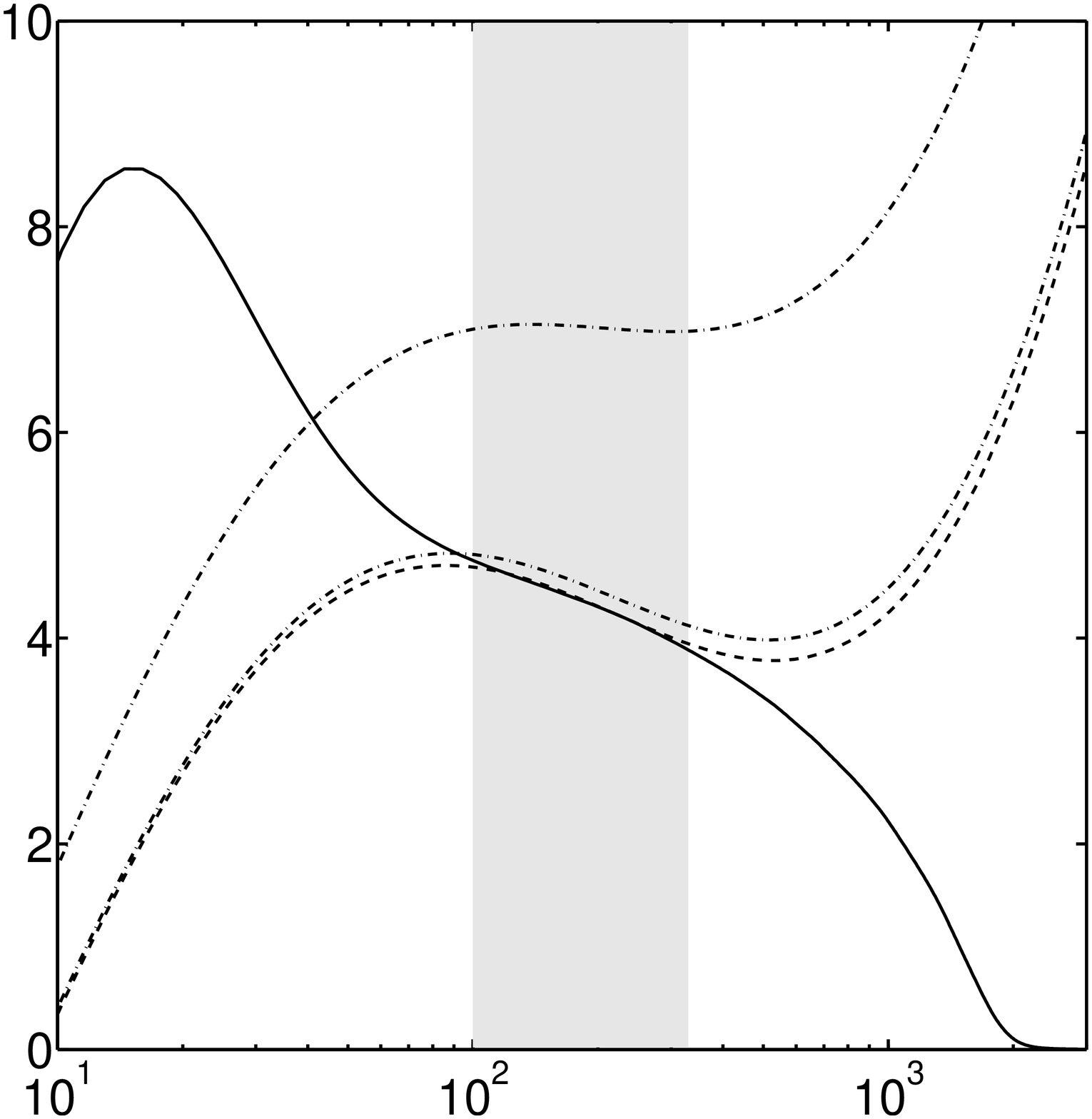}}
{$y^+$}{.5mm}{\begin{rotate}{90} $\tau^{11+}$ \end{rotate}}{1mm}
\end{minipage}
\hfill
\begin{minipage}[c]{.48\linewidth}
\centering
\FigureXYLabel{\includegraphics[width=.91\textwidth]{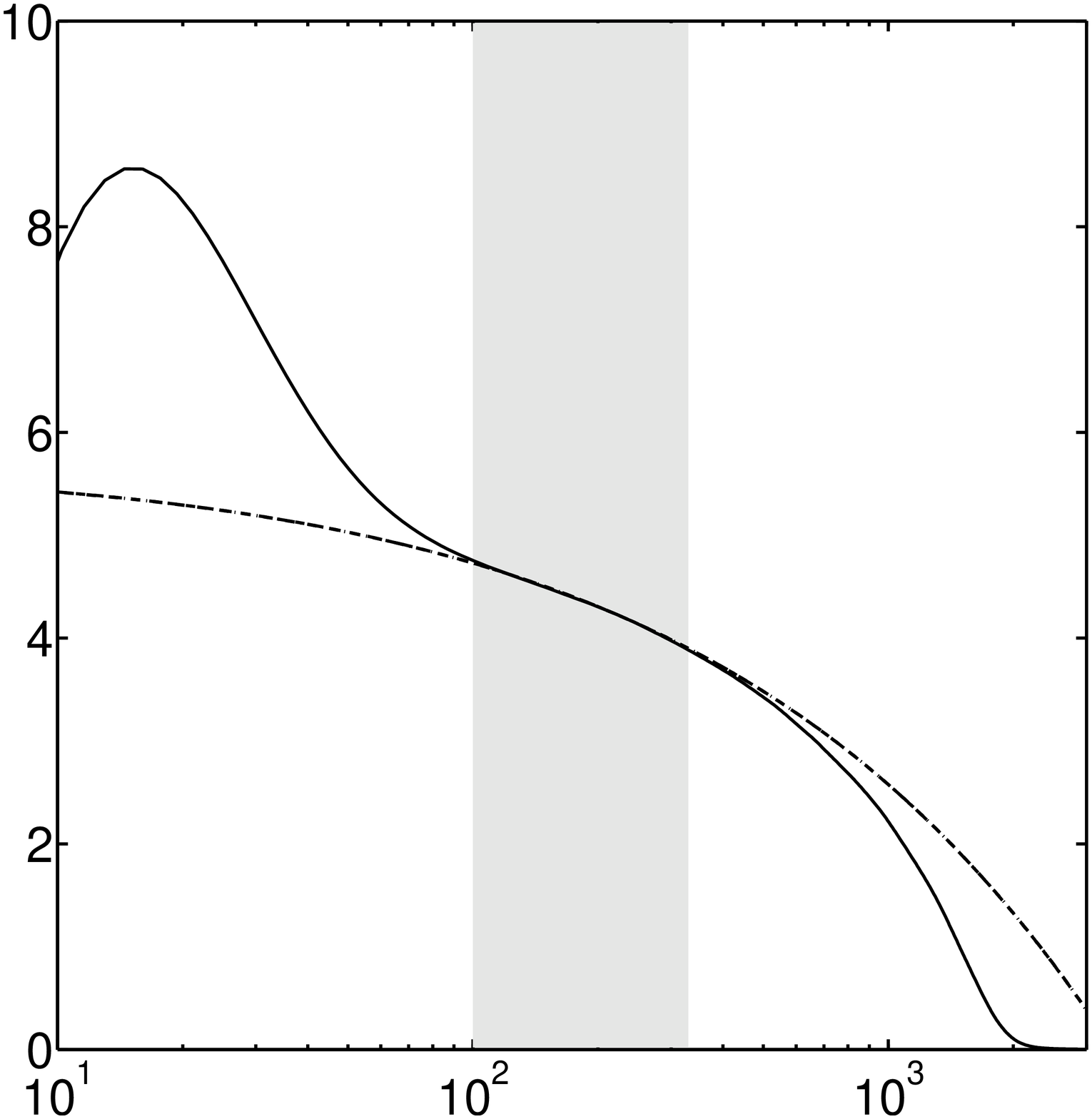}}
{$y^+$}{.5mm}{\begin{rotate}{90} $\hat{\tau}^{11+}$
\end{rotate}}{1mm}
\end{minipage}
\caption{{\it Left plot:} Best-fit of $\tau^{11+}$ according to
the new statistical scaling law (\ref{131203:1022}) using
different fitting precisions. It shows the high sensitivity
against small perturbations in the parameters values, when going
from six digits (lower and best matching fit), to four digits
(middle fit) down to three digits (upper fit). The corresponding
parameters are given in the left part of the table below.
\\
{\it Right plot:} Best-fit of $\hat{\tau}^{11+}$ according to the
power-law (\ref{131203:1020}), without the unphysical
log-squared-term $U^{+2}_{\text{exp}}$ appearing in the new
scaling law (\ref{131203:1022}). In contrast to the fit in the
left plot it shows complete insensitivity against small
perturbations in the parameters values when going from six down to
three digits. The corresponding parameters are given in the right
part of the~table~below.}
\begin{center}
\vspace{-0.75em}
\begin{tabular}{c|c|c|c||c|c|c}
Digits &  $\alpha_H ^{11} \cdot 10^{-3}$
       &  $\beta_H ^{11} \cdot 10^{-3}$
       &  $\gamma $
       &  ${\hat\alpha}_H^{11} \cdot 10^{-1}$
       &  $\hat{\beta}_H^{11}$
       &  ${\hat\gamma}$\\[3pt]
\hline \hline
       $6$ & $\; -0.321261 \;$ & $\;  0.302558 \;$ & $\; 0.144571 \;$
           & $\;  0.589339 \;$ & $\; -0.134415 \;$ & $\; 0.463731 \;$\\
       $4$ & $-0.3213\phantom{00}$    & $0.3026\phantom{00}$
           & $0.1446\phantom{00}$     & $0.5893\phantom{00}$
           & $-0.1344\phantom{00}$    & $0.4637\phantom{00}$\\
       $3$ & $-0.321\phantom{000}$    & $0.303\phantom{000}$
           & $0.145\phantom{000}$     & $0.589\phantom{000}$
           & $-0.134\phantom{000}$    & $0.464\phantom{000}$\\
\end{tabular}
\\[0.5em]
{\small Parameters for Figure 4.}
\end{center}
\label{fig:tau11_S}\vspace{-1em}
\end{figure}

When excluding these two from the group
$\mathsf{T}\circ\mathsf{S}_\mathsf{1}\circ\mathsf{S}_\mathsf{2}
\circ\mathsf{Q}_\mathsf{1}\circ\mathsf{Q}_\mathsf{2}$, the
physical subgroup
$\mathsf{T}\circ\mathsf{S}_\mathsf{1}\circ\mathsf{S}_\mathsf{2}$
is obtained, from which then, if the scaling in the mean velocity
profile $U$ (\ref{130820:2209}) is not forced to be broken, only
pure power laws as given by (\ref{131203:1020}) can be induced as
invariant functions, thereby explicitly demonstrating that the
functional shift $U^{+2}_{\text{exp}}$ in (\ref{131203:1022}) has
its origin solely in the unphysical ``new statistical symmetries"
$\mathsf{Q}_\mathsf{1}$ and $\mathsf{Q}_\mathsf{2}$. However, the
reader should note that the power-law as specifically defined in
(\ref{131203:1020}) is not a genuine invariant function of the
considered physical subgroup
$\mathsf{T}\circ\mathsf{S}_\mathsf{1}\circ\mathsf{S}_\mathsf{2}$,
due to the constant offset $\hat{\alpha}_H^{11}$ we explicitly
included in order to ensure a same level of competitiveness as for
function (\ref{131203:1022}) which features three open parameters.
In other words, although (\ref{131203:1020}) {\it cannot} be
identified as an exact invariant function of
$\mathsf{T}\circ\mathsf{S}_\mathsf{1}\circ\mathsf{S}_\mathsf{2}$,
it is nevertheless motivated through this physical transformation
group to take the structural form of an invariant power-law,
providing for (\ref{131203:1020}) thus a far more stronger
physical background than
$\mathsf{T}\circ\mathsf{S}_\mathsf{1}\circ\mathsf{S}_\mathsf{2}
\circ\mathsf{Q}_\mathsf{1}\circ\mathsf{Q}_\mathsf{2}$ can provide
for scaling law (\ref{131203:1022}) --- a difference which becomes
easily visible now when determining and comparing their best fits.

Although the quality of both fits for (\ref{131203:1022}) and
(\ref{131203:1020}) is comparable relative to the statistical
measure $\delta_\text{fit}$\footnote[2]{The dimensionless quality
measure $\delta_\text{fit}$ is the normalized root mean squared
error relative to the maximum value inside the considered domain
to be fitted.}, we clearly see however that the fit in Figure
\ref{fig:tau11}L for (\ref{131203:1022}) definitely has a far more
{\it unnatural} functional behavior than the fit in Figure
\ref{fig:tau11}R for (\ref{131203:1020}). In other words, although
the fit on the left is of nearly equal quality as that on the
right regarding its residual $\delta_\text{fit}$, the scaling law
(\ref{131203:1020}), which {\it excludes} the unphysical log-term
$U^{+2}_{\text{exp}}$, follows the DNS data more naturally on the
right than the scaling law (\ref{131203:1022}) on the left of
Figure~\ref{fig:tau11}, which {\it includes} the log-term
$U^{+2}_{\text{exp}}$. This is also quantitatively expressed by
the fact that the two fitting parameters $\alpha_H^{11}$ and
$\beta_H^{11}$ in the left plot resulted in far higher values than
$\hat{\alpha}_H^{11}$ and $\hat{\beta}_H^{11}$ in the plot on the
right, showing that in order to optimally fit an unnatural
behavior is at the expense of having to use large (unnatural)
values in the parameters.

Additionally the fit in Figure~\ref{fig:tau11}L is by far more
sensitive to small changes in the parameters than the fit in
Figure~\ref{fig:tau11}R, which, in contrast to the left plot,
shows a very robust behavior against small perturbations in the
parameter values. In order to reproduce Figure~\ref{fig:tau11}L at
least six digits of precision are necessary in the values for the
fitting parameters: $\alpha^{11}_H=-0.321261\cdot 10^3$,
$\beta^{11}_H=0.302558\cdot 10^3$, $\gamma=0.144571$. To reproduce
however Figure~\ref{fig:tau11}R only three digits of precision
turn out to be already sufficient. This very strong dependence
upon small changes in the parameters demonstrates the artificial
behavior of the new scaling law (\ref{131203:1022}), as its
corresponding fit in Figure \ref{fig:tau11}L can only be generated
under great effort in maintaining a high (unnatural) degree of
precision. Figure \ref{fig:tau11_S} explicitly shows and compares
this behavior, thus explicitly displaying that (\ref{131203:1022})
essentially cannot be regarded as a physically useful scaling law.

Another measure is $\chi^2_\text{red}$, which in Figure
\ref{fig:tau11}L is greater than one and by one magnitude larger
than in Figure \ref{fig:tau11}R, indicating that the mixed scaling
law (\ref{131203:1022}) of Oberlack et al. underfits the data more
strongly than an alternative scaling law (\ref{131203:1020}) with
a pure power-law behavior would do, hence ultimately indicating
again that relative to the underlying DNS error the pure power-law
in Figure \ref{fig:tau11}R fits the DNS data more naturally. For
both fits the local error field was taken to be the zero-field
$\tau^{13+}$.

\begin{figure}
\centering
\begin{minipage}[c]{.48\linewidth}
\FigureXYLabel{\includegraphics[width=.91\textwidth]{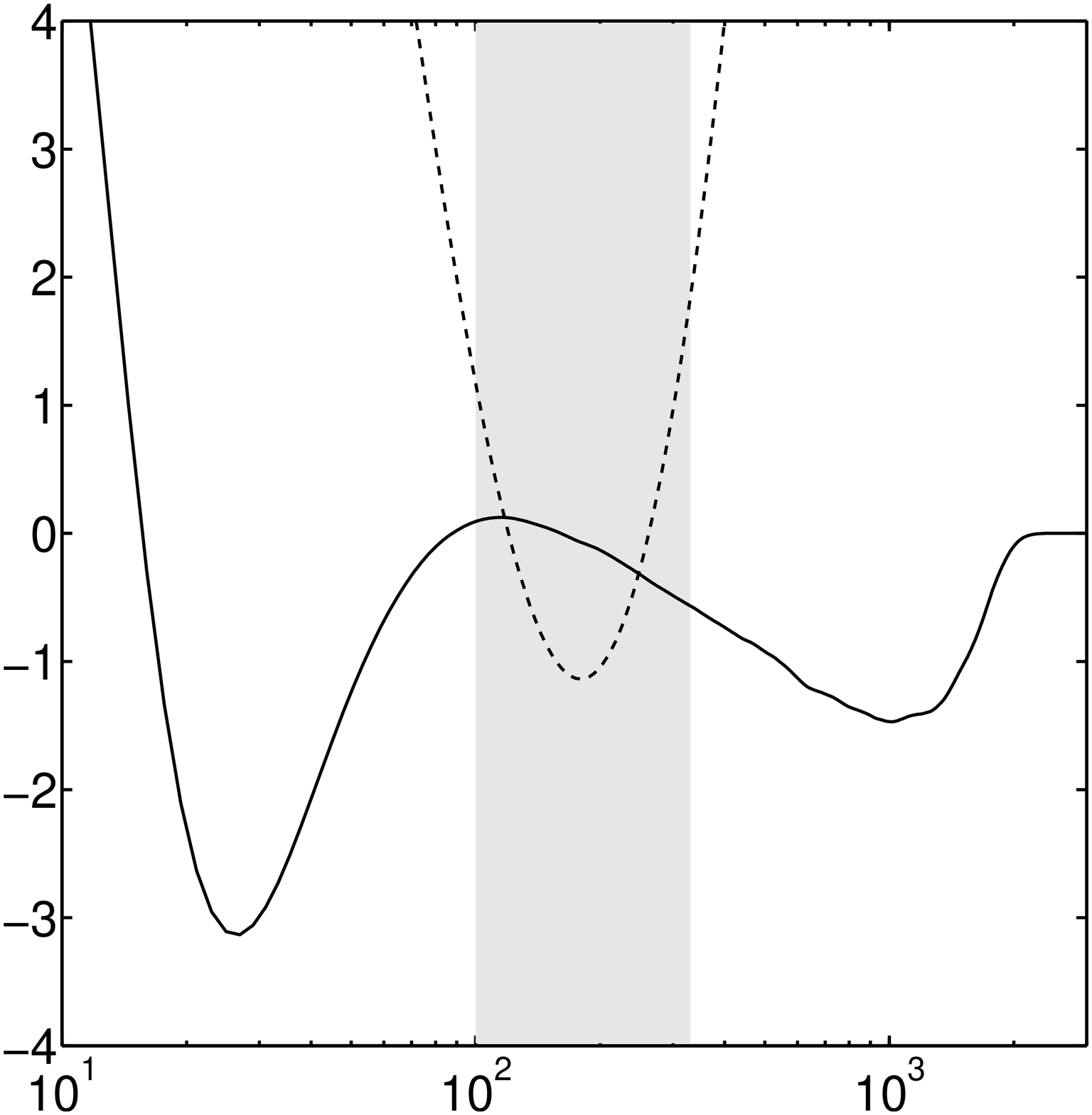}}
{$y^+$}{.5mm}{\begin{rotate}{90} $T^{111+}$
\end{rotate}}{1.5mm}
\end{minipage}
\hfill
\begin{minipage}[c]{.48\linewidth}
\centering
\FigureXYLabel{\includegraphics[width=.91\textwidth]{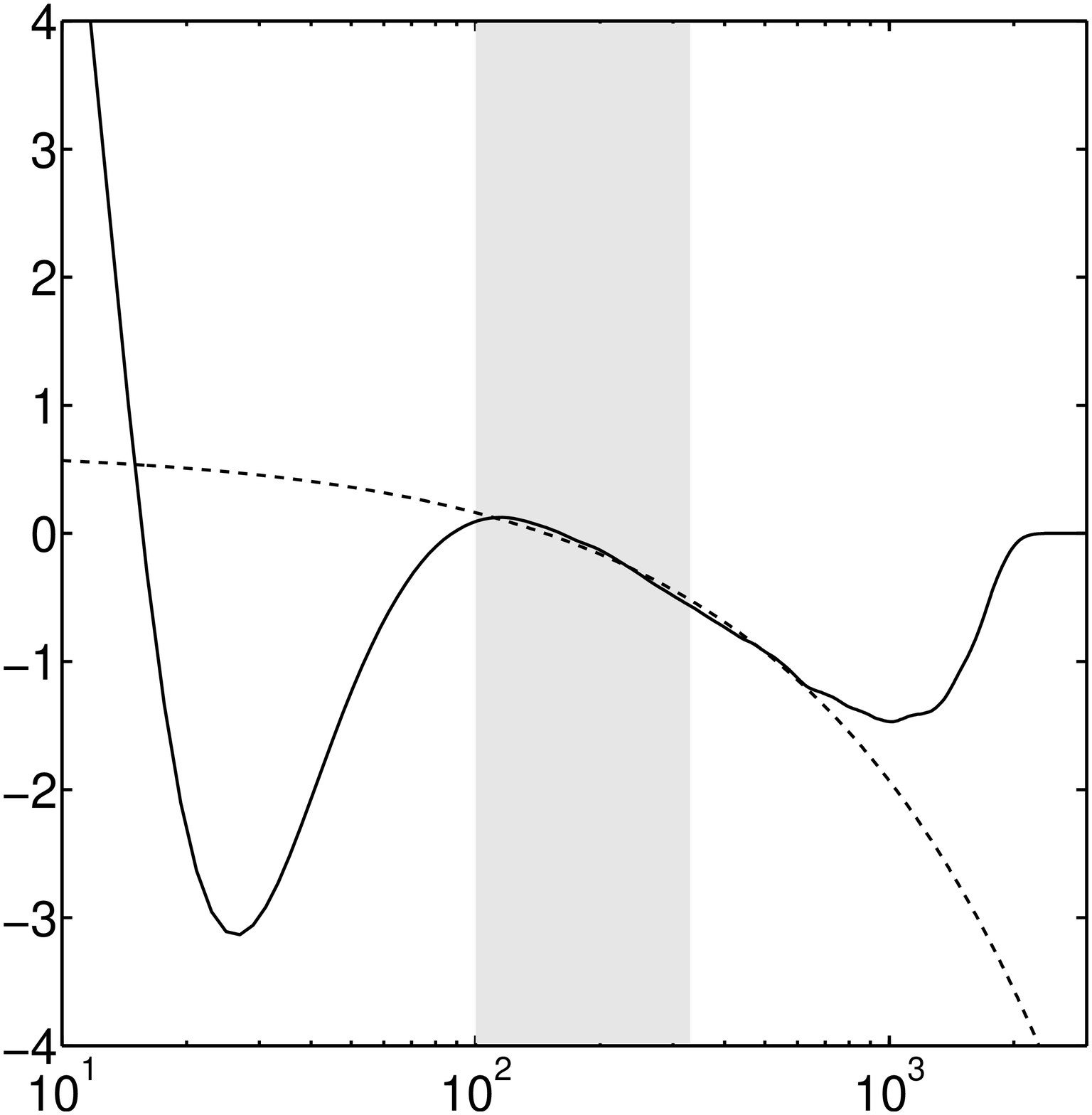}}
{$y^+$}{.5mm}{\begin{rotate}{90} $\hat{T}^{111+}$
\end{rotate}}{1.5mm}
\end{minipage}
\caption{{\it Left plot:} Best-fit of $T^{111+}$ according to the
new statistical scaling law (\ref{131101:1806}). Based on the
result $\gamma=0.145$ from Figure \ref{fig:tau11}L, the best
fitted parameters are $\alpha_H^{111}=-2755$,
$\beta_H^{111}=1913$, with the quality measures
$\delta_{\text{fit}}= 1.8$, $\chi^2_\text{red}= 7.8\cdot 10^{6}$.
\\
{\it Right plot:} Best-fit of $\hat{T}^{111+}$ according to the
pure power-law (\ref{131206:2034}), excluding both physically
inconsistent log-terms appearing in the new scaling law
(\ref{131101:1806}). Based on the result $\hat{\gamma}=0.464$ from
Figure \ref{fig:tau11}R, the corresponding parameters are
$\hat{\alpha}_H^{111}=0.711$, $\hat{\beta}_H^{111}=-0.021$, with
the quality measures $\delta_{\text{fit}}= 6.6\cdot 10^{-2}$,
$\chi^2_\text{red}= 1.7\cdot 10^{4}$.} \label{fig:tau111}
\end{figure}

This fitting procedure is then repeated for the streamwise triple
velocity moment $T^{111+}$ shown in Figure \ref{fig:tau111}L and
Figure \ref{fig:tau111}R. Without even referring to any
statistical measures one clearly sees that the corresponding mixed
scaling law of Oberlack et al. (\ref{140929:2347})
\begin{equation}
T^{111+} = \alpha_H^{111}+\beta_H^{111}\cdot(y^+
+c_{\text{exp}})^{2\gamma} -3U^+_{\text{exp}}\cdot\tau^{11+}-U^{+
3}_{\text{exp}}, \label{131101:1806}
\end{equation}
shown in Figure \ref{fig:tau111}L, is completely incapable to
predict the corresponding DNS data in the shaded inertial region
correctly. To guarantee for a consistent fit of
(\ref{131101:1806}) the parameters $\gamma$ and those of
$\tau^{11+}$ were taken as determined in the previous fit for
Figure \ref{fig:tau11}L, thus essentially dealing only with a
2-parametric fitting function for $\alpha_H^{111}$ and
$\beta_H^{111}$. In stark contrast, Figure \ref{fig:tau111}R shows
a huge improvement in the fitting results as soon as the two
unphysical log-terms appearing in the above scaling law are
removed, and, instead of (\ref{131101:1806}), the following pure
power-law is used
\begin{equation}
\hat{T}^{111+} = \hat{\alpha}_H^{111}+\hat{\beta}_H^{111}\cdot(y^+
+c_{\text{exp}})^{3\hat{\gamma}/2}. \label{131206:2034}
\end{equation}
As before for the Reynolds stress $\hat{\tau}^{11+}$
(\ref{131203:1020}), this is again motivated by us to compare the
newly proposed function (\ref{131101:1806}) to an alternative
scaling law which is not linked to the unphysical ``new
statistical symmetries" $\mathsf{Q}_\mathsf{1}$
(\ref{140110:1748}) and $\mathsf{Q}_\mathsf{2}$
(\ref{140110:1749}), as they form the origin of the peculiar
functional shift $(-3U^+_{\text{exp}}\cdot\tau^{11+}-U^{+
3}_{\text{exp}})$ in (\ref{131101:1806}). Carefully note here
again, that although the power-law for the triple moment
(\ref{131206:2034}) with exponent $3\hat{\gamma}/2$ has been
consistently generated from the physical subgroup
$\mathsf{T}\circ\mathsf{S}_\mathsf{1}\circ\mathsf{S}_\mathsf{2}\subset
\mathsf{T}\circ\mathsf{S}_\mathsf{1}\circ\mathsf{S}_\mathsf{2}
\circ\mathsf{Q}_\mathsf{1}\circ\mathsf{Q}_\mathsf{2}$
(\ref{140303:1246})-(\ref{140110:1749}) in accord with the
power-law for the double moment (\ref{131203:1020}) with exponent
$\hat{\gamma}$, the complete scaling function as specifically
defined in (\ref{131206:2034}) is not a genuine invariant function
of the considered physical subgroup
$\mathsf{T}\circ\mathsf{S}_\mathsf{1}\circ\mathsf{S}_\mathsf{2}$
itself. This is due to the constant offset $\hat{\alpha}_H^{111}$
which we included again to ensure for both functions
(\ref{131101:1806}) and (\ref{131206:2034}) the same size of
parameter space, as only this will allow for an equal and fair
comparison.

\begin{figure}
\begin{minipage}[c]{.48\linewidth}
\centering
\FigureXYLabel{\includegraphics[width=.91\textwidth]{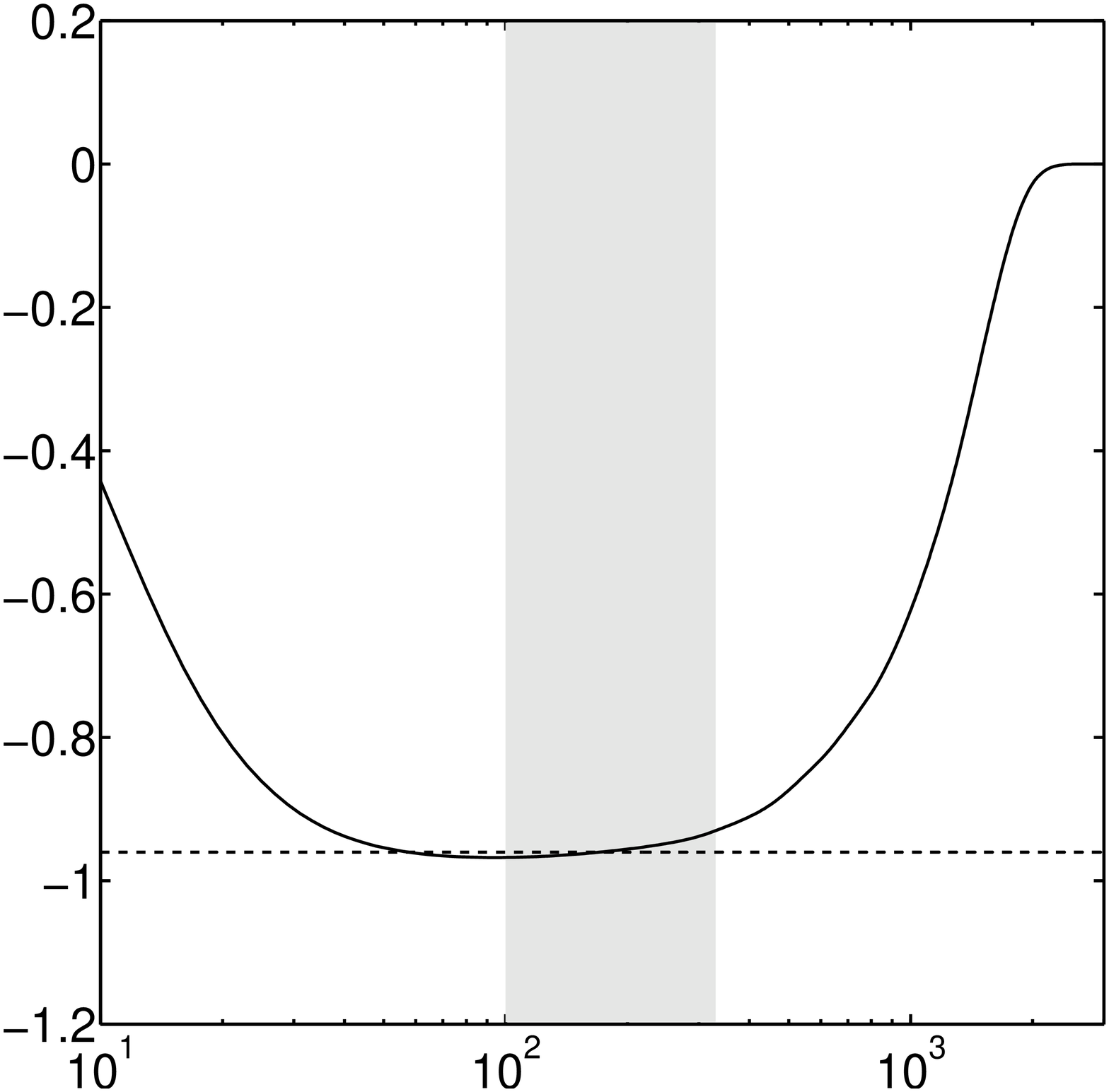}}
{$y^+$}{.5mm}{\begin{rotate}{90} $\tau^{12+}$ \end{rotate}}{1mm}
\end{minipage}
\hfill
\begin{minipage}[c]{.48\linewidth}
\centering
\FigureXYLabel{\includegraphics[width=.91\textwidth]{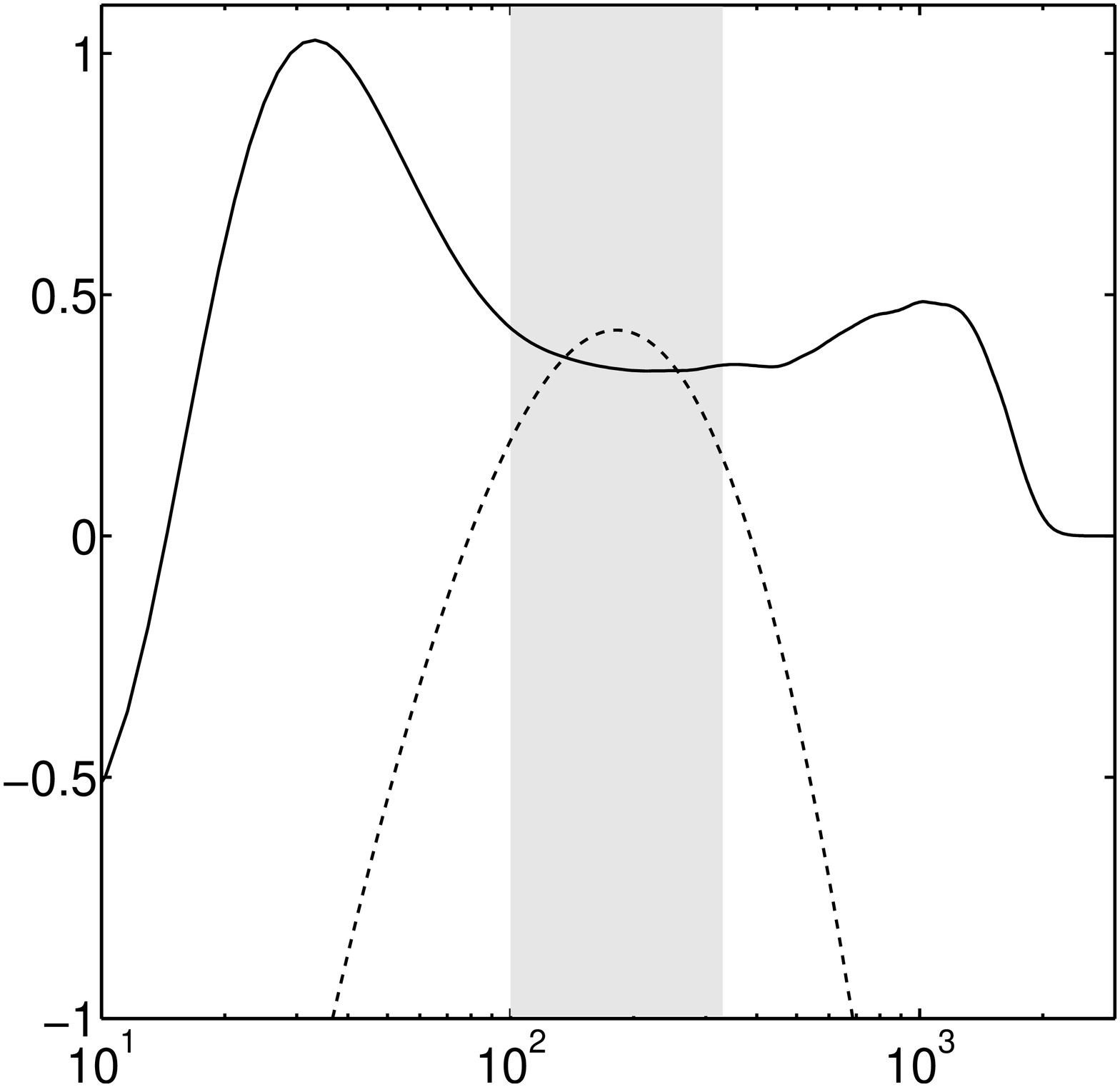}}
{$y^+$}{.5mm}{\begin{rotate}{90} $T^{112+}$
\end{rotate}}{1.5mm}
\end{minipage}
\caption{{\it Left plot:} Best-fit of $\tau^{12+}$
(\ref{131206:2100}). The fitted parameters are
$\alpha_H^{12}=-0.961$ and $\beta_H^{12}=0$, with the quality
measures $\delta_{\text{fit}}= 1.4\cdot 10^{-2}$,
$\chi^2_\text{red}= 1.5\cdot 10^{2}$.
\\
{\it Right plot:} Best-fit of $T^{112+}$ according to the new
statistical scaling law (\ref{131206:2310}). Based on the result
$\tau^{12+}=-0.961$ from the left plot in this figure, the best
fitted parameters are $\alpha_H^{112}=-16.38$,
$\beta_H^{112}=-3.851$, with the quality measures
$\delta_{\text{fit}}= 2.2\cdot 10^{-1}$, $\chi^2_\text{red}=
4.9\cdot 10^{6}$.} \label{fig:tau12}
\end{figure}

Referring to $\chi^2_\text{red}$ in Figure \ref{fig:tau111}, the
improvement in choosing the pure power-law (\ref{131206:2034})
instead of the mixed law (\ref{131101:1806}) spans nearly three
orders of magnitude. Of course, to guarantee for a consistent fit
in this case, too, the parameter $\hat{\gamma}$ was taken from the
previous result of Figure \ref{fig:tau11}R. For both fits in
Figure \ref{fig:tau111}L and Figure \ref{fig:tau111}R the local
error field was approximated by the second moment zero-field
$(\tau^{13+})^{3/2}$, as unfortunately no zero-fields for third
order velocity moments were generated during the DNS.

\begin{figure}
\begin{minipage}[c]{.48\linewidth}
\centering
\FigureXYLabel{\includegraphics[width=.91\textwidth]{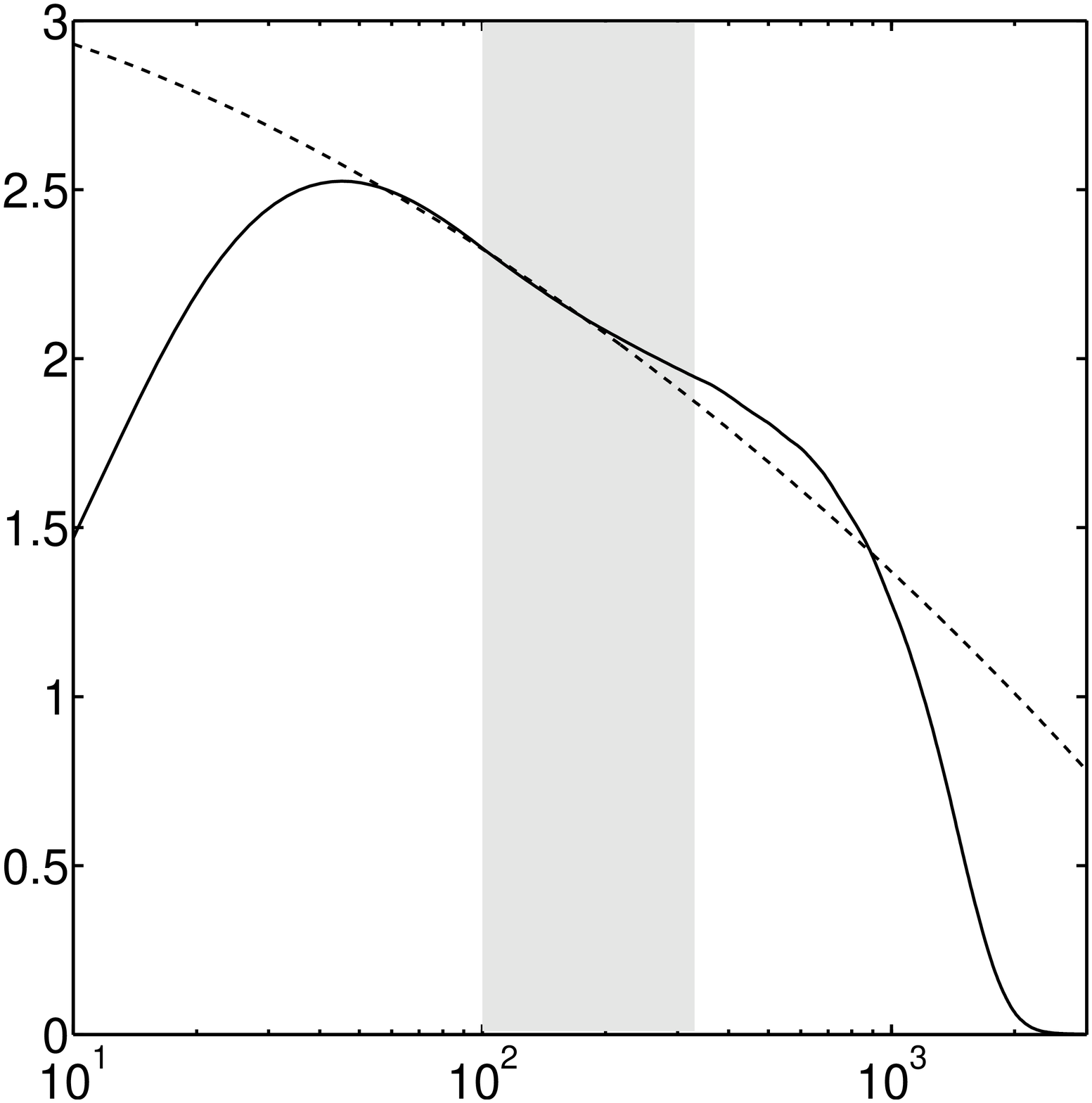}}
{$y^+$}{.5mm}{\begin{rotate}{90} $\tau^{33+}$ \end{rotate}}{1mm}
\end{minipage}
\hfill
\begin{minipage}[c]{.48\linewidth}
\centering
\FigureXYLabel{\includegraphics[width=.91\textwidth]{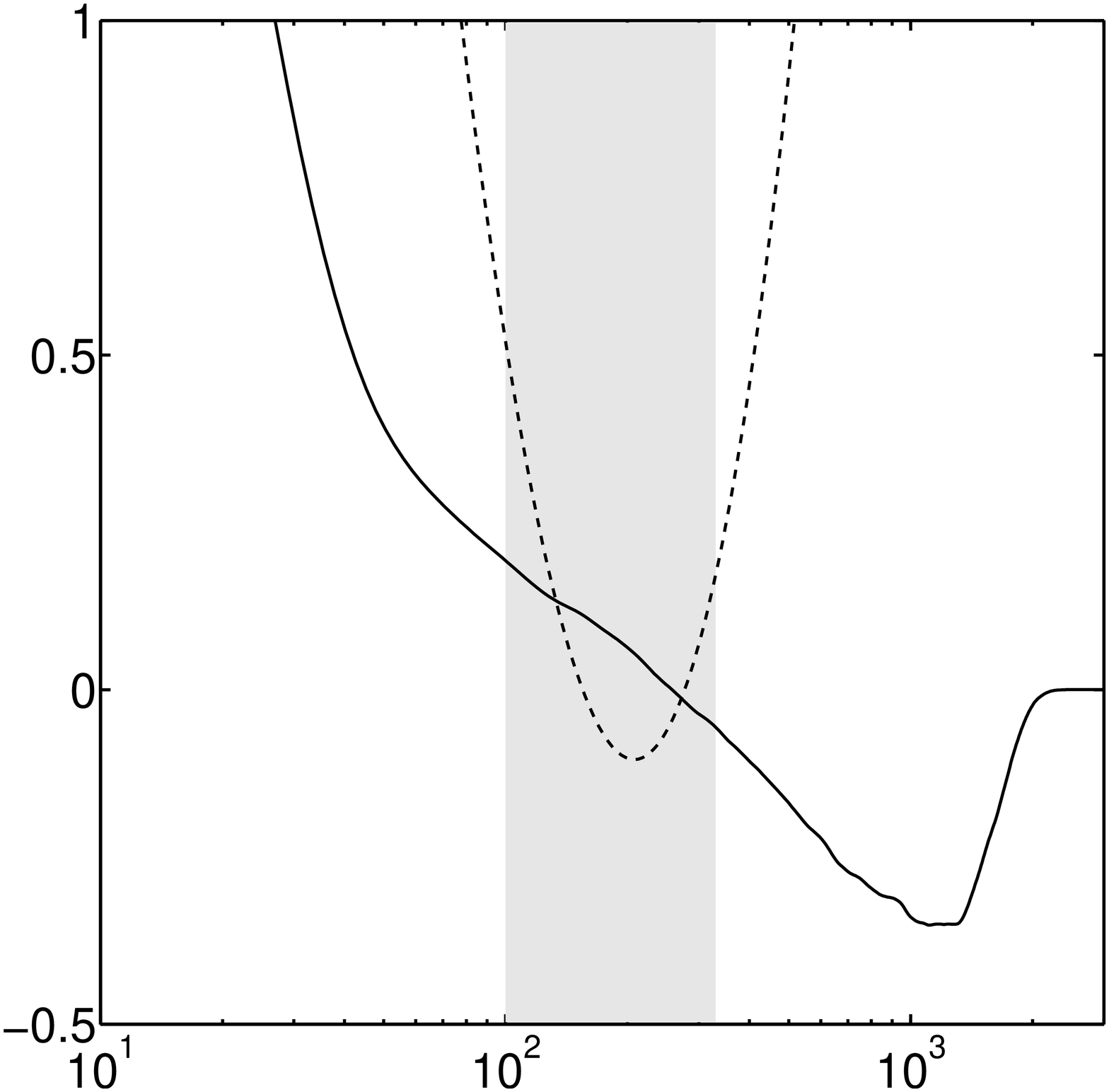}}
{$y^+$}{.5mm}{\begin{rotate}{90} $T^{331+}$
\end{rotate}}{1.5mm}
\end{minipage}
\caption{{\it Left plot:} Best-fit of $\tau^{33+}$ according to
(\ref{131212:1719}). Based on the result $\gamma=0.145$ from
Figure \ref{fig:tau11}L, the parameters are $\alpha_H^{33}=4.798$,
$\beta_H^{33}=-1.262$, with the quality measures
$\delta_{\text{fit}}= 1.4\cdot 10^{-2}$, $\chi^2_\text{red}=
1.7\cdot 10^{2}$.
\\
{\it Right plot:} \hspace{-0.089cm} Best-fit of $T^{331+}$
according to the new statistical scaling law (\ref{131212:1721}).
Based on the result $\tau^{33+}$ from the left plot in this
figure, the best fitted parameters are $\alpha_H^{331}=43.56$,
$\beta_H^{331}=-1.315$, with the quality measures
$\delta_{\text{fit}}= 7.3\cdot 10^{-1}$, $\chi^2_\text{red}=
6.2\cdot 10^{5}$.} \label{fig:tau33}
\end{figure}

Figure \ref{fig:tau12}L shows the {\it best-fit} for the
off-diagonal stress component (\ref{140929:2347})
\begin{equation}
\tau^{12+}  = \alpha_H^{12}+\beta_H^{12}\cdot(y^+
+c_{\text{exp}})^{\gamma}, \label{131206:2100}
\end{equation}
where the local error field was taken to be the zero-field
$\tau^{23+}$. This result was then absorbed into the next fit
shown in Figure \ref{fig:tau12}R, which again clearly demonstrates
the poor prediction ability of an Oberlack et al. proposed scaling
law of mixed type. In particular, Figure \ref{fig:tau12}R shows
the {\it best-fit} of the triple velocity moment
(\ref{140929:2347})
\begin{equation}
T^{112+}  = \alpha_H^{112}+\beta_H^{112}\cdot(y^+
+c_{\text{exp}})^{2\gamma} -2U^+_{\text{exp}}\cdot\tau^{12+},
\label{131206:2310}
\end{equation}
with the fixed parameter $\gamma=0.145$, which was already
consistently determined for $\tau^{11+}$ in Figure
\ref{fig:tau11}L. Here the local error field was approximated by
the zero-field $(\tau^{23+})^{3/2}$.

Finally, Figure \ref{fig:tau33}L shows the {\it best-fit} for the
Reynolds-stress component (\ref{140929:2347})
\begin{equation}
\tau^{33+}=\alpha^{33}_H+\beta^{33}_H\cdot(y^+ +
c_{\text{exp}})^\gamma, \label{131212:1719}
\end{equation}
and Figure \ref{fig:tau33}R the {\it best-fit} for the third order
velocity moment (\ref{140929:2347})
\begin{equation}
T^{331+}=\alpha_H^{331}+\beta_H^{331}\cdot(y^+
+c_{\text{exp}})^{2\gamma} -U^+_{\text{exp}}\cdot\tau^{33+},
\label{131212:1721}
\end{equation}
facing again the same poor quality issues. The zero-fields
$\tau^{23+}$ and $(\tau^{23+})^{3/2}$ were respectively chosen
again as the underlying DNS error.

\newgeometry{left=3cm,right=3cm,top=2.1cm,bottom=1.75cm,headsep=1em}
\section{Concluding remarks\label{Sec6}}

The investigation done in the previous section could independently
demonstrate that the ``new statistical scaling laws", which in
particular are based on the ``new scaling symmetry"
$\mathsf{Q}_\mathsf{E}$ (\ref{140530:2112}) recently proposed in
\cite{Oberlack10}, are by no means useful and, in our opinion,
should be discarded in future work as they clearly fail to fulfil
the most basic predictive requirements of a scaling law. This is
addressed in particular to those scalings which are of mixed type,
containing next to a power-law also a log-law, where the latter,
due to using for the statistical moments an overall inconsistent
invariance analysis  in not properly incorporating the underlying
deterministic theory, arise as unphysical terms proportional to
the mean streamwise velocity profile.

Section \ref{Sec4} explicitly revealed this failure. When taking
the perspective of the equivalence transformation
$\mathsf{Q}_\mathsf{E}$ (\ref{140530:2112}) for the $n$-point
velocity correlations, the reason is twofold, depending on the
particular level of the statistical description:

\begin{itemize}
\item[i)] The lower level {\it equivalence}
$\mathsf{Q}_\mathsf{E}$ (\ref{140530:2112}) is {\it induced} by
the higher level {\it symmetry} $\mathsf{Q}$ (\ref{140124:1833}),
which itself is physically inconsistent. By considering the fine-
to coarse-grained transition rule (\ref{140126:1607}), this
inconsistency in $\mathsf{Q}$ (\ref{140124:1833}) is exposed as a
violation of cause and effect, in that the system on the
statistical higher level would experience an effect without a
corresponding cause.
\item[ii)] The invariance $\mathsf{Q}_\mathsf{E}$
(\ref{140530:2112}) is {\it admitted} by a specific system of
equations (\ref{140601:1140}) whose statistical representation
hides essential information about the underlying deterministic
system. Since nonlinear aspects of turbulence theory are not
revealed, $\mathsf{Q}_\mathsf{E}$ (\ref{140530:2112}) misleadingly
represents itself as an invariance which only linear systems can
admit.
\end{itemize}

Both reasons now turn the invariance $\mathsf{Q}_\mathsf{E}$
(\ref{140530:2112}) into an unphysical transformation. Reason i),
because the physical inconsistency on the higher level is directly
transferred onto the lower level. Reason ii), because when
revealing the hidden nonlinear information, the invariance gets
broken as shown in (\ref{140127:1405}), since the linear scaling,
in which all system variables scale uniformly and independently
from its coordinates, is fully incompatible with any nonlinear
structure.

This very same conclusion also applies to the second ``new
statistical symmetry" (\ref{140110:1749}), first proposed in
\cite{Oberlack10} and recently again in \cite{Oberlack14.1}.
Admitted by the unclosed system (\ref{140601:1140}) as a
translation equivalence, in which again the coordinates stay
invariant and only the system variables get transformed due to the
equation's oversimplified and thus misleading representation as a
linear system of gradient-type, it straightforwardly can be
exposed as a further unphysical invariance just by using the same
procedure developed in this study.

Thus the claim made in \cite{Oberlack10} and \cite{Oberlack14.1},
in dealing with a first-principle construction method to generate
scaling laws for wall-bounded turbulent flows does not comply, all
the more so as the previous section clearly demonstrated a
mismatch between theory and numerical experiment. For the third
order moments this was more strongly pronounced than for the
second order moments, and would most probably continue to decline
in quality if the order of the moments is increased further. In
fact, the quality of these fits relative to the measure
$\chi^2_\text{red}$ declined drastically by several orders of
magnitude as the moments increased to the next higher order. This
just reflects the physical inconsistency of the invariance
$\mathsf{Q}_\mathsf{E}$ (\ref{140530:2112}), which, as proven in
(\ref{140127:1405}), manifestly intensifies as the order of the
transformed moments increase. The same is true for second
unphysical invariance (\ref{140110:1749}).

Important to note in this respect is that although the comparison
was explicitly based only for the case of a ZPG turbulent boundary
layer flow over a flat plate, it is obvious that all conclusions
and results generalize and transfer in a one-to-one manner when
comparing this ``new scaling theory" to any other wall-bounded
flow configuration, for example as to a channel-, pipe- or to a
Couette-type of flow. For example, the result recently
formu-\linebreak lated\hfill in\hfill \cite{Oberlack14}\hfill
for\hfill a\hfill more\hfill sophisticated\hfill
wall-bounded\hfill flow\hfill heavily\hfill relies\pagebreak[4]
\restoregeometry
\newgeometry{left=3cm,right=3cm,top=2.1cm,bottom=1.9cm,headsep=1em}
\noindent on those unphysical ``new statistical symmetries". Thus,
generating a useful statistical scaling law within
\cite{Oberlack14} for any higher order ve\-lo\-ci\-ty moment,
which goes beyond the mean velocity profile, is predetermined to
fail when considering all facts discussed~herein. Indeed, this
failure is confirmed in \cite{Frewer16.3}.

To conclude, we finally want to stress again in a brief historical
outline that using the terminology `invariant solution' in the
theory of turbulence is more than misleading.

For example, the classical von Kármán log-law was definitely not
derived as a solution of the (unclosed) statistical wall-bounded
Navier-Stokes equations, but only as a self-similar candidate
function upon pure dimensional arguments \citep{Monin71}. In other
words, although the classical von Kármán log-law is based on a
dimensional scaling symmetry, it is definitely not a first
principle solution; it only performs as an invariant function
which {\it only possibly but not necessarily} can serve as a
useful turbulent scaling function inside the inertial region. In
this region also other different functions exist which all scale
equally well, sometimes even better than the classical log-law
\citep{Barenblatt93a,Barenblatt93b,Barenblatt14}.~The reason for
this non-uniqueness is clearly to be seen as the consequence of
the statistical closure problem of turbulence, in that no true
solutions or true invariant solutions can be analytically
constructed by just employing the method of Lie symmetry groups
alone \citep{Frewer14.2,Frewer16.4}. Hence, also a so-called
`advanced' or `modern' invariance theory still faces the same
problems as von Kármán had at his~time.

Another prominent historical example which shows the complexity in
the statistical description of turbulence is that of anomalous
scaling and the breaking of global self-similarity, which both
interdependently can be attributed to the complex property of
intermittency \citep{Frisch85,Frisch95}. For example in the
inertial range of homogenous isotropic turbulence the results
clearly show that the flow cannot be globally invariant under
scaling, neither in a deterministic nor in a statistical sense.
Inertial range intermittency when measured with the longitudinal
multi-point structure functions show a clear lack of global
statistical self-similarity \citep{Frisch95,Biferale03}. The point
we want to stress here is that even if we would only consider a
highly idealized turbulent flow as that of a homogeneous isotropic
turbulent flow, the statistical solutions, in particular the
higher order correlations, are still by far more complicated than
we currently can imagine and that it's actually unrealistic to
believe that this complicated behavior can be captured by some
{\it global} scaling symmetries. In particular, when realizing the
fact that intermittency is essentially a property which rather
breaks a symmetry than statistically restoring it.

\appendix
\titleformat{\section}
{\large\bfseries}{Appendix \thetitle.}{0.5em}{}
\numberwithin{equation}{section}

\section{Invariant solutions for underdetermined systems\label{SecA}}

Since the strong property of a Lie {\it symmetry} transformation
only applies for fully determined equations (or overdetermined
systems of equations) it can be exploited to construct invariant
{\it solutions} of the considered equations. The reader should
note that we emphasize the word {\it `solution'}. Because, for
unclosed and thus underdetermined equations, for which only Lie
{\it equivalence} transformations can be generated, this situation
is different, as we want to explicitly  demonstrate in the
following.

For fully- or overdetermined systems of equations the word
`solution' is clearly defined. However, for {\it under}determined
equations the word `solution' is defined in a broader context. It
is defined as a mathematical expression which directly solves the
{\it under}determined equation without having to solve it again if
more information is added to this equation, and, of course, that
the solution must be {\it in principle} constructible without
having to initially model the equation.

\subsection{Basic illustrative examples\label{SecA.1}}

First consider the following algebraic but underdetermined
(unclosed) equation for $x$
\begin{equation}
x^2-y=0.\label{r15}
\end{equation}
\restoregeometry
\newgeometry{left=3cm,right=3cm,top=2.1cm,bottom=2.1cm,headsep=1em}
\noindent Formally this equation can be solved for the desired
variable $x$, to give the general solution
\begin{equation}
x=\pm \sqrt{y},\label{r16}
\end{equation}
as a set of infinitely many (non-unique) possible solutions, where
the solution set itself has the geometrical structure of the
parabola $y=x^2$. Note that at this stage all infinite solutions
along this parabola are equally privileged, i.e. no preferred
solution exists. But this situation of course will change as soon
as we will get additional information about equation \eqref{r15}.
Imagine to get the following additional information, either
\begin{itemize}
\item[{\bf i)}] that the unclosed value $y$ can be uniquely
constructed from some existing but still unknown equation
$z(y)=0$, i.e. that $y$ is determined by a specific but
analytically non-accessible process $z$ which as a hidden process
uniquely acts in the background, \item[or] \item[{\bf ii)}] that
the unclosed variable $y$ is showing some uniquely existing but
still unknown substructure $y=y(x)$, i.e. that $y$ is an arbitrary
but fixed function of $x$.
\end{itemize}

\vspace{1em}\noindent It's clear that although we give additional
information as i) or ii), in both cases equation \eqref{r15}
remains underdetermined and thus unclosed. But, for its solution
manifold the situation completely changed:
\begin{itemize}
\item[{\bf a)}] For information i) the general (infinitely
non-unique) solution \eqref{r16} turns into the unique (up to the
option of choosing either plus or minus) but still unknown
relation
\begin{equation}
x=\pm \sqrt{y_0},\;\;\text{with $\; z(y_0)\equiv 0$},\label{r17}
\end{equation}
which, from all possible solutions \eqref{r16} which lie on the
parabola $y=x^2$, is now a privileged solution on this parabola.
That means, once the underlying process $z$ in \eqref{r17} is
known, all other (infinitely many) solutions lying on this
parabola with $y\neq y_0$ must be discarded then as they no longer
satisfy equation \eqref{r17} anymore.
\item[{\bf b)}] For information ii) the general solution
\eqref{r16} turns into an own unclosed equation for $x$
\begin{equation}
x=\pm \sqrt{y(x)},\label{r18}
\end{equation}
which needs to solved again for $x$, depending of course on the
specification of the function $y(x)$.
\end{itemize}

\vspace{1em}\noindent In both cases we therefore cannot determine
or construct a solution for $x$ without modelling the process $z$
or without modelling the substructure $y(x)$. Using the word
`solution' for the unmodelled expressions \eqref{r17} or
\eqref{r18} would thus be completely misleading.

Hence, for underdetermined equations as given in \eqref{r15} we
can either construct non-unique and equally privileged (mostly
infinitely many) solutions \eqref{r16}, or no solution at all if
the unclosed term emerges from an underlying but unmodelled
process \eqref{r17}, or if it shows an existing but unmodelled
substructure \eqref{r18}.

The same reasoning also holds for differential equations. For
example, consider the following simple but underdetermined
(unclosed) first-order ODE for $f$
\begin{equation}
\frac{d}{dx} f(x)-g(x)=0,\label{r19}
\end{equation}
which, if the arbitrary function $g$ is integrable, can be
generally solved as
\begin{equation}
f(x)=f(x_0)+\int_{x_0}^x g(x^\prime)\, dx^\prime,\label{r20}
\end{equation}
to give a solution set with infinitely many possible and equally
privileged solutions for $f=f_g$ (infinite dimensional solution
manifold), depending in each case on the particular specification
of the arbitrary function $g$.

Now, important to note in the context we are considering in this
study is that this infinite set of formal solutions can also
include those functions which will stay invariant under a given
invariance (equivalence) group. For example, consider the
following Lie equivalence (scaling) transformation
\begin{equation}
\tilde{x}=e^{-\alpha} x,\qquad \tilde{f}=e^{\alpha}f,\qquad
\tilde{g}=e^{2\alpha}g, \label{r21}
\end{equation}
which leaves the unclosed equation \eqref{r19} form-invariant
\begin{equation}
\frac{d}{d\tilde{x}} \tilde{f}(\tilde{x})-\tilde{g}(\tilde{x})=0.
\label{r22}
\end{equation}
Then, from \eqref{r21}, the following two invariant functions can
be constructed
\begin{equation}
f(x)=\frac{c_f}{x},\qquad g(x)=\frac{c_g}{x^2},\label{r23}
\end{equation}
which, when choosing the integration constants as $c_g=-c_f$, will
satisfy \eqref{r20} and thus will form a solution of \eqref{r19}.
In other words, among the infinite possible and equally privileged
solutions of \eqref{r19}, we have picked out one specific solution
\eqref{r23} which has the additional property of staying invariant
under the arbitrarily chosen transformation group \eqref{r21}. But
careful, there is no reason at all that solution \eqref{r23}
should be identified as a privileged or preferred solution among
the infinite set of all other possible solutions \eqref{r20}.
Relative to their corresponding equation \eqref{r19} {\it all}
solutions including \eqref{r23} are still {\it equally}
privileged~--- solution \eqref{r23} only has some special
additional transformational property, that's all!

In this sense we can define, in contrast to the strong form of a
Lie-symmetry-based invariant solution, the opposite weak form of
an invariant solution when based on a Lie {\it equivalence}
transformation, namely as only being a particular and
non-privileged solution within an infinite set of other possible
and equally privileged solutions.

But, if we now would get the additional information that the
arbitrary function $g(x)$ in (\ref{r19}) either i) can be uniquely
determined from some existing but still unknown process $h$, e.g.
from a functional relation of the form $h[g(x)]=0$, or, ii) that
it shows some existing but unknown substructure, e.g. in the
functional form $g(x)=g[f(x),x]$, then the general solution
\eqref{r20} looses its status of being an explicit and
constructible solution. Because, for i) the general non-unique
solution \eqref{r20} will turn into the unique and privileged
relation
\begin{equation}
f(x)=f(x_0)+\int_{x_0}^x g_0(x^\prime)\, dx^\prime, \;\;\text{with
$\; h[g_0(x)]\equiv 0$},\label{r24}
\end{equation}
while for ii) it will turn itself into an underdetermined
(unclosed) integral equation
\begin{equation}
f(x)=f(x_0)+\int_{x_0}^x g[f(x^\prime),x^\prime]\, dx^\prime,
\label{r25}
\end{equation}
for which in both cases no solution and thus also no invariant
solution for $f(x)$ can be determined or constructed; of course,
only as long as the functional $h$ in \eqref{r24} or the kernel
$g$ in \eqref{r25} stays unspecified. In other words, no solution
and thus also no invariant solution can be determined or
constructed without invoking a modelling procedure for the
underlying process $h$ or the substructure $g$.

Hence, for an underdetermined (unclosed) differential equation as
\eqref{r19} we thus have the same situation as before for the
unclosed algebraic equation \eqref{r15}, in that either infinitely
many and equally privileged solutions (including all possible
invariant solutions) can be constructed, or in that, depending on
whether the unclosed term underlies an unique but analytically
non-accessible process \eqref{r24} or on whether it shows an
existing but analytically unknown substructure \eqref{r25}, no
solutions and thus also no invariant solutions can be determined
without a prior modelling assumption in both cases. It's clear
that all arguments given in this example can be easily extended
also to underdetermined (unclosed) PDEs.

\subsection{Examples from turbulence theory\label{SecA.2}}

First consider the underdetermined (unclosed) differential system
(\ref{140529:2236}) of Example~2 in Section \ref{Sec2}. We
obviously face both issues i) and ii) as defined in the beginning
of the previous Section \ref{SecA.1}. First of all, the unclosed
second moment $\vT$ definitely shows a substructure since it can
be uniquely determined from the underlying instantaneous
(fluctuating) velocity field $\vu$. Furthermore, since the mean
velocity field $\L \vu\R$ is the most basic element which can be
constructed from $\vu$, the second moment $\vT$ is mostly assumed
to be in first approximation a functional of the following form
\citep{Pope00,Davidson04}
\begin{equation}
\vT=\vT\big[\L\vu\R,\nabla\otimes\L\vu\R\big].
\end{equation}
Hence, if this arbitrary functional $\vT$ stays unspecified, then
no solution and thus also no invariant solution can be determined,
which thus corresponds to situation ii). Even if we would suppress
the substructure and would only demand a dependence on the
coordinates
\begin{equation}
\vT=\vT\big[\vx,t\big],\label{140921:1922}
\end{equation}
which then, according to system (\ref{140529:2236}), would allow
(similar to \eqref{r20}) for a formal construction of infinitely
many and equally privileged solutions for the mean velocity field
$\L\vu\R$, the usage of the word `solution' would nonetheless be
misleading in this case because we basically reside in situation
i). The reason for it is twofold: Not only because the unclosed
second moment (\ref{140921:1922}) suffices an own unclosed
one-point transport equation $\boldsymbol{\mathcal{E}}[\vT]=\v0$
\citep{Pope00,Davidson04}, which is structurally different to
system (\ref{140529:2236}), but also because since the second
moment (\ref{140921:1922}) is uniquely determined by the
underlying instantaneous (fluctuating) velocity field $\vu$, there
can be only one physical (privileged) realization for $\vT$. That
means that all other solutions within this infinite dimensional
solution manifold have to be discarded as unphysical, once this
physical solution is determinable. But, the probability to find
this particular specification (\ref{140921:1922}) which belongs to
this one physical solution (also within only a locally
pre-specified spatiotemporal range) is practically zero. Even more
unlikely is the case if this one particular specification and its
corresponding physical solution would additionally stay invariant
e.g.~under the global scaling group $\mathsf{E}_\mathsf{2}$
(\ref{140529:2336}). Hence, for this reason we claim that without
a prior modelling assumption for the unclosed system
(\ref{140529:2236}) the determination of its solutions and thus
also of its invariant solutions is misleading and essentially
ill-defined.

A further, more general example is the infinite differential chain
of $n$-point moment equations based on the full instantaneous
velocity and pressure fields of the incompressible Navier-Stokes
equations as presented in \cite{Oberlack10}, and also recently in
\cite{Oberlack14.1}
\begin{gather}
\frac{\partial H_{i_{\{n\}}}}{\partial t}+ \sum_{l=1}^n\Bigg[
\frac{\partial H_{i_{\{n+1\}}[i_{(n+1)}\mapsto
k_{(l)}]}[\vx_{(n+1)}\mapsto\vx_{(l)}] }{\partial
x_{k_{(l)}}}\text{\hspace{4.75cm}}\nonumber\\
\text{\hspace{3.5cm}}+\frac{\partial I_{i_{\{n-1\}}[l]}}{\partial
x_{i_{(l)}}}-\nu\frac{\partial^2 H_{i_{\{n\}}}}{\partial
x_{k_{(l)}}\partial x_{k_{(l)}}}\Bigg]=0,\quad n=1,\dotsc ,\infty,
\label{r1}
\end{gather}
\begin{equation}
\left.\begin{aligned}& \frac{\partial H_{i_{\{n\}}[i_{(l)}\mapsto
k_{(l)}]}}{\partial x_{k_{(l)}}}=0,\;\; \text{for}\;\,
l=1,\ldots , n\\
&\frac{\partial I_{i_{\{n-1\}}[k][i_{(l)}\mapsto
m_{(l)}]}}{\partial x_{m_{(l)}}}=0,\;\; \text{for}\;\,
k,l=1,\ldots , n\;\,\text{and}\;\, k\neq l,
\end{aligned}
~~~ \right \}\label{r2}
\end{equation}
where \eqref{r1} are the transport equations for the (equal-time)
multi-point velocity correlation functions (\ref{140530:1156}) of
tensor order $n\geq 1$
\begin{equation}
H_{i_{\{n\}}}:=H_{i_{(1)}i_{(2)}\ldots i_{(n)}}:= \big\L
u_{i_{(1)}}(\vx_{(1)})\cdot\ldots\cdot
u_{i_{(n)}}(\vx_{(n)})\big\R , \label{140109:1130}
\end{equation}
and the $n$-point pressure-velocity correlation functions of
tensor order $(n-1)$
\begin{multline}
I_{i_{\{n-1\}}[l]}:=\\
\big\L u_{i_{(1)}}(\vx_{(1)})\cdot\ldots\cdot
u_{i_{(l-1)}}(\vx_{(l-1)})\cdot p(\vx_{(l)})\cdot
u_{i_{(l+1)}}(\vx_{(l+1)})\cdot\ldots\cdot
u_{i_{(n)}}(\vx_{(n)})\big\R ,\label{140923:1340}
\end{multline}
along with \eqref{r2} as the two continuity constraints.

In the following it will be helpful to briefly introduce the
notation of Oberlack et al. The first index of ${i_{\{n\}}}$ in
(\ref{140109:1130}) defines the tensor character of the quantity
$\vH$, while the second index in braces denotes its tensor order.
The curly brackets point out that not an index of a tensor but an
enumeration is meant. On the other hand, the spatial component
index ${i_{(n)}}$ runs in general from $1$ to $3$ for all points
$n\geq 1$. For $n=1$ one has the connection to the mean velocity
field according to
\begin{equation}
H_{i_{\{ 1\}}}=H_{i_{(1)}}=\big\L u_{i_{(1)}}(\vx_{(1)})\big\R
=:U_{i_{(1)}}(\vx_{(1)}),
\end{equation}
and to the mean scalar pressure
\begin{equation}
I_{i_{\{0\}}[1]}=:I_{[1]}=:P(\vx_{(1)}).
\end{equation}
To note is that for convenience and a better readability we
denote, in contrast to \cite{Oberlack10} and \cite{Oberlack14.1},
the instantaneous fields by small and all averaged fields by
capital (latin) letters and, as well, we let all indices run from
$1$ and not from $0$ upwards. Also note that in all definitions to
follow the explicit time dependence in all functions will be
suppressed. Next, if the list of indices gets interrupted by one
or more other indices it is pointed out by attaching the replaced
value in brackets to the index
\begin{multline}
H_{i_{\{n\}}[i_{(l)}\mapsto k_{(l)}]}:=\\
\big\L u_{i_{(1)}}(\vx_{(1)})\cdot\ldots\cdot
u_{i_{(l-1)}}(\vx_{(l-1)})\cdot u_{k_{(l)}}(\vx_{(l)})\cdot
u_{i_{(l+1)}}(\vx_{(l+1)})\cdot\ldots\cdot
u_{i_{(n)}}(\vx_{(n)})\big\R ,
\end{multline}
which is further extended by
\begin{align}
H_{i_{\{n+1\}}[i_{(n+1)}\mapsto
k_{(l)}]}[\vx_{(n+1)}\mapsto\vx_{(l)}]:= &\; \big\L
u_{i_{(1)}}(\vx_{(1)})\cdot\ldots\cdot
u_{i_{(n)}}(\vx_{(n)})\cdot u_{k_{(l)}}(\vx_{(l)})\big\R\label{140923:1319}\\
=: &\; \widehat{H}_{i_{\{n+1\}}[l]}, \label{140110:1544}
\end{align}
where not only the index $i_{(n+1)}$ is replaced by $k_{(l)}$, but
also the independent variable $\vx_{(n+1)}$ is replaced by
$\vx_{(l)}$. If it is clear from the context, quantity
(\ref{140923:1319}) will be constantly abbre\-via\-ted as
(\ref{140110:1544}).

Now, although infinite in dimension, the hierarchal system
\eqref{r1}-\eqref{r2} is by construction unclosed and thus
underdetermined, where, due to that
$\widehat{H}_{i_{\{n+1\}}[l]}\neq
H_{i_{\{n+1\}}}$,\footnote[2]{Note that although {\it all}
components of $\widehat{H}_{i_{\{n+1\}}[l]}$ (\ref{140110:1544})
can be uniquely constructed from the higher dimensional moments
$H_{i_{\{n+1\}}}$ once they are known, which can be formally
written as the process
$\widehat{H}_{i_{\{n+1\}}[l]}=\lim_{\vx_{(n+1)}\to\vx_{(l)}}
H_{i_{\{n+1\}}}$, the necessary inverse construction, however,
fails.} the lower dimensional moments
$\widehat{H}_{i_{\{n+1\}}[l]}$ (\ref{140110:1544}) are to be
identified as the unclosed terms since they do not {\it directly}
enter the system's next higher order correlation equation (for a
more detailed explanation we refer to Appendix \ref{SecC}).

Exactly as in the example before, system \eqref{r1}-\eqref{r2}
again represents a specific underdetermined system of equations
for which no solutions can be determined endogenously. The obvious
reason is that we again face both issues i) and ii) as defined in
the beginning of Section \ref{SecA.1}. First of all, every
unclosed term $\widehat{H}_{i_{\{n+1\}}[l]}$ \eqref{140110:1544}
definitely shows a substructure as they all can be uniquely
determined from a single instantaneous (fluctuating) velocity
field $\vu$, and therefore will show at each order $n$ a high
correlation at least among its neighboring orders, i.e. in first
approximation they will at least be functionals of the form (for
all $k=1,\dotsc ,n$):
\begin{equation}
\widehat{H}_{i_{\{n+1\}}[l]}=\widehat{H}_{i_{\{n+1\}}[l]}\Big[H_{i_{\{n\}}};\:
\widehat{H}_{i_{\{n+1\}}[k\neq
l]};\:{\textstyle\lim_{(\vx_{(n+2)},\vx_{(n+1)})\to\vx_{(k)}}}
H_{i_{\{n+2\}}}\Big].\label{r26}
\end{equation}
Hence, if these arbitrary functionals
$\widehat{H}_{i_{\{n+1\}}[l]}$ stay unspecified, then no solutions
and thus also no invariant solutions can be determined. Secondly,
even if we would suppress the substructure and would only demand a
dependence on the coordinates
\begin{equation}
\widehat{H}_{i_{\{n+1\}}[l]}=\widehat{H}_{i_{\{n+1\}}[l]}
\big[\vx_{(k)},t\big],\label{r27}
\end{equation}
which then, according to system \eqref{r1}-\eqref{r2}, would allow
(similar again to \eqref{r20}) for a formal construction of
infinitely many and equally privileged solutions $H_{i_{\{n\}}}$
for each order $n$, the usage of the word `solution' would be
again still misleading.

As before, the reason is again twofold: Not only because the
unclosed terms \eqref{r27}, defined by (\ref{140923:1319}),
suffice own unclosed transport equations
$\mathcal{E}_n[\widehat{H}_{i_{\{n+1\}}[l]}]=0$
\citep{Pope00,Davidson04}, which are structurally different to
system \eqref{r1}-\eqref{r2} due to the non-commuting property of
the zero-correlation-length limit (see Appendix \ref{SecC}), but
also because since all arbitrary functions \eqref{r27} are
uniquely determined by the underlying instantaneous (fluctuating)
velocity field $\vu$, there can be only one physical (privileged)
realization for each order $n$. That means that all other
solutions within this infinite dimensional solution manifold have
to be discarded as unphysical, once this physical solution is
determinable. But, the probability again to find this particular
specification \eqref{r27} which belongs to this one physical
solution (also within only a locally pre-specified spatiotemporal
range) is practically zero. Even more unlikely is the case if this
one particular specification and its corresponding physical
solution would additionally stay invariant e.g.~under the global
scaling group $\mathsf{Q}_\mathsf{E}$ (\ref{140530:2112}). Hence,
for this reason we again claim that without any prior modelling
assumption for the unclosed system \eqref{r1}-\eqref{r2}, the
determination of its solutions and thus also of its invariant
solutions is misleading and essentially ill-defined.

\section{Formal derivation of the Friedmann-Keller hierarchy\label{SecB}}

Following the procedure from \cite{Fursikov99}, we briefly revisit
the formal derivation of the integro-differential Friedmann-Keller
chain of equations for incompressible and spatially unbounded
Navier-Stokes turbulence.

The starting point is the deterministic Navier-Stokes equation in
differential form (\ref{140529:1736}). The pressure gradient can
be eliminated by acting to both sides of (\ref{140529:1736}) with
the operator $\boldsymbol{\mathcal{P}}$ of orthoprojection to
solenoidal vector fields. As a result we then get the following
equations being absent of the gradient pressure field
\begin{equation}
\partial_t\vu - \boldsymbol{\mathcal{P}}\left[\nu\Delta\vu\right]
+
\boldsymbol{\mathcal{P}}\left[(\vu\cdot\nabla)\vu\right]=\v0,\quad\text{with}\;\;
\nabla\cdot\vu=0,
\end{equation}
which in a more compact form can also be written as
\begin{equation}
\partial_t\vu +\vA\cdot\vu+\vB(\vu)=\v0,
\end{equation}
where $\vA$ and $\vB$ is the corresponding linear and nonlinear
(quadratic) operator respectively. The explicit expression for
$\boldsymbol{\mathcal{P}}$ acting on any arbitrary vector-field
$\vf$ has the form
\begin{equation}
\boldsymbol{\mathcal{P}}[\vf]=\int
d^3\vx^\prime\,\vrho(\vx-\vx^\prime)\cdot \vf(\vx^\prime),
\end{equation}
where the kernel $\vrho$ is given by
\begin{align}
\vrho(\vx-\vx^\prime)&=\delta^3(\vx-\vx^\prime)\cdot
\boldsymbol{1}
+\nabla\otimes\nabla\left(\frac{1}{4\pi\vert\vx-\vx^\prime\vert}\right)\nonumber\\
& = \frac{2}{3}\delta^3(\vx-\vx^\prime)\cdot \boldsymbol{1}
-\frac{1}{4\pi\vert\vx-\vx^\prime\vert^3}\left[\boldsymbol{1}-
\frac{3(\vx-\vx^\prime)\otimes(\vx-\vx^\prime)}{\vert\vx-\vx^\prime\vert^2}\right].
\end{align}
By construction, the projection properties of
$\boldsymbol{\mathcal{P}}$ are
\begin{equation}
\left. \begin{aligned}
\boldsymbol{\mathcal{P}}[\vf]=\vf,\quad\text{if}\;\; \nabla\cdot
\vf=0,\hspace{1.4cm}\\
\boldsymbol{\mathcal{P}}[\vf]=\v0,\quad\text{if}\;\; \nabla\times
\vf=\v0, \;\;\text{e.g. if}\;\; \vf =\nabla\phi .
\end{aligned}
~~~\right \}
\end{equation}
With these definitions at hand, it is now straightforward to show
that the corresponding dynamical equations for the $n$-point
velocity moments $\vH_n$ (\ref{140530:1156}) can be iteratively
organized into the following infinite hierarchy of {\it linear}
equations, known as the (integro-differential) Friedmann-Keller
equations:
\begin{equation}
\partial_t \vH_n+\boldsymbol{\mathcal{A}}_n\cdot \vH_n +
\boldsymbol{\mathcal{B}}_n\cdot \widehat{\vH}_{n+1}=\v0,\quad
n\geq 1, \label{140605:1413}
\end{equation}
where the two spatial linear operators are defined as
\begin{align}
\boldsymbol{\mathcal{A}}_n \cdot\vH_n & = \sum_{i=1}^n
\boldsymbol{\mathcal{P}}_{\vx_i}\left[-\nu\Delta_{\vx_i}\vH_n\right],
\label{140922:1616}\\
\boldsymbol{\mathcal{B}}_n \cdot\widehat{\vH}_{n+1} & =
\sum_{i=1}^n \boldsymbol{\mathcal{P}}_{\vx_i}
\left[\nabla_{\vx_i}\cdot\widehat{\vH}_{i,n+1}\right],
\quad\text{with}\;\;
\widehat{\vH}_{i,n+1}=\vH_{n+1}\big\vert_{\vx_{n+1}=\vx_i},\label{140922:1617}
\end{align}
in which the projection operator
$\boldsymbol{\mathcal{P}}_{\vx_i}$, as well as the differential
operators $\Delta_{\vx_i}$ and $\nabla_{\vx_i}$ are to be
evaluated for each summand at the specific point $\vx=\vx_i$, for
all $i=1,\dotsc ,n$.

Two important things should be noted here. Firstly, since
$\widehat{\vH}_{n+1}\neq \vH_{n+1}$, i.e. since
$\widehat{\vH}_{n+1}$ is not a $(n+1)$-point function but only a
lower level $n$-point function of $(n+1)$-th moment, the hierarchy
of equations (\ref{140605:1413}), although infinite in dimension,
is not formally closed. For more details we refer to Appendix
\ref{SecC}. Hence, the hierarchy of equations (\ref{140605:1413})
is always a permanently underdetermined system, for which any
invariance analysis only will generate the weaker class of
equivalence transformations.

Secondly, great care has to be taken when performing a systematic
invariance analysis upon the linear system (\ref{140605:1413}).
Because, as no natural or additional physical constraints come
along with this system, it can lead to misleading and ultimately
to unphysical invariance results when not revealing the actual
nonlinear and nonlocal structure behind the oversimplified
notation for the symbols $\vH_n$ and $\widehat{\vH}_{n+1}$. In
other words, system (\ref{140605:1413}) represents itself as a
formal linear system only because an oversimplified notation is
being used, which hides essential information about the underlying
deterministic system. In this sense it admits formal equivalences
which also every underdetermined linear system of gradient-type
will admit, but as soon as one unfolds the oversimplified
notation, as thoroughly discussed in Section \ref{Sec4} at the
example of a scaling equivalence, these equivalences lead to
physical inconsistencies.

\section{Formally closed and unclosed infinite systems\label{SecC}}

For this discussion let us consider all three statistical and
infinite dimensional approaches to turbulence:

\vspace{0.5em} {\bf Approach 1:} The functional Hopf equation
(HEq)
\begin{equation}
\frac{\partial\Phi}{\partial t}=\int\! d^3\vx\, \alpha_k \left(
i\frac{\partial}{\partial x_l}
\frac{\delta^2}{\delta\alpha_k\delta\alpha}_l
+\nu\Delta\frac{\delta}{\delta\alpha_k}\right)\Phi,
\label{140922:1409}
\end{equation}
for the probability density functional
\begin{equation}
P[\vu(\vx);t]=\int\! \Phi[\valpha(\vx);t]\, e^{-i\int\! d^3\vx\,
\valpha(\vx)\cdot\vu(\vx)}\mathcal{D}\valpha(\vx)\geq 0,
\label{140922:1435}
\end{equation}
of the velocity field sampled at infinitely many, non-denumerable
(continuum) number of points \citep{Hopf52,McComb90,Shen91}. The
evolution equation (\ref{140922:1409}) goes along with the three
physical constraints
\begin{equation}
\Phi^*[\valpha(\vx);t]=\Phi[-\valpha(\vx);t],\quad
\Phi[0;t]=1,\quad\big\vert\Phi[\valpha(\vx);t]\big\vert\leq 1,
\label{140922:1412}
\end{equation}
in order to guarantee for a physical solution for all times $t$.

\vspace{0.5em} {\bf Approach 2:} The integro-differential
Lundgren-Monin-Novikov (LMN) hierarchy
\begin{multline}
\left[\partial_t +\sum_{i=1}^n \vv_i\cdot\vnabla_i\right]f_n
=-\sum_{i=1}^n\frac{\partial}{\partial \vv_i}\bigg[
\lim_{\vx_{n+1}\to \vx_i}\nu\Delta_{n+1}\int d^3\vv_{n+1}\vv_{n+1}f_{n+1}\\
 -\int
d^3\vx_{n+1}d^3\vv_{n+1}\left(\vnabla_i\frac{1} {4\pi\vert
\vx_i-\vx_{n+1}\vert}\right)
(\vv_{n+1}\cdot\vnabla_{n+1})^2f_{n+1}\bigg], \label{140922:1418}
\end{multline}
for the $n$-point probability density function (PDF)
\begin{equation}
f_n=f_n(\vx_1,\vv_1;\dotsc ;\vx_n,\vv_n;t)\geq 0,
\end{equation}
of the velocity field sampled at a finite number of points, being
thus the discrete version of the above Hopf equation
\citep{Lundgren67,Monin67,Friedrich12}. Apart from the usual
continuity constraints, the evolution equation (\ref{140922:1418})
goes along with infinitely many physical constraints in order to
guarantee for a physical solution for all times $t$. The most
cited ones are i) the reduction or normalization constraint
\begin{gather}
\int\! d\vv_1\, f_1(\vx_1,\vv_1;t)=1,\nonumber\\
\int\! d\vv_{n+1}\, f_{n+1}(\vx_1,\vv_1;\dotsc
;\vx_{n+1},\vv_{n+1};t)=f_n(\vx_1,\vv_1;\dotsc ;\vx_n,\vv_n;t),
\;\; n\geq 1, \label{140922:1454}
\end{gather}
ii) the coincidence constraint for $n\geq 2$, and $1\leq (i,j)\leq
n$, $i\neq j$
\begin{multline}
\int \! d\vx_i\, \delta(\vx_i-\vx_j)f_{n}(\vx_1,\vv_1;\dotsc
;\vx_{n},\vv_{n};t)\sim\lim_{\vert \vx_i-\vx_j\vert\to
0}f_n(\vx_1,\vv_1;\dotsc
;\vx_n,\vv_n;t)\\=f_{n-1}(\vx_1,\vv_1;\dotsc
;\vx_{i-1},\vv_{i-1};\vx_{i+1},\vv_{i+1};\dotsc;\vx_n,\vv_n;t)
\cdot\delta(\vv_{i}-\vv_j),\label{140922:1455}
\end{multline}
and iii) the separation property for  $n\geq 2$, and $1\leq
(i,j)\leq n$, $i\neq j$, with $\vert \vx_j\vert <\infty$
\begin{multline}
\lim_{\vert \vx_i-\vx_j\vert\to\infty}f_n(\vx_1,\vv_1;\dotsc
;\vx_n,\vv_n;t)\\= f_{n-1}(\vx_1,\vv_1;\dotsc
;\vx_{i-1},\vv_{i-1};\vx_{i+1},\vv_{i+1};\dotsc;\vx_n,\vv_n;t)
\cdot f_1(\vx_i,\vv_i;t). \label{140922:1456}
\end{multline}
Yet the reader should note that besides these three usually
mentioned LMN constraints as `normalization', `coincidence' and
`separation', there exists a fourth and even more strong
constraint which unfortunately is not mentioned anymore in the
recent literature, as e.g. in \cite{Friedrich12}. We are talking
about the additional constraint first derived in \cite{Ievlev70}
(listed therein as constraint (2.6)), and also presented in
\cite{Monin75} as constraint (19.139).

\vspace{0.5em} {\bf Approach 3,1:} The differential
Friedmann-Keller or multi-point correlation (MPC) hierarchy
\begin{multline}
\qquad\frac{\partial H_{i_{\{n\}}}}{\partial t}+
\sum_{l=1}^n\Bigg[ \frac{\partial H_{i_{\{n+1\}}[i_{(n+1)}\mapsto
k_{(l)}]}[\vx_{(n+1)}\mapsto\vx_{(l)}] }{\partial x_{k_{(l)}}}\\
+\frac{\partial I_{i_{\{n-1\}}[l]}}{\partial
x_{i_{(l)}}}-\nu\frac{\partial^2  H_{i_{\{n\}}}}{\partial
x_{k_{(l)}}\partial x_{k_{(l)}}}\Bigg]=0,\quad n\geq 1,\quad
\label{140922:1519}
\end{multline}
for the $n$-point velocity moments (\ref{140109:1130})
\begin{equation}
H_{i_{\{n\}}}:=H_{i_{(1)}i_{(2)}\ldots i_{(n)}}:= \big\L
u_{i_{(1)}}(\vx_{(1)})\cdot\ldots\cdot
u_{i_{(n)}}(\vx_{(n)})\big\R , \label{140922:1522}
\end{equation}
and the $n$-point pressure-velocity moments as defined in
(\ref{140923:1340}), where both sets of moments are based on the
full instantaneous fields of the incompressible Navier-Stokes
equations \citep{Oberlack10,Oberlack14.1}. Note that
\begin{equation}
\widehat{H}_{i_{\{n+1\}}[l]}:=H_{i_{\{n+1\}}[i_{(n+1)}\mapsto
k_{(l)}]}[\vx_{(n+1)}\mapsto\vx_{(l)}]=\lim_{\vx_{(n+1)}\to\vx_{(l)}}
H_{i_{\{n+1\}}},\label{140922:2050}
\end{equation}
is not a $(n+1)$-point moment, but only a $n$-point moment of
$(n+1)$th order, i.e.
$\widehat{H}_{i_{\{n+1\}}[l]}~\neq~H_{i_{\{n+1\}}}$.

\vspace{0.5em} {\bf Approach 3,2:} The integro-differential
Friedmann-Keller or multi-point correlation (MPC) hierarchy
\begin{equation}
\partial_t \vH_n+\boldsymbol{\mathcal{A}}_n\cdot \vH_n +
\boldsymbol{\mathcal{B}}_n\cdot \widehat{\vH}_{n+1}=\v0,\quad
n\geq 1, \label{140922:1554}
\end{equation}
for the $n$-point velocity moments
\begin{equation}
\vH_n= \big\L \vu(\vx_1,t)\otimes\cdots \otimes\vu(\vx_n,t)\big\R
,\quad n\geq 1, \label{140922:1556}
\end{equation}
based on the instantaneous velocity field $\vu$
\citep{Fursikov99}. The integral operators
$\boldsymbol{\mathcal{A}}_n$ and $\boldsymbol{\mathcal{B}}_n$ are
defined in (\ref{140922:1616}) and (\ref{140922:1617})
respectively. Note that
\begin{equation}
\widehat{\vH}_{n+1}=
\lim_{\vx_{n+1}\to\vx_{n}}\vH_{n+1},\label{140922:2054}
\end{equation}
is not a $(n+1)$-point moment, but only a $n$-point moment of
$(n+1)$th order, i.e. $\widehat{\vH}_{n+1}~\neq~\vH_{n+1}$.

\vspace{1em}\noindent In the discussion to follow, it is essential
to recognize that in contrast to the Hopf equation and the LMN
hierarchy, the MPC equations do {\it not} go along with additional
physical constraints (besides the usual continuity constraints),
neither in the differential nor in the integro-differential form.

Now, before we investigate these three infinite-dimensional
approaches on formal closure, it is necessary to distinguish the
terminology of {\it `closed'} from {\it `formally closed'}. The
latter is obviously a much more weaker concept than the former.

First of all, all three statistical approaches to turbulence are
{\it unclosed}, because in a {\it practical} sense one either has
to discretize the continuous functional formulation or to truncate
the infinite discrete hierarchy of equations. Discretizing the
Hopf equation leads to the LMN equations and truncating these at a
certain order then leads to unclosed terms (arbitrary functions)
which need to be modelled in order to close the equations in each
case. The same situation we face for the MPC equations in either
form.

{\it But}, if we would {\it not} discretize the Hopf equation or
truncate either system of LMN and MPC, i.e. if we would {\it
formally} consider for each approach the full continuous and
infinite formulation, then from a formal point of view, for
example from the viewpoint of an invariance analysis, the Hopf and
LMN equations act completely different than the MPC equations: In
contrast to the MPC equations, the Hopf and LMN equations act in a
{\it formally closed} manner. In other words, on the {\it formal
(non-truncated)} level the Hopf equation and the LMN equations can
be regarded as formally closed systems, while the MPC equations
not.

The key point here is that since the LMN system is just the
discrete version of the functional Hopf equation, which for itself
undoubtedly acts as a formally closed system, the LMN system will
thus induce in its {\it non-truncated} form an infinite
dimensional but functionally {\it unique} solution manifold. This
is not the case for the MPC system, which in its {\it
non-truncated} form would induce infinitely many functionally
different and thus {\it non-unique} solutions manifolds, each
being itself of course infinite dimensional and equally
privileged.

This different {\it formal} behavior can only be understood when
directly comparing the LMN system (which in its well-known form
only holds for spatially {\it unbounded} flow configurations) with
the MPC system: Next to the usual continuity constraints, the LMN
system goes along with several additional and independent physical
constraints, which all will naturally restrict the general
solution space down to a physical (unique) solution space (like
boundary conditions restricting the general solution space for
usual PDEs), while the MPC system is completely free of such
constraints (up to the usual continuity constraints). In other
words, the infinite many {\it unphysical} solution manifolds of
the MPC system cannot be separated from the {\it physical}
solution manifold as it automatically happens in the LMN system.

This restriction in the solution manifold of the LMN system can
already be observed when performing any invariance analysis upon
them. Because, for such an analysis the LMN constraints are
obstructive, as they all favor the mechanism of symmetry breaking.
In strong contrast of course to a performed invariance analysis of
the MPC system, where, due to the absence of physical constraints,
no invariance breaking mechanism exists; {\it all} admitted
invariant transformations of the MPC system, physical or
unphysical, thus agglomerate to the invariant solution manifold.

In order to demonstrate the difference between `formally unclosed'
and `formally closed' in an explicit manner, let's consider e.g.
the following infinite chain of second order PDEs for the
different $n$-dimensional scalar functions
$f_n:=f_n(x_1,x_2,\dotsc ,x_n)$
\begin{equation}
\mathcal{Z}[f_n]:=\frac{\partial^2 f_n}{\partial
x_n^2}+\frac{\partial}{\partial x_n}\left[\lim_{x_{n+1}\to x_n}
f_{n+1}\right]=0,\;\;\text{for}\;\; n=1,\dotsc ,\infty,\label{r60}
\end{equation}
which {\it in principle} should mimic the basic behavior of the
MPC equations (\ref{140922:1519}) in a very primitive form: The
first term in \eqref{r60} stands for the dissipative term and the
second one for the convective term.

First of all, the infinite hierarchy \eqref{r60} is unclosed,
because if we would truncate this system at an arbitrary but fixed
order $n=n_0$, we would explicitly gain the unclosed and thus
arbitrary function $f_{n_0+1}$, which then needs to be modelled in
order to close this system at order $n_0$.

But now, since \eqref{r60} is free of any constraints, it is also
{\it formally unclosed}, i.e. even if we would {\it not} truncate
system \eqref{r60} and therefore would formally consider {\it all}
(infinite) equations, system \eqref{r60} still has to be regarded
as unclosed. The simple reason is that infinitely many disjoint
and thus different solution manifolds can be generated, i.e.
system \eqref{r60} does not induce a {\it unique} (infinite
dimensional) solution manifold. In other words, also on the formal
(non-truncated) level system \eqref{r60} is still underdetermined.

That \eqref{r60} really induces infinitely many different and
independent solution manifolds can be easily seen. For example
consider the following {\it special} solution set of \eqref{r60}
\begin{equation}
f_n=2^{-\frac{1}{2}n^2+\frac{5}{2}n-2+c}\cdot
e^{-\frac{x_n}{2^{n-2}}-\sum_{i=1}^{n-1}\frac{x_i}{2^{i-1}}},\;\;
c\in\mathbb{R},\;\;\text{and for all $n\geq 1$},\label{r61}
\end{equation}
which will be part of a more {\it general} solution manifold, say,
of $S_1$, where {\it all} functions $f_n$ will be non-zero:
\begin{equation}
S_1=\Big\{\mathcal{Z}[f_n]=0\;\Big\vert\; f_n\neq 0,\;\;\text{for
all $n\geq 1$}\Big\}.\label{140922:1827}
\end{equation}
To construct for \eqref{r60} a from $S_1$ disjoint and thus
independent general solution manifold can now be easily achieved.
By choosing at an arbitrary but fixed order $n=n_0$ a specific
functional relation for the next higher order variable
$f_{n_0+1}=f^0_{n_0+1}$, one can iteratively determine all other
(infinitely many) possible functions $f_n$ for $n\leq n_0$ as well
as for $n>n_0$. Now, since the choice of $n_0$ and the choice for
$f^0_{n_0+1}$ are arbitrary one can consequently construct
infinitely many independent general solution manifolds $S_n$. For
example, if we would choose $n_0=1$ and $f^0_2=0$, we obtain the
following to \eqref{r61} different {\it special} solution
\begin{equation}
f_1=c_1\cdot x_1+c_2,\quad\; f_n=0, \;\;\text{for}\;\; n\geq 2,
\label{r62}
\end{equation}
which will be part of a more {\it general} solution manifold
$S_2$, where the function $f_{n_0+1}=f_{n=2}$ is permanently zero
and all functions $f_n$ below $n=2$ are strictly non-zero, while
all functions $f_n$ beyond $n=2$ remain unrestricted in this
regard:\footnote[2]{The explicit form is: $S_2= \big\{
f_1=c_1\cdot x_1+c_2\neq 0,\, f_2=0,\, f_3=f_3(x_1,x_3-x_2),\,
f_4(x_1,x_2,x_3,x_4),\, \dotsc \big\}$.}
\begin{equation}
S_2 = \Big\{\mathcal{Z}[f_n]=0\;\Big\vert\; f_1\neq 0,\; f_2=
0\;\;\text{and}\;\; (f_n\neq 0\;\:\text{or}\;\: f_n=0),\;\text{for
$n\geq 3$}\Big\}.\label{140922:1841}
\end{equation}
It's clear that this (infinite) solution manifold $S_2$ is
independent and disjoint to the previous (infinite) solution
manifold $S_1$ (\ref{140922:1827}). If we would choose $n_0=2$ and
$f_3^0=0$, then we will obtain the following to \eqref{r62}
different {\it special} solution
\begin{equation}
\left. \begin{aligned} f_1 & =-\int_{x_1} dz_1 \int_{z_1}
f_2(z_2,z_2)\, dz_2+c_1\cdot x_1+c_2,\\ f_2 & =F_1(x_1)\cdot
x_2+F_2(x_1),\quad\; f_n=0, \;\;\text{for}\;\; n\geq 3,
\end{aligned}
~~~\right \}\label{r63}
\end{equation}
which now will be part of a more {\it general} solution manifold
$S_3$, where now the function $f_{n_0+1}=f_{n=3}$ is permanently
zero and all functions $f_n$ below $n=3$ are strictly non-zero,
while all functions $f_n$ beyond $n=3$ remain unrestricted again:
\begin{equation}
S_3=\Big\{\mathcal{Z}[f_n]=0\;\Big\vert\; f_1\neq 0,\; f_2\neq
0,\; f_3= 0\;\;\text{and}\;\; (f_n\neq 0\;\:\text{or}\;\:
f_n=0),\;\text{for $n\geq 4$} \Big\}.\label{140922:1849}
\end{equation}
The solution manifold $S_3$ is then independent and disjoint to
$S_2$ (\ref{140922:1841}) and $S_1$ (\ref{140922:1827}). This
process can then be continued to infinity to gain infinitely many
{\it independent} general solution manifolds $S_n$.

But what is actually the deeper reason that \eqref{r60} is
formally underdetermined, although each equation is linked to the
next higher order one. The main reason is of course that the
system \eqref{r60} is infinite dimensional. But another reason is
also that the limit appearing in \eqref{r60} is artificial. The
limit pretends a formal closure although there is none
\citep{Frewer15.1}. The term in brackets is a $n$-dimensional
function for which no {\it direct} equation corresponds to. It
thus presents for each order the unclosed term. For example,
consider (for simplicity) the case $n=1$. The limit in \eqref{r60}
then takes the explicit form
\begin{equation}
\widehat{f}_2(x_1):=\lim_{x_2\to x_1} f_2(x_1,x_2)=f_2(x_1,x_1).
\label{r64}
\end{equation}
The problem is that from a given $2$-dimensional function
$f_2(x_1,x_2)$, i.e. from `above', one can uniquely construct the
corresponding lower $1$-dimensional function $f_2(x_1,x_1$), but
not vice versa, i.e. from `below', that is, from a given
$1$-dimensional function $\widehat{f}_2(x_1)$ one {\it cannot}
uniquely construct the corresponding higher $2$-dimensional
function $f_2(x_1,x_2)$. But this latter process is exactly what
happens when writing the limit in \eqref{r60}, namely that on the
lower $n$-dimensional level, i.e. from `below', a higher
$(n+1)$-dimensional function is identified. In other words, the
limit in \eqref{r60} artificially forces a lower $n$-dimensional
function $\widehat{f}_{n+1}$ into a higher $(n+1)$-dimensional
function $f_{n+1}$.

Hence, for a system as \eqref{r60} the formal level of
unclosedness (degree of underdeterminedness) even is {\it higher}
than, for example, if we would consider instead of \eqref{r60} the
following, also formally unclosed system
\citep{Frewer15.1,Frewer15.2}
\begin{equation}
\frac{\partial^2 f_n}{\partial x_n^2}+\frac{\partial
f_{n+1}}{\partial x_n} =0,\;\;\text{for}\;\; n=1,\dotsc
,\infty,\label{141007:1441}
\end{equation}
which, in contrast to \eqref{r60}, {\it directly} and thus {\it
uniquely} links each equation to the next higher order one.
Because, when choosing an arbitrary but fixed order $n=n_0$ for
(\ref{141007:1441}), one only has two variable options to generate
a solution: either to specify $f_{n_0}$ or to specify $f_{n_0+1}$,
from which, in each case, {\it all} (infinite) remaining functions
can be formally determined then. For system \eqref{r60}, however,
the two corresponding specifications, either  $f_{n_0}$ or
$\widehat{f}_{n_0+1}:=\lim_{x_{n_0+1}\to x_{n_0}} f_{n_0+1}$, are
not sufficient, because all functions for $n>n_0$ cannot be {\it
uniquely} determined anymore without exogenously also specifying
the functions $f_{n_0+1}$, $f_{n_0+2}$, etc. Thus, explicitly
writing the limit in \eqref{r60} does not provide a positive
contribution to formally close this system.

Now, regarding the original MPC system
(\ref{140922:1519})-(\ref{140922:2054}), we see that within
these~equations in either form, too, the lower dimensional
unclosed terms $\widehat{H}_{i_{\{n+1\}}[l]}$ (\ref{140922:2050})
and $\widehat{\vH}_{n+1}$ (\ref{140922:2054}) can be explicitly
written as a limit of higher dimensional functions in order to
apparently establish a connection between a lower order equation
and the next higher one. But, as was just explained above, this
connection is artificially enforced, since the formal closure
problem is not eliminated by explicitly writing this limit; the
closure problem still exists independently of whether this limit
is being written or not. In other words, this `lim' notation is
misleading as it suggests a formal closure for the MPC system
although in reality there is none \citep{Frewer15.1}.

Note that the LMN equations (\ref{140922:1418}) also involve such
an artificial connection between the lower and higher order
functions, but, in contrast to the MPC equations, the LMN
equations give something in return in that they come along with
additional physical constraints in order to constitute themselves
as a {\it formally} closed system.~In fact, these additional
constraints will restrict the infinitely many possible solution
manifolds of the LMN evolution equations down to a physical
solution manifold.

To explicitly demonstrate such a restriction, let us mimic for
example the normalization property of the LMN equations
(\ref{140922:1454}), by demanding next to our infinite chain of
emulated equations \eqref{r60} the following e.g. half-sided
`normalization' constraint
\begin{equation}
\int_0^\infty f_1 dx_1=1,\;\; \text{and}\;\;\int_0^\infty
f_{n+1}dx_{n+1}=f_n, \;\;\text{for all}\;\; n\geq 1, \label{r65}
\end{equation}
which is a well-defined constraint for a differential system of
type \eqref{r60}. If we regard \eqref{r65} as a physical
constraint, then the {\it special} solutions \eqref{r62} and
\eqref{r63} must be regarded as unphysical solutions and thus have
to be discarded, due to the non-convergence of the integral if
$c_1\neq0$ or $c_2\neq 0$. But, not only the special solutions,
even the {\it general} solution manifolds $S_2$ and $S_3$
themselves have to be discarded as unphysical, because the
functional break $f_{n_0}\neq f_{n_0+1}=0$ at the chosen order
$n=n_0$ is not compatible with \eqref{r65}. Also the {\it special}
closed form solution \eqref{r61} must be discarded in this sense
as unphysical, since also its functional form is not compatible
with the constraint \eqref{r65}. However, the general solution
manifold $S_1$ itself must not be discarded, because there may
still exist different special closed form solutions which in
contrast to \eqref{r61} are compatible with \eqref{r65}.

Thus we see that the more physical constraints go along with an
infinite hierarchy of equations, the more the general non-unique
solution manifold gets restricted down to a unique (physical)
solution manifold. In other words, through a sufficient number of
constraints a formally underdetermined (formally unclosed) system
can turn into a formally fully determined (formally closed)
system. And exactly this is the case for the LMN equations as they
are just the discrete version of the functional Hopf equation,
which itself, after all, represents a formally closed equation.

Only due to the fact that the LMN equations go along with
additional physical constraints, they constitute in contrast to
the MPC equations a formally (non-truncated) closed system. Hence,
the LMN system in its {\it non-truncated} form thus constitutes a
more physical system than the corresponding {\it non-truncated}
MPC equations, as already said by \cite{Ievlev70}: ``However, the
equations for the probability distributions (the LMN equations)
yield a more complete and compact statistical description of
turbulence than do the usual moment equations (the Friedman-Keller
equations) and apparently permit an easier formulation of the
approximate conditions closing the equations."

Now, in the case of the MPC equations, what would be the
appropriate procedure to turn them into a {\it formally} closed
system? The only answer is to extend the MPC equations at each
order with the lower order moment equations of the corresponding
unclosed terms. But this is a non-manageable task, as the
lower-order moment equations cannot be condensed anymore into a
single hierarchy as it is the case for the MPC equations, neither
in the differential form (\ref{140922:1519}) nor in the integral
form (\ref{140922:1554}). The reason for this jump in complexity
is that the above artificial limit does not commute with any
differential operator, e.g. as in the relevant expression of
(\ref{140922:1519})
\begin{equation}
\frac{\partial}{\partial
\vx_{(l)}}\left[\lim_{\vx_{(n+1)}\to\vx_{(l)}}
H_{i_{\{n+1\}}}\right]\neq \lim_{\vx_{(n+1)}\to\vx_{(l)}}
\frac{\partial}{\partial \vx_{(l)}}H_{i_{\{n+1\}}},\label{r68}
\end{equation}
nor with any integral operator as in (\ref{140922:1554})
\begin{equation}
\boldsymbol{\mathcal{B}}_n\cdot\left[\lim_{\vx_{n+1}\to\vx_n}\vH_{n+1}\right]\neq
\lim_{\vx_{n+1}\to\vx_n}\boldsymbol{\mathcal{B}}_n\cdot
\vH_{n+1},\label{r69}
\end{equation}
which everyone working in fluid mechanics may have already
experienced, when writing the transport equations for the lower
$n$-point order moments as a limit from the higher $(n+1)$-point
equations. Due to this non-commuting property, the number of
unclosed terms increases, while at the same time the number
constraints decreases. Here a small explicit example when
considering the limit of the two-point continuity constraint
\eqref{r2}, which just reduces to a non-useful zero-identity due
to the continuity constraint of the instantaneous velocity field:
\begin{center}
\begin{tabular}{r c l}
${\displaystyle 0\quad =\quad \lim_{\vx_{(1)}\to
\vx_{(0)}}\frac{\partial H_{k_{(0)}i_{(1)}}}{\partial
x_{k_{(0)}}}}$ & \multicolumn{1}{@{${}\neq{}$}}{} &
${\displaystyle \frac{\partial }{\partial
x_{k_{(0)}}}\left[\lim_{\vx_{(1)}\to
\vx_{(0)}}H_{k_{(0)}i_{(1)}}\right]}$\\[1.25em]
${\displaystyle =\lim_{\vx_{(1)}\to \vx_{(0)}}\frac{\partial
}{\partial x_{k_{(0)}}}\Big\L
u_{k_{(0)}}(\vx_{(0)})u_{i_{(1)}}(\vx_{(1)})\Big\R}$ &
\multicolumn{1}{c|}{}$\;\;\,$ & ${\displaystyle =\frac{\partial
}{\partial
x_{k_{(0)}}}H_{k_{(0)}i_{(0)}}}$\\[1.25em]
${\displaystyle = \lim_{\vx_{(1)}\to
\vx_{(0)}}\Big\L\Big(\,\frac{\partial }{\partial
x_{k_{(0)}}}u_{k_{(0)}}(\vx_{(0)})\Big)
u_{i_{(1)}}(\vx_{(1)})\Big\R}$ & \multicolumn{1}{c|}{}$\;\;\,$ &
${\displaystyle =\frac{\partial }{\partial x_{k_{(0)}}}\Big\L
u_{k_{(0)}}(\vx_{(0)})u_{i_{(0)}}(\vx_{(0)})\Big\R}$\\[1.25em]
${\displaystyle =\Big\L\Big(\,\frac{\partial }{\partial
x_{k_{(0)}}}u_{k_{(0)}}(\vx_{(0)})\Big)
u_{i_{(0)}}(\vx_{(0)})\Big\R}$ & \multicolumn{1}{c|}{}$\;\;\,$ &
${\displaystyle =\frac{\partial}{\partial x_{k}}\L u_{k}u_{i}\R
\;\;\neq\;\; 0}$\\[1.25em]
${\displaystyle =\frac{\partial}{\partial x_{k}}\L u_{k}u_{i}\R
-\Big\L\Big(u_{k}\frac{\partial }{\partial
x_{k}}\Big)u_{i}\Big\R\;\;\equiv\;\; 0}$ &
\multicolumn{1}{c|}{}$\;\;\,$ &
\end{tabular}
\end{center}

\begin{flushright}
\addtocounter{equation}{1} \vspace{-3.25em}\phantom{x}\hfill
(C.\arabic{equation})
\end{flushright}

\vspace{0.5em}\noindent Hence, the overall degree of
unterdeterminism of the MPC equations even increases when trying
to formally close them; and exactly in this sense we can say that
the MPC equations (as they are standardly used by Oberlack et al.)
are underdetermined in that they involve more unknowns than
determining equations. Thus, as a consequence, any invariance
analysis performed upon them will only result into weak {\it
equivalence} transformations and {\it not} into a strong
invariance relation as that of a true {\it symmetry}
transformation.

\subsection{The non-equivalent relation between LMN and MPC\label{SecC.1}}

Regarding the above discussion, the reader should finally note
that the following statement made in \cite{Friedrich12}
\emph{``... It can be shown that the LMN approach is completely
equivalent to the statistical description of turbulence by moment
equations. ..."} can be misleading if not carefully drawn from the
context.

This cited {\it ``equivalence"} refers to the fact that not only
{\it if} a $n$-point PDF is known {\it then} the $n$-point moments
can be determined from its expectation value, but also that {\it
if} {\it all} $n$-point moments are known {\it then} the $n$-point
PDF can be reconstructed. The latter `inverse' construction is
achieved by making use of Taylor series of the PDFs characteristic
functions (see e.g. \cite{Monin67}).

The problem now is that in order to perform such a construction
either the $n$-point PDF or the $n$-point moments must be known
before the other one can be determined. For that one needs to
solve the underlying evolution or transport equations, either for
the PDFs or for the moments. While the transport equations for the
moments (MPC) can be uniquely determined from the evolution
equations of the PDFs (LMN) (see e.g. \cite{Monin67}), the reverse
cannot be established. In other words, the LMN equations cannot be
{\it uniquely} determined from the MPC equations. \cite{Ievlev70}
clearly has shown, which is also mentioned in \cite{Monin75} (the
statements following the constraint (19.139)), that at least two
{\it different} evolution equations for the PDFs can be
constructed which all precisely result into the {\it same} MPC
equations. Careful, although the two evolution equations for the
PDFs differ by the fact that one is an approximation of the other
one, this approximation is {\it not} transferred down to the MPC
equations, which is clearly shown in Ievlev's proof in section 4.5
on page~89 \citep{Ievlev70}.

That the LMN equations cannot be uniquely constructed from the MPC
equations can also be easily understood from a different
perspective, which already has been discussed in detail before:
The additional physical constraints which go along with the LMN
equations have no counterpart in the MPC equations and thus, in
turn, are unable to uniquely induce the full LMN equations
(including {\it all} possible constraint equations).

Hence, the above cited {\it ``equivalence"} only refers to the
defining relations of the PDFs and its moments (in \cite{Monin67}
given by (3.1)-(3.3)), but definitely not to their underlying
evolution equations. The LMN equations imply the MPC equations,
but not oppositely, which is also clear from the aspect that a PDF
formulation always operates on a higher statistical level than a
formulation of the moments.

\section{Additional comments on the proof for inconsistency\label{SecD}}

The proof (\ref{140127:1405}) reads:
\begin{align}
\tilde{\vH}_1=e^q \vH_1 \; & \Rightarrow\;
\big\L\tilde{\vu}(\tilde{\vx}_1,\tilde{t})\big\R=e^q\big\L
\vu(\vx_1,t)\big\R,\;\text{for all points
$\vx_1=\vx$}\label{35}\\
& \Rightarrow\;
\big\L\tilde{\vu}(\tilde{\vx}_k,\tilde{t})\big\R=e^q\big\L
\vu(\vx_k,t)\big\R,\;\text{for all}\; k\geq 1\label{36}\\
& \Rightarrow\;
\big\L\tilde{\vu}(\tilde{\vx}_k,\tilde{t})\big\R=\big\L e^q
\vu(\vx_k,t)\big\R,\;\text{for all possible configurations $\vu$}\label{37} \\
& \Rightarrow\;\tilde{\vu}(\tilde{\vx}_k,\tilde{t})=e^q
\vu(\vx_k,t)\label{38}\\
& \Rightarrow\;\tilde{\vu}(\tilde{\vx}_1,\tilde{t})\otimes\cdots
\otimes\tilde{\vu}(\tilde{\vx}_n,\tilde{t})=e^{n\cdot q}
\vu(\vx_1,t)\otimes\cdots \otimes\vu(\vx_n,t)\label{39}\\
&
\Rightarrow\;\big\L\tilde{\vu}(\tilde{\vx}_1,\tilde{t})\otimes\cdots
\otimes\tilde{\vu}(\tilde{\vx}_n,\tilde{t})\big\R=\big\L e^{n\cdot
q}
\vu(\vx_1,t)\otimes\cdots \otimes\vu(\vx_n,t)\big\R\label{40}\\
&
\Rightarrow\;\big\L\tilde{\vu}(\tilde{\vx}_1,\tilde{t})\otimes\cdots
\otimes\tilde{\vu}(\tilde{\vx}_n,\tilde{t})\big\R=e^{n\cdot
q}\big\L
\vu(\vx_1,t)\otimes\cdots \otimes\vu(\vx_n,t)\big\R\quad\label{41}\\
& \Rightarrow\; \tilde{\vH}_n=e^{n\cdot q} \vH_n .\label{42}
\end{align}

\subsection{Comment No.1\label{SecD.1}}

In the first step \eqref{35} we identify
$\tilde{\vH}_1=\widetilde{\L\vu_1\R}$ as $\L \widetilde{\vu}_1\R
:= \big\L \tilde{\vu}(\tilde{\vx}_1,\tilde{t})\big\R$. This
conclusion is based on the simple fact that the transformation of
$\tilde{\vH}_1$ is a trivial one, in which all values of $\vH_1$
just get globally scaled by a constant factor $e^q$.

In the {\it general case}, however, a careful distinction must be
made between the two transformed expressions $\widetilde{\L\vu\R}$
and $\L\widetilde{\vu}\R$, since the former directly refers to the
transformed mean velocity field while the latter refers to the
transformed instantaneous (fluctuating) velocity field which is
then averaged, and thus, {\it in general}, is mathematically
distinct from the former expression. But here we are not
considering the case of such a {\it general} variable (point)
transformation
\begin{equation}
\widetilde{t}=\widetilde{t}\,(t,\vx_k,\vH_l),\qquad
\widetilde{\vx}_n=\widetilde{\vx}_n (t,\vx_k,\vH_l),\qquad
\widetilde{\vH}_n=\widetilde{\vH}_n (t,\vx_k,\vH_l),\qquad
k,l=1,\dotsc , n ,\label{62}
\end{equation}
between the independent variables $(t,\vx_n)$ and the dependent
variables $\vH_n$, but only, as given by (\ref{140530:2112}), the
far more simpler {\it specific case} of a globally uniform scaling
in the dependent variables
\begin{equation}
\widetilde{t}=t,\qquad \widetilde{\vx}_n=\vx_n, \qquad
\widetilde{\vH}_n=e^q\vH_n,\label{63}
\end{equation}
which, when written for example for the one-point moment at
$\vx_1=\vx$
\begin{equation}
\widetilde{t}=t,\qquad \widetilde{\vx}=\vx, \qquad
\widetilde{\L\vu\R}=e^{q}\L\vu\R ,\label{64}
\end{equation}
acts as a trivial subset of \eqref{62}. Note that in the following
we only investigate the mathematical property of the
transformation \eqref{64} {\it itself}, i.e.~whether it
additionally represents an equational invariance or not is
irrelevant. In other words, we will investigate \eqref{64} very
generally, solely as a transformation of variables detached from
any underlying transport equations.

Now, it is straightforward to recognize that particularly in this
trivial case \eqref{64}, the two above mentioned transformed
one-point expressions $\widetilde{\vH}_1=\widetilde{\L\vu\R}$ and
$\L\widetilde{\vu}\R$ are identical
\begin{equation}
\widetilde{\L\vu\R}\equiv\L\widetilde{\vu}\R .\label{65}
\end{equation}
This conclusion is based on the following argument, in that we can
write
\begin{align}
\widetilde{\vH}_1=e^q \vH_1 \;\; \Leftrightarrow \;\;
\widetilde{\L
\vu \R}& =e^q \L \vu\R\label{d43}\\
& =\L e^q\vu\R \label{d44}\\
&\underset{\text{def.}}{=:}\L \vu^*\R,\label{d45}
\end{align}
due to the fact that any constant factor as $e^q$ commutes with
every averaging operator~$\L,\R$. Hence one is able to define a
unique transformation relation $\vu\rightarrow \vu^*$ on the
instantaneous level having the {\it same} transformational
structure
\begin{equation}
\vu^*=e^q\vu,\label{66}
\end{equation}
as its averaged value given in \eqref{64}, namely a simple
multiplication of a constant factor $e^q$ on some field
values.\footnote[2]{Note that if \eqref{64} would be additionally
admitted as a symmetry of some mean field transport equations,
then we may {\it not} conclude that \eqref{66} is a symmetry, too,
of the underlying instantaneous (fluctuating) equations. Because,
on the mean field level one can have a symmetric structure which
on the fluctuating level must not exist.} This, then, uniquely
allows us to identify
\begin{equation}
\vu^*=\widetilde{\vu}.\label{140921:1033}
\end{equation}
In other words, since the symbol $\vu^*$ on the left-hand side of
\eqref{66} is defined by the mathematical operation on the
right-hand side (a simple multiplication of a constant factor
$e^q$), and since this mathematical operation is exactly identical
to the right-hand side of the initial transformation \eqref{64},
one can therefore uniquely identify the transformed symbol on the
left-hand side of \eqref{66} with the same transformation symbol
as it's used on the left-hand side of \eqref{64}, i.e.
$\ast\,=\,\sim$.

Again, the reason is that \eqref{64} and \eqref{66} show exactly
the same transformation structure on their right-hand sides,
namely a simple multiplication of a constant factor $e^q$ on some
field values, which then define their left-hand sides. But since
we are dealing here with the same transformational process in
\eqref{66} as in \eqref{64}, we should also explicitly display it,
namely by using $\widetilde{\vu}$ and not $\vu^*$, which would
only unnecessarily overload the notation. Exactly this fact was
implicitly assumed when writing the first line of
(\ref{140127:1405}).

But, as soon as we would consider a more complicated
transformation than \eqref{64}, as for example
\begin{equation}
\widetilde{t}=t,\qquad \widetilde{\vx}=\vx, \qquad
\widetilde{\L\vu\R}=e^{q(\vx)}\L\vu\R ,\label{67}
\end{equation}
where, instead of a globally constant scaling exponent $q$, we now
would have a {\it local} scaling exponent $q(\vx)$ which
explicitly depends on the spatial coordinates, the identification
\eqref{65}, of course, generally no longer holds and becomes
invalid. The reason simply is that in contrast to \eqref{64} the
scaling factor in \eqref{67} is no longer a global constant
anymore which can commute with every averaging operator $\L,\R$.
In other words, since generally
\begin{equation}
\widetilde{\L\vu\R}=e^{q(\vx)}\L\vu\R\neq \L e^{q(\vx)}\vu\R,
\label{68}
\end{equation}
we are no longer able to define a corresponding transformation
relation $\vu\rightarrow \widetilde{\vu}$ on the instantaneous
level which has the {\it same} transformational structure as its
averaged value \eqref{67}. On the contrary, its {\it real}
corresponding transformation rule $\vu\rightarrow \vu^{**}$ will
rather show a far more complex functional structure than given by
\eqref{67}, which in the first instance also cannot be
mathematically determined in a straightforward manner.

Hence, since the situation of proof (\ref{140127:1405}) is not
dealing with a complex situation like \eqref{67}, but only with a
trivial one as \eqref{64}, the notation used throughout
(\ref{140127:1405}) is correct and not misleading.

Note that already from an intuitive point of view the
identification \eqref{65} must be valid if we consider a simple
transformation as \eqref{64}. Because, since in \eqref{64} only
the field values and not the coordinates get transformed we can
perform the following thought experiment: Imagine we have an
ensemble of DNS results for the instantaneous velocity field $\vu$
(hereby it is irrelevant from which specific equations this data
set was numerically generated). From the field $\vu$ we now
construct the mean field $\L \vu\R$ (either as an ensemble average
over a set of different $\vu$, or, if we have a statistically
homogenous direction, over an integral of a single $\vu$ in this
direction). Thus we then obtain all functional values of~$\L
\vu\R$, which we now collectively multiply with a same constant
factor, say by $e^q=2$, to get the new transformed values
$\widetilde{\L\vu\R}$ of \eqref{64}.

Now, the critical question: How should the underlying DNS data for
the instantaneous field $\vu$ be transformed in order to generate
with the same corresponding averaging process the just previously
constructed values $\widetilde{\L\vu\R}$? The intuitive and
correct answer is that all DNS data must be {\it coherently}
multiplied with the same factor $e^q=2$. Only then will the new
transformed data $\widetilde{\vu}$, if it emerges from the
operation $\widetilde{\vu}:=2\cdot \vu$ and if it's
correspondingly averaged to $\L \widetilde{\vu}\R$, give the
ability to reconstruct the functional values
$\widetilde{\L\vu\R}$. Since there is no other option, we hence
obtain within this process the unique result
$\L\widetilde{\vu}\R=\widetilde{\L\vu\R}$. Of course, this
reasoning is only valid for a global (coherent) transformation as
given by \eqref{64}; for a more complicated (local) transformation
as \eqref{67} this reasoning no longer holds.

\subsection{Comment No.2\label{SecD.2}}

The conclusion \eqref{38} is based on the relation \eqref{37}
which goes along with the explicit comment that this relation, by
definition, must hold for {\it all} possible configurations or
functional realizations of the fluctuating velocity field $\vu$,
and not only for any certain functional specification
$\vu=\vu_0(\vx,t)$. In this case, of course, conclusion \eqref{38}
would {\it not} be correct, because for a certain specification
$\vu=\vu_0$, we generally have the situation that
$\L\widetilde{\vu}_0\R=\L \vu_0\R$ although $\widetilde{\vu}_0\neq
\vu_0$.

Now, for the reason that \eqref{37} must hold for {\it all}
possible configurations $\vu$, it is important to recognize that
\eqref{37} is not an equation to be solved for, but that it
represents a definition. And exactly this is the argument when
going from the third \eqref{37} to fourth line \eqref{38}. The
third line
\begin{equation}
\L \widetilde{\vu}(\tilde{\vx}_k,\tilde{t})\R= \L e^q
\vu(\vx_k,t)\R,\label{88}
\end{equation}
does not stand for an equation but for a definition (as it's the
case for any variable transformation in mathematics)
\begin{equation}
\L \widetilde{\vu}(\tilde{\vx}_k,\tilde{t})\R
\,\underset{\text{def.}}{:=}\,\L e^q \vu(\vx_k,t)\R,\label{89}
\end{equation}
since the left-hand side (transformed side) is defined by the
mathematical expression and operation given on the right-hand
side. Now, since the right-hand side by definition must hold {\it
for all} possible (functional) configurations of the instantaneous
velocity field $\vu$, and since both functions $e^q\vu$ and
$\widetilde{\vu}$ undergo the {\it same} operation of averaging,
we thus can only conclude that both functions themselves must be
identical. In other words, definition \eqref{89} implies the
definition
\begin{equation}
\widetilde{\vu}(\tilde{\vx}_k,\tilde{t})
\,\underset{\text{def.}}{:=}\, e^q \vu(\vx_k,t),\label{90}
\end{equation}
in that the transformed instantaneous velocity field
$\widetilde{\vu}$ is defined by the expression $e^q\vu$. This then
gives the fourth line \eqref{38}.

Two things should be noted here. Firstly, the above conclusion is
similar to the arguments which are standardly used in fluid
mechanics when deriving a differential conservation law from its
corresponding integral version. The similarity is given in so far
as the argument for the validity of the integral conservation law
is also based on the requirement {\it `for all'}, however here,
for {\it all} possible volumes or surfaces. Hence the integral
operator from the integral conservation law can be dropped and the
integrand itself is identified as the corresponding conservation
law on the differential level.

Secondly, according to the arguments given in Comment No.1, we are
not obliged to make the direct conclusion \eqref{38} from relation
\eqref{37}. Conclusion \eqref{38} can already be directly obtained
from \eqref{35} by considering the result (\ref{140921:1033}),
i.e. we can directly conclude that
\begin{equation}
\widetilde{\vH}_1=e^q\vH_1 \;\;\Rightarrow\;\;
\widetilde{\vu}(\tilde{\vx}_k,\tilde{t})= e^q
\vu(\vx_k,t),\label{56}
\end{equation}
which just explicitly expresses the fact again that result
(\ref{140921:1033}) was uniquely obtained (for all $\vx_k$ within
the physical space $\vx$) as the induced transformation rule
$\vu\rightarrow\vu^*=\widetilde{\vu}$ (the cause on the
fluctuating level) from the transformation rule of the mean
velocity $\vH_1\rightarrow\widetilde{\vH}_1$ (the effect on the
averaged level).

\subsection{Comment No.3\label{SecD.3}}

In \eqref{42} we identify the transformed $n$-point correlation
function $\widetilde{\vH}_n$ as the transformed expression
$\big\L\widetilde{\vu}(\tilde{\vx}_1,\tilde{t})\otimes\cdots
\otimes\widetilde{\vu}(\tilde{\vx}_n,\tilde{t})\big\R$. This is
just the obvious consequence in knowing the fact that only from
the transformed velocity field $\widetilde{\vu}$, as it is defined
in (\ref{140921:1033}) and thus in \eqref{38}, {\it all}
transformed correlation functions $\widetilde{\vH}_n$ can be
uniquely defined and constructed without inducing contradictions
and without violating the principle of causality. In other words,
making the conclusion
\begin{equation}
\widetilde{\vH}_1=e^q\vH_1 \;\;\Rightarrow\;\;
\widetilde{\vH}^\prime_n=e^{n\cdot q}\vH_n,\label{57}
\end{equation}
where $\widetilde{\vH}^\prime_n$ represents the mean product of
$n$ spatial coordinate evaluations of the transformed and for all
points $\vx_n$ unique instantaneous velocity field
$\widetilde{\vu}=\widetilde{\vu}(\tilde{\vx},\tilde{t})$, and in
which it then gets identified as the transformed $n$-point
correlation function $\widetilde{\vH}_n$, as done in \eqref{42},
\begin{equation}
\widetilde{\vH}_n:=\widetilde{\vH}^\prime_n=\big\L\widetilde{\vu}
(\tilde{\vx}_1,\tilde{t})\otimes\cdots
\otimes\widetilde{\vu}(\tilde{\vx}_n,\tilde{t})\big\R , \label{59}
\end{equation}
is the {\it only} possible conclusion without running into any
contractions and without violating the principle of cause and
effect.

\subsection{Comment No.4\label{SecD.4}}

Note that already from an intuitive point of view only the
conclusion
\begin{equation}
\widetilde{\vH}_1=e^q\vH_1 \;\;\Rightarrow\;\;
\widetilde{\vH}_n=e^{n\cdot q}\vH_n,\label{60}
\end{equation}
given through \eqref{35}-\eqref{42}, makes physically sense, while
the conclusion induced by (\ref{140530:2112})
\begin{equation}
\widetilde{\vH}_1=e^q\vH_1 \;\;\Rightarrow\;\;
\widetilde{\vH}_n=e^{ q}\vH_n,\label{61}
\end{equation}
is physically senseless. This can be seen for example by making
again the following small thought experiment: Imagine we have the
following arbitrary but fixed mean velocity profile $\vH_1=\L
\vu\R$ based on some instantaneous (fluctuating) velocity field
$\vu$. Now, according to the left-hand side of \eqref{60} or
\eqref{61}, if we scale this mean profile $\vH_1$ by, say, a
constant factor $e^q=2$, we will get the two times amplified mean
velocity profile $\widetilde{\vH}_1$. Hereby it should be noted
that this scaling is performed globally, i.e. for {\it all} points
in the considered physical space $\vx$ the mean velocity values
$\vH_1$ are scaled uniformly by a constant factor two.

Intuitively it's clear that a {\it globally} two times higher
amplitude in the mean profile can only go along with a {\it
globally} two times higher amplitude in the instantaneous
velocity. In other words, in order to account for a global scaling
$\widetilde{\vH}_1=2\vH_1$ on the averaged level (the effect), the
underlying instantaneous velocity must transform accordingly
$\widetilde{\vu}=2\vu$ on the fluctuating level (the cause),
otherwise we would not manage to reproduce this coherent
amplification of a factor two on the averaged level.

But now, if the instantaneous velocity $\vu$ globally scales (i.e.
for all points $\vx_n$ in physical space $\vx$) by a factor two,
then e.g. the two-point correlation function $\vH_2$ will globally
scale with a factor $e^{2q}=4$ as given in \eqref{60}, and {\it
not} as in \eqref{61} with the same factor $e^q=2$ as the mean
velocity $\vH_1$ is scaling. Hence the conclusion \eqref{61} is
obviously unphysical.

\section{Statistical scaling of the nonlinear Schrödinger equation\label{SecE}}

This section will demonstrate that the unphysical statistical
scaling invariance $\mathsf{Q}_\mathsf{E}$ (\ref{140530:2112}) is
not specific to the incompressible Navier-Stokes equation
(\ref{140529:1736}), or (\ref{140529:1657}), when transcribed into
its statistical form for the $n$-point velocity correlation
moments $\vH_n$ (\ref{140530:1156}), respectively, either into its
local differential form (\ref{r1})-(\ref{r2}) or into its nonlocal
integro-differential form (\ref{140605:1413}). Because, as a
representative example, the unphysical statistical scaling
invariance $\mathsf{Q}_\mathsf{E}$ (\ref{140530:2112}) is also
admitted~e.g.~by the following cubic nonlinear Schrödinger
equation (see e.g. \cite{Sulem99})
\begin{equation}
i\partial_t \psi +\Delta\psi+\kappa\vert\psi\vert^2\psi=0,
\label{141109:1504}
\end{equation}
when transcribed into its statistical form (in the thermodynamical
sense) for the corresponding (equal-time) multi-point correlation
moments
\begin{equation}
\left. \begin{aligned}
H_1 &=\L\psi_1\R,\\
H_{2n-1} &=\L \psi_1\cdots
\psi_n\cdot\psi^*_{n+1}\cdots\psi^*_{2n-1}\R ,\; n\geq 2,\;
n\in\mathbb{N},
\end{aligned}
~~~\right \} \label{141109:1523}
\end{equation}
of the full instantaneous (fluctuating) scalar wave function
$\psi=\psi(t,\vx)$, where, respectively, $\psi_n$ and $\psi^*_n$
stand for the evaluation of the field and its complex conjugate at
the specific point $\vx=\vx_n$ within the single physical domain
$\vx\in\mathbb{R}^3$. Note that in (\ref{141109:1504}) the
parameter $\kappa$ is defined as a real constant, and that
$\vert\psi\vert^2=\psi\cdot\psi^*$ stands for the square modulus
of the wave function. Further note that the main difference
between the deterministic Navier-Stokes equation
(\ref{140529:1736}) and the deterministic nonlinear Schrödinger
equation (\ref{141109:1504}) is that the former involves a
quadratic nonlinearity while the latter shows a cubic
nonlinearity, thus having the effect that the complete Schrödinger
hierarchy of multi-point moments (\ref{141109:1523}) is ordered by
odd numbers only.

To derive the corresponding transport equations for the
multi-point moments (\ref{141109:1523}) we proceed as given in
\cite{Oberlack10} for the Navier-Stokes equation. In the first
step the necessary deterministic equations for the wave function
and its complex conjugate will be abbreviated as
\begin{equation}
\left. \begin{aligned}
N_\psi & =i\partial_t \psi +\Delta\psi+\kappa\cdot\psi\cdot\psi\cdot\psi^*=0,\\
N_{\psi^*} & =-i\partial_t \psi^*
+\Delta\psi^*+\kappa\cdot\psi\cdot\psi^*\cdot\psi^*=0,
\end{aligned}
~~~\right \} \label{141109:1541}
\end{equation}
in order to then allow, in the second step, the construction of
the statistical transport equations for all moments
(\ref{141109:1523})
\begin{align}
n=1: & \quad\L N_{\psi_1}\R=0,\\
n\geq 2: &\quad \L N_{\psi_1}\cdot
\psi_2\cdots\psi_n\cdot\psi^*_{n+1}\cdots\psi^*_{2n-1}\R + \L
\psi_1\cdot N_{\psi_2}\cdot
\psi_3\cdots\psi_n\cdot\psi^*_{n+1}\cdots\psi^*_{2n-1}\R\nonumber\\
&\quad\hspace{2.25cm} + \cdots + \L \psi_1\cdots\psi_{n-1}\cdot
N_{\psi_n}\cdot\psi^*_{n+1}\cdots\psi^*_{2n-1}\R\nonumber\\
&\quad - \L \psi_1\cdots\psi_n\cdot N_{\psi^*_{n+1}}\cdot
\psi^*_{n+2}\cdots\psi^*_{2n-1}\R - \L \psi_1\cdots\psi_n\cdot
\psi^*_{n+1}\cdot
N_{\psi^*_{n+2}}\cdot\psi^*_{n+3}\cdots\psi^*_{2n-1}\R\nonumber\\
&\quad\hspace{2.25cm} - \cdots - \L \psi_1\cdots\psi_n\cdot
\psi^*_{n+1}\cdots \psi^*_{2n-2}\cdot N_{\psi^*_{2n-1}}\R =0,
\end{align}
which can be condensed and written as the following infinite
hierarchy of linear equations
\begin{align}
n=1: &\quad i\partial_t H_1+\Delta H_1 +\kappa \widehat{H}_3=0,\label{141109:1806}\\
n\geq 2: &\quad i\partial_t H_{2n-1}+\sum_{k=1}^n
\Delta_{\vx_k}H_{2n-1}-\sum_{k=n+1}^{2n-1}
\Delta_{\vx_k}H_{2n-1}\nonumber\\
&\quad\hspace{1.5cm} +\kappa \sum_{k=1}^n \widehat{H}_{k;\,
2(n+1)-1}-\kappa\sum_{k=n+1}^{2n-1} \widehat{H}_{k;\,
2(n+1)-1}=0.\label{141109:1807}
\end{align}
Exactly as in the case of the Navier-Stokes hierarchy, either as
shown in its local differential form (\ref{r1})-(\ref{r2}), or as
in its nonlocal integro-differential form (\ref{140605:1413}), the
corresponding lower dimensional Schrödinger moments
$\widehat{H}_3$ in (\ref{141109:1806}) and $\widehat{H}_{k;\,
2(n+1)-1}$~in~(\ref{141109:1807}) can be uniquely constructed from
the corresponding higher dimensional Schrödinger moments $H_3$ and
$H_{2(n+1)-1}$, too, namely as
\begin{equation}
\widehat{H}_3  = \lim_{\vx_3\to\vx_1,\,\vx_2\to\vx_1}H_3,
\label{141109:2220}
\end{equation}
and for all $n\geq 2$ as
\begin{equation}
\widehat{H}_{k;\, 2(n+1)-1}  = \lim_{\vx_{2n+1}\to\vx_{n+1}}
\left[\lim_{\vx_{2n}\to \vx_{k},\, \vx_{n+1}\to \vx_{k}}
H_{2(n+1)-1}\right],\;\, \text{for $k=1,\dotsc ,2n-1$.}
\label{141109:2221}
\end{equation}
However, since of course the inverse construction
fails,~i.e.~since the higher dimensional moments $H_3$ and
$H_{2(n+1)-1}$ cannot be uniquely constructed from the lower
dimensional moments $\widehat{H}_3$ and $\widehat{H}_{k;\,
2(n+1)-1}$, and therefore, since these latter moments do not {\it
directly} enter the next higher correlation equation in the
hierarchy (\ref{141109:1807}), they have to be identified as
unclosed terms. In particular, since this infinite hierarchy of
Schrödinger moments also does not come along with additional
constraint equations in order to realize the single physical
solution (\ref{141109:1523}), which uniquely emerges from the
underlying deterministic equation (\ref{141109:1504}) as a
multiple evaluation of the single instantaneous (fluctuating) wave
function $\psi=\psi(t,\vx)$, the system of equations
(\ref{141109:1806})-(\ref{141109:1807}) is permanently
underdetermined and therefore unclosed if {\it no} prior modelling
assumption is invoked beforehand (see Appendix~\ref{SecC} for a
more detailed explanation, which, in a one-to-one way, can be
carried over from the Navier-Stokes to the present Schrödinger
case). The unclosed Schrödinger moments $\widehat{H}_3$
(\ref{141109:2220}) and $\widehat{H}_{k;\, 2(n+1)-1}$
(\ref{141109:2221}) thus play the same role as the unclosed
Navier-Stokes moments
$\widehat{H}_{i_{\{n+1\}}[l]}=\lim_{\vx_{n+1}\to\vx_l}H_{i_{\{n+1\}}}$
(\ref{140110:1544}) in the local differential formulation, or
equivalently
$\widehat{\vH}_{i,n+1}=\lim_{\vx_{n+1}\to\vx_i}\vH_{n+1}$
(\ref{140922:1617}) in the nonlocal integro-differential
formulation. However, note that the sequence in the limit process
of (\ref{141109:2221}) is crucial as it's not interchangeable,
i.e.~the inner and outer limits have to be taken only in this
order as given in (\ref{141109:2221}), otherwise the limit process
would give a different result.

Now, since the infinite hierarchy of linear equations
(\ref{141109:1806})-(\ref{141109:1807}) is unclosed it can at most
only admit equivalence transformations. It is straightforward to
see that this system (\ref{141109:1806})-(\ref{141109:1807})
admits the same unphysical scaling equivalence
$\mathsf{Q}_\mathsf{E}$ (\ref{140530:2112}) for the Schrödinger
moments as for the Navier-Stokes moments.~For the Schrödinger
moments (\ref{141109:1523}) the unphysical invariance
$\mathsf{Q}_\mathsf{E}$ thus reads
\begin{align}
\mathsf{Q}_\mathsf{E}: & \;\;\; \tilde{t}=t,\;\;\;
\tilde{\vx}_n=\vx_n,\;\;\;\tilde{H}_{2n-1}=e^q H_{2n-1},\quad
n\geq 1, \label{141110:1427}
\end{align}
which, again, only arises due to the fact that for the underlying
{\it nonlinear} deterministic equation (\ref{141109:1504}) the
corresponding statistical hierarchy of moments
(\ref{141109:1806})-(\ref{141109:1807}) is misleadingly formulated
as a {\it linear} system.~In other words, as was thoroughly
explained and demonstrated in this study at the example of the
Navier-Stokes moments, the linear statistical formulation
(\ref{141109:1806})-(\ref{141109:1807}) for the Schrödinger
moments is misleading too when employing an invariance analysis
upon them, since one directly obtains the inconsistent and
unphysical result (\ref{141110:1427}).

Note again that as the invariance (\ref{141110:1427}) only scales
the dependent variables while leaving the coordinates unchanged,
the transformation rule for the lower dimensional (unclosed)
moments $\widehat{H}_3$ (\ref{141109:2220}) and $\widehat{H}_{k;\,
2(n+1)-1}$ (\ref{141109:2221}) is identical to the transformation
rule (\ref{141110:1427}) for the corresponding higher dimensional
moments $H_3$ and $H_{k;\, 2(n+1)-1}$, i.e.
\begin{align}
\widetilde{\widehat{H}}_3 &=e^q \widehat{H}_3,\\
\widetilde{\widehat{H}}_{k;\, 2(n+1)-1} &= e^q \widehat{H}_{k;\,
2(n+1)-1},\;\text{for $k=1,\dotsc ,2n-1$, and $n\geq 2$},
\end{align}
as it was also respectively derived in more detail for the
corresponding lower dimensional (unclosed) Navier-Stokes moments
in (\ref{140607:2309}).~Finally note that when decomposing the
instantaneous wave function into its mean and purely fluctuating
field $\psi=\L\psi\R+\psi^\prime$, the equivalence transformation
for the instantaneous moments (\ref{141110:1427}) will then
bijectively change to the corresponding more detailed
representation (\ref{140222:1123}), from which, then, the
artificial transformation behavior of (\ref{141110:1427}) is
immediately recognized.

To conclude, this demonstrating example makes it clear that the
scaling invariance $\mathsf{Q}_\mathsf{E}$ (\ref{140530:2112}) for
the statistical moments is neither specific to the Navier-Stokes
nor specific to any nonlinear Schrödinger equation. It is just
(besides its unphysical nature) only a non-specific and thus a
non-significant equivalence transformation, since it will be
admitted by any nonlinear deterministic system which necessitates
a statistical description (in the thermodynamical sense) for its
solution manifold when considering a certain hierarchy of
multi-point moments.

\section{Generating scaling laws from invariant transformations\label{SecF}}

According to \cite{Oberlack10} the key invariant Lie-point
transformations to generate useful statistical scaling laws for
wall-bounded flows in the inertial region are, at first, the
physical translation symmetry in the space-time coordinates
\begin{align}
\mathsf{T}: & \;\;\; \tilde{t}=t+c_{0,0},\;\;\; \tilde{\vx}_n =
\vx_n+\vc_{0,n}, \;\;\; \tilde{\vH}_n= \vH_n,\label{140303:1246}
\end{align}
and the two physical scaling symmetries of the inviscid Euler
equations, which inherently translate to the instantaneous
multi-point functions in their most general form as
\begin{align}
\mathsf{S}_\mathsf{1}: & \;\;\; \tilde{t}=e^{a_1}t,\;\;\;
\tilde{\vx}_n = \vx_n, \;\;\;\hspace{0.45cm}
\tilde{\vH}_n=e^{-n\cdot a_1}
\vH_n,\label{140110:1759}\\
\mathsf{S}_\mathsf{2}: & \;\;\; \tilde{t}=t,\;\;\;\hspace{0.47cm}
\tilde{\vx}_n = e^{a_2}\vx_n, \;\;\; \tilde{\vH}_n=e^{n\cdot
a_2}\vH_n,
\end{align}
and then finally the two new {\it unphysical} invariant
transformations, which are not reflected by the deterministic
Euler or Navier-Stokes equations (\ref{140529:1736}), as they only
apply to the notationally oversimplified statistical transport
equations (\ref{140601:1140})
\begin{align}
\phantom{xxxx}\mathsf{Q}_\mathsf{1}: & \;\;\; \tilde{t}=t,\;\;\;\;
\tilde{\vx}_n = \vx_n,\;\;\;\; \tilde{\vH}_n=e^{q}\vH_n,
\label{140110:1748}\\
\mathsf{Q}_\mathsf{2}: & \;\;\; \tilde{t}=t,\;\;\;\; \tilde{\vx}_n
= \vx_n,\;\;\;\; \tilde{\vH}_n=\vH_n+\vc_{1,n},
\label{140110:1749}
\end{align}
where only the globally invariant {\it scaling}
$\mathsf{Q}_\mathsf{1}=\mathsf{Q}_\mathsf{E}$ (\ref{140530:2112})
has been explicitly exposed in this study as an unphysical
transformation. However, by using the procedure developed in
Section \ref{Sec4}, in particular (\ref{140127:1405}), it is
straightforward to also expose the globally invariant {\it
translation} $\mathsf{Q}_\mathsf{2}$ (\ref{140110:1749}) as an
unphysical transformation. Up to now, no higher level functional
Hopf-symmetry has been found which can induce this translation
invariance via (\ref{140127:1615}), as it is the case for the
scaling invariance $\mathsf{Q}_\mathsf{1}$ (\ref{140110:1748})
which is induced by the Hopf-symmetry $\mathsf{Q}$
(\ref{140124:1833}). It is the non-holonomic constraint in
(\ref{140124:2059}) which mostly breaks every symmetry, as it is
too restrictive to find a symmetry which is compatible to it for
all times $t\geq 0$. Note that in all five transformations
(\ref{140303:1246})-(\ref{140110:1749}) we suppressed the
transformation rule for all pressure correlations, in contrast to
the listing in \cite{Oberlack10}, since we only consider the
solenoidal (nonlocal) statistical transport equations
(\ref{140601:1140}), based on the deterministic form
(\ref{140529:1657}) in which the pressure got eliminated by the
continuity equation. Indeed, both $\mathsf{Q}_\mathsf{1}$
(\ref{140110:1748}) and $\mathsf{Q}_\mathsf{2}$
(\ref{140110:1749}) are admitted by the underlying (unclosed)
Friedmann-Keller equations (\ref{140601:1140}) as equivalence
transformations.

Already these few invariant Lie-group transformations in their
most general form (\ref{140303:1246})-(\ref{140110:1749}) allow
now to generate a vast range of turbulent scaling laws. Our
interest here however is to focus only on the considerably smaller
set of scaling laws for one-point statistics. To generate these
scalings, the above transformations for multi-point correlations
has to be reduced to one-point statistics, as was also done in
\cite{Oberlack10} by performing the smooth and regular limit
$(\vx_n-\vx_1)\rightarrow \v0$, for all $n\geq 2$.

Furthermore, since we are also only interested in the scaling
behavior of geometrically simple wall-bounded flows in the
inertial region, we will additionally perform next to the
one-point limit the following approximation: According to
\cite{Oberlack01,Oberlack10} and \cite{Lindgren04} the flow in the
inertial region between the inner (near-wall) and outer region is
approximated as a {\it stationary inviscid parallel shear flow}.
Under this assumption the wall-normal velocity component can be
neglected, and all fields only depend on one independent
coordinate, the wall-normal coordinate $y$, described then by an
inviscid set of balance equations.

If we now consider the specific flow configuration of a ZPG
turbulent boundary layer flow, in order to compare to the DNS
results in Section \ref{Sec5}, the invariant transformations
(\ref{140303:1246})-(\ref{140110:1749}) will then simplify even
further. Due to spanwise homogeneity and a spanwise reflection
symmetry in the flow, the mean spanwise velocity as well as all
moments involving an uneven number of spanwise velocity fields
vanish.

In this approximation within the one-point limit for ZPG turbulent
boundary layer flow, the above transformations
(\ref{140303:1246})-(\ref{140110:1749}) reduce to
\begin{equation}
\left. \begin{aligned}
\mathsf{T}: & \;\;\;  \tilde{y}=y+c_0^2,\;\;\; \tilde{U}=U,\;\;\;
\tilde{H}^{ij}=H^{ij},\;\;\;
\tilde{H}^{ijk}=H^{ijk},\\
\mathsf{S}_\mathsf{1}: & \;\;\;  \tilde{y}=y,\;\;\;
\tilde{U}=e^{-a_1}U,\;\;\; \tilde{H}^{ij}=e^{-2a_1}H^{ij},\;\;\;
\tilde{H}^{ijk}=e^{-3a_1}H^{ijk},\\
\mathsf{S}_\mathsf{2}: & \;\;\; \tilde{y}=e^{a_2}y,\;\;\;
\tilde{U}=e^{a_2}U,\;\;\; \tilde{H}^{ij}=e^{2a_2}H^{ij},\;\;\;
\tilde{H}^{ijk}=e^{3a_2}H^{ijk},\\
\mathsf{Q}_\mathsf{1}: & \;\;\; \tilde{y}=y,\;\;\;
\tilde{U}=e^{q}U,\;\;\; \tilde{H}^{ij}=e^{q}H^{ij},\;\;\;
\tilde{H}^{ijk}=e^{q}H^{ijk},\\
\mathsf{Q}_\mathsf{2}: & \;\;\; \tilde{y}=y,\;\;\;
\tilde{U}=U+c_1^1,\;\;\; \tilde{H}^{ij}=H^{ij}+c_1^{ij},\;\;\;
\tilde{H}^{ijk}=H^{ijk}+c_1^{ijk},\\
\end{aligned}
~~~\right \} \label{140603:1900}
\end{equation}
where we considered the chain only up to third moment $(n\leq 3)$,
and where the number of superscript indices on each term denotes
the tensor rank of the corresponding variable, i.e. $\vx=(x^i)$,
$\vH_1=(H^i)$, $\vH_2=(H^{ij})$, $\vH_3=(H^{ijk})$, etc., with the
short-hand notation $x^2=y$ and $H^1=U$.

Hence the corresponding invariant surface condition
\citep{Olver93,Ibragimov94,Bluman96} to generate invariant
functions from (\ref{140603:1900}) reads
\begin{align}
\frac{dy}{a_2\cdot y+c_0^2} & = \frac{dU}{(-a_1+a_2+q)U+c_1^1}
\nonumber\\
&= \frac{dH^{ij}}{(-2a_1+2a_2+q)H^{ij}+c_1^{ij}}
=\frac{dH^{ijk}}{(-3a_1+3a_2+q)H^{ijk}+c^{ijk}_1}.
\label{130820:2209}
\end{align}
To note is that in the above expression all indices are open and
not contracted, i.e. they are not being summed over, and, due to
the symmetries in the flow, only those indexed moments give a
contribution which involve an even number of spanwise velocity
fields.

According to \cite{Oberlack01,Oberlack10} and \cite{Lindgren04}
equation (\ref{130820:2209}) provokes  a further critical
assumption. The line of argumentation is, since the friction
velocity at the wall $U_\tau\sim\sqrt{\partial_yU\vert_{y=0}}$ can
be seen as an external parameter or boundary condition which
inhibits the free scaling of the streamwise velocity $U$, the
corresponding scaling invariance must get broken, i.e. the scaling
coefficient of $U$ must be put to zero: $-a_1+a_2+q=0$. As in
\cite{Oberlack10} we choose solving for $a_1=a_2+q$. Inserting
this restriction and dividing the equation by $a_2\neq 0$ gives
instead of (\ref{130820:2209}) the more simplified zero-surface
condition
\begin{equation}
\frac{dy}{y+ \bar{c}_0^2}  = \frac{dU}{\bar{c}_1^1} =
\frac{dH^{ij}}{-\bar{q}\cdot H^{ij}+\bar{c}_1^{ij}}
=\frac{dH^{ijk}}{-2\bar{q}\cdot H^{ijk}+\bar{c}^{ijk}_1},
\label{131227:1036}
\end{equation}
where all overbared constants represent the original constants
relative to $a_2$. Finally, solving these equations lead to the
following set of invariant scaling laws in the \emph{inertial
region} on the basis of the full \emph{instantaneous}-fields
\begin{equation}
\left. \begin{aligned} U(y)= \alpha_U\cdot\ln(y+c)+\beta_U,\hspace{2.75cm}\\
H^{ij}(y)=\alpha_H^{ij}+\beta_H^{ij}\cdot(y+c)^\gamma,\;\;
H^{ijk}(y)=\alpha_H^{ijk}+\beta_H^{ijk}\cdot(y+c)^{2\gamma},
\end{aligned}
~~~\right \} \label{130821:1043}
\end{equation}
where all the $\beta$'s are integration constants, whereas the
rest of the parameters comprise the group constants relative to
$a_2$ as: $\gamma=-\bar{q}$, $c=\bar{c}_0^2$,
$\alpha_U=\bar{c}^1_1$, $\alpha_H^{ij}=\bar{c}^{ij}_1/\bar{q}$,
and $\alpha_H^{ijk}=\bar{c}^{ijk}_1/2\bar{q}$. Note the strong
dependence of the scaling laws (\ref{130821:1043}) on the two {\it
unphysical} invariance transformations $\mathsf{Q}_\mathsf{1}$
(\ref{140110:1748}) and $\mathsf{Q}_\mathsf{2}$
(\ref{140110:1749}). Except for the parameter $\bar{c}^2_0$, all
other parameters stem from $\mathsf{Q}_\mathsf{1}$ and
$\mathsf{Q}_\mathsf{2}$.

Finally, using the Reynolds decomposition as given in
(\ref{140115:1947}) and (\ref{140603:2126}) will give the
transformation rule for the corresponding moments of the {\it
fluctuating} fields
\begin{equation}
\left. \begin{aligned} U(y) & = \alpha_U\cdot\ln(y+c)+\beta_U,\\
\tau^{ij}(y) & =
\alpha_H^{ij}+\beta_H^{ij}\cdot(y+c)^{\gamma}-\delta^{1i}\delta^{1j}U(y)^2,\\
T^{111} & = \alpha_H^{111}+\beta_H^{111}\cdot(y+c)^{2\gamma}
-3U(y)\cdot\tau^{11}(y)-U(y)^3,\\
T^{112} & = \alpha_H^{112}+\beta_H^{112}\cdot(y+c)^{2\gamma}
-2U(y)\cdot\tau^{12}(y),\\
T^{ij1} & = \alpha_H^{ij1}+\beta_H^{ij1}\cdot(y+c)^{2\gamma}
-U(y)\cdot\tau^{ij}(y),\;\text{for}\;
(i,j)=(2,2),(3,3),\\
T^{ij2} & =
\alpha_H^{ij2}+\beta_H^{ij2}\cdot(y+c)^{2\gamma},\;\text{for}\;
(i,j)=(2,2),(3,3).
\end{aligned}
~~~ \right \} \label{130821:1404}
\end{equation}
We see that the above set of invariant scaling functions is lead
by a generalized log-law in the mean velocity profile $U$, which
differs from the classical von Kármán log-law by the presence of a
constant shift $c$. This non-classical log-law was first obtained
in \cite{Oberlack01}, however, on basis of a different invariance
procedure as developed in \cite{Oberlack10} and as presented
herein.\footnote[2]{To note is that the Lie-group based derivation
of the generalized log-law as given in (\ref{130821:1404}) is not
only heavily misleading in \cite{Oberlack10}, but also in
\cite{Oberlack01}. While the derivation in \cite{Oberlack10} is
based on an unphysical invariance, the derivation in
\cite{Oberlack01} is based on an incorrectly concluded invariance
\citep{Frewer14.2}.} In \cite{Lindgren04} it is referred to as the
{\it modified log-law} extending the predictability in the lower
end of the inertial region. By construction all parameters are
completely independent of the Reynolds number, which obviously
stems from the strong approximation that the inertial region was
identified as an inviscid parallel shear layer.

Important to note in (\ref{130821:1404}) are the new results for
the streamwise Reynolds stress $\tau^{11}$ and for several triple
moments $T^{ijk}$, which, due to the presence of the mean
streamwise velocity profile $U$, all scale in the inertial region
as a combination of a log-law and a power-law. It should be clear
that this peculiar scaling behavior simply and only has its
origins in the unphysical invariance transformations
$\mathsf{Q}_\mathsf{1}$ (\ref{140110:1748}) and
$\mathsf{Q}_\mathsf{2}$ (\ref{140110:1749}), which again are due
to having performed an invariance analysis on the notational
oversimplified set of equations~(\ref{140601:1140}), which
misleadingly represent themselves as a linear gradient-type set of
equations due to not incorporating the underlying deterministic
equations into the analysis. It should also be clear that this
unphysical scaling in the streamwise direction is not restricted
to one-point quantities only, but essentially is incorporated in
all multi-point~quantities.

Finally also note again that, independent of the fact that the
scaling laws (\ref{130821:1043}) or (\ref{130821:1404}) are based
on two unphysical invariant transformations, it would be more than
misleading to identify these invariant functions as a set of
privileged \emph{solutions} to the underlying unmodelled system
(\ref{140601:1140}). As it was already discussed in Section
\ref{Sec2.1}, the reason is that the unclosed system
(\ref{140601:1140}) is not arbitrarily underdetermined, but
underdetermined in the sense that all its unknown terms can be
physically and uniquely determined from the underlying but yet
analytically not accessible instantaneous (fluctuating) velocity
field~$\vu$ according to (\ref{140530:1156}).

Hence, the result (\ref{130821:1043}) or (\ref{130821:1404}) can
only be interpreted as a set of functional relations which stay
invariant under the global group transformations
(\ref{140303:1246})-(\ref{140110:1749}), and which {\it possibly
but not necessarily} could scale the region of the inertial layer
within ZPG turbulent boundary layer flow. But, since there can be
only {\it one} physical realization for {\it all} moments, which
{\it all} are driven by the {\it same} single deterministic
velocity field $\vu$ according to (\ref{140530:1156}), and since
the performed statistical invariance analysis in \cite{Oberlack10}
did not appropriately involve this underlying deterministic layer
of description, the chances are extremely low that exactly this
determined set (\ref{130821:1043}) or (\ref{130821:1404}) of
invariant functions should represent the statistically correct and
thus for all correlation orders consistent solution to the complex
inertial scaling problem of incompressible wall-bounded turbulence
(aside from the fact, of course, that these functions are
additionally based on unphysical reasoning and thus being
physically void).~The obvious reason for this negative result is
that currently no method (including the invariant Lie-group
method) exists to establish a profound and at the same time an
{\it analytically} accessible and correct connection between the
deterministic and the statistical description of the wall-bounded
Navier-Stokes~theory.

\bibliographystyle{jfm}
\bibliography{BibDaten}

\end{document}